\documentclass[12p]{iopart}

\usepackage[english]{babel}
\usepackage[utf8x]{inputenc}
\usepackage[T1]{fontenc}
\usepackage{braket}

\usepackage[a4paper,top=3cm,bottom=2cm,left=3cm,right=3cm,marginparwidth=1.75cm]{geometry}

\usepackage{amssymb}
\usepackage{bm}
\usepackage{graphicx}
\usepackage{xcolor}
\usepackage{epstopdf}
\usepackage{multirow}
\usepackage{hyperref}

\newcommand{\nuc}[2]{\ensuremath{^{#1}\mathrm{#2}}}

\newcommand{\NNLO}{NNLO}

\newcommand{\comm}[2]{\ensuremath{[{#1},{#2}]}}

\newcommand{\op}[1]{\ensuremath{#1}}
\newcommand{\adj}[1]{\ensuremath{{{#1}}^{\dag}}}

\newcommand{\aO}{\ensuremath{\op{a}}}
\newcommand{\aaO}{\ensuremath{\adj{\op{a}}}}

\newcommand{\SpR}[1]{\ensuremath{\mathrm{Sp}( #1,\mathbb{R} )}}
\newcommand{\betb}{\begin{tabular}{p{4.0cm}p{9.0cm}}}
\newcommand{\entb}{\end{tabular}}

\newcommand{\red}[1]{{#1}}
\begin{document}

\title{White paper: From bound states to the continuum}

\author{Calvin W. Johnson$^1$, Kristina D. Launey$^2$, Naftali Auerbach$^3$, Sonia Bacca$^4$, Bruce R. Barrett$^5$,  
Carl Brune$^6$,
Mark A.~Caprio$^7$,
Pierre Descouvemont$^8$, W. H. Dickhoff$^9$,
Charlotte Elster$^6$,
Patrick J.~Fasano$^7$,
Kevin Fossez$^{11}$,
Heiko Hergert$^{11,12}$,
Morten Hjorth-Jensen$^{11,12}$,
Linda Hlophe$^{11}$,
Baishan Hu$^{13}$,
Rodolfo M. Id Betan$^{14}$,
Andrea Idini$^{15}$,
Sebastian K\"onig$^{16,26}$,
Konstantinos Kravvaris$^{17}$,
Dean Lee$^{11,12}$,
Jin Lei$^{6}$,
Alexis Mercenne$^{2}$,
Rodrigo Navarro Perez$^{1}$,
Witold Nazarewicz$^{10,12}$,
Filomena M. Nunes$^{11,12}$,
Marek P{\l}oszajczak$^{18}$,
Jimmy Rotureau$^{11,15}$,
Gautam Rupak$^{20}$,
Andrey M. Shirokov$^{21}$,
Ian Thompson$^{19}$,
James P. Vary$^{22}$,
Alexander Volya$^{17}$,
Furong Xu$^{13}$,
Remco G.T. Zegers$^{11,12,23}$,
Vladimir Zelevinsky$^{10,11,12}$,
Xilin Zhang$^{24,25}$}
\address{$^1$ Department of Physics, San Diego State University, San Diego, CA 92182, USA} 
\address{$^2$Department of Physics and Astronomy, Louisiana State University, Baton Rouge, LA 70803, USA} 
\address{$^3$School  of  Physics  and  Astronomy, Tel  Aviv  University,  Tel  Aviv 69978,  Israel}
\address{$^4$Institut f\"ur Kernphysik and PRISMA Cluster of Excellence, Johannes Gutenberg-Universit\"at Mainz, 55128 Mainz, Germany}
\address{$^5$Department of Physics, University of Arizona, Tucson, Arizona 85721}
\address{$^6$Institute of Nuclear and Particle Physics,  and
Department of Physics and Astronomy,  Ohio University, Athens, OH 45701,
USA}
\address{$^7$Department of Physics, University of Notre Dame, Notre Dame, IN 46556, USA}
\address{$^8$Physique Nucleaire Theorique et Physique Mathematique, C.P. 229, Universite Libre de Bruxelles (ULB), B 1050 Brussels, Belgium}
\address{$^9$Department of Physics, Washington University in St. Louis, MO 63130 USA}
\address{$^{10}$FRIB Laboratory, Michigan State University, East Lansing, MI 48824}
\address{$^{11}$National Superconducting Cyclotron Laboratory, Michigan State University, East Lansing, MI 48824}
\address{$^{12}$Department of Physics and Astronomy, Michigan State University, East Lansing, MI 48824}
\address{$^{13}$State Key Laboratory of Nuclear Physics and Technology, School of Physics, Peking University, Beijing 100871, China}
\address{$^{14}$Physics Institute of Rosario, S2000EZP Rosario, Argentina}
\address{$^{15}$Division of Mathematical Physics, Department  of  Physics, LTH, Lund  University,  P.O.  Box  118,  S-22100  Lund,  Sweden}
\address{$^{16}$Inst. f. Kernphysik, TU Darmstadt, 64289 Darmstadt, Germany}
\address{$^{17}$Department of Physics, Florida State University, Tallahassee, FL 32306, USA}
\address{$^{18}$GANIL, CEA/DRF-CNRS/IN2P3, BP 55027, F-14076 Caen Cedex, France}
\address{$^{19}$Lawrence Livermore National Laboratory, P.O. Box 808, L-414, Livermore, California 94551, USA}
\address{$^{20}$Department  of Physics \& Astronomy and HPC Center for Computational  Sciences, Mississippi  State  University,  Mississippi  State,  MS  39762,  USA}
\address{$^{21}$Skobeltsyn Institute of Nuclear Physics, Moscow State University, Moscow 119991, Russia}
\address{$^{22}$Department of Physics and Astronomy, Iowa State University, Ames, IA 50011, USA}
\address{$^{23}$Joint Institute for Nuclear Astrophysics, Center for the Evolution of the Elements, East Lansing, MI 48824}
\address{$^{24}$Physics Department, University of Washington, Seattle, WA 98195, USA}
\address{$^{25}$Department of Physics, the Ohio State University}
\address{$^{26}$Department of Physics, North Carolina State University,
Raleigh, NC 27695, USA}

\begin{abstract}
This white paper reports on the discussions 
of the 2018 Facility for Rare Isotope Beams Theory Alliance (FRIB-TA) topical program ``From bound states to the continuum: Connecting bound state calculations with scattering and reaction theory''. One of the biggest and most important frontiers in nuclear theory today is to construct better and stronger bridges 
between bound state calculations and calculations in the continuum, especially scattering  and reaction theory, as well as teasing out the influence of the continuum on states near threshold.  This is particularly 
challenging as  many-body structure calculations typically use a bound state basis, while reaction calculations  more commonly utilize 
few-body continuum approaches. The many-body bound state and few-body continuum   methods  use different language and emphasize different 
properties.  To build better foundations for these bridges, we present an overview of several bound state and continuum methods and, 
where possible, point to current and possible future connections. 
\end{abstract}

\maketitle

\noindent
\emph{
The presentations and discussions at 2018 Facility for Rare Isotope Beams Theory Alliance (FRIB-TA) topical program ``From bound states to the continuum: Connecting bound state calculations with scattering and reaction theory'' encompassed many, but certainly not all,  topics and current challenges related to the workshop scope. Many topics were subjects of active debates, and the text presents specific viewpoints that 
may not necessarily reflect the views of all authors. 
This only implies that future theoretical and experimental work is necessary.
}

\section{Motivation and context}

Rutherford discovered the atomic nucleus through scattering. 
Today we obtain data on nuclei through scattering and reactions 
experiments, either with other nuclei or electromagnetic probes, 
and via decays. 
Many, if not most, experiments measure  {cross sections},
  dictating the critical need for reliable and consistent 
  quantum mechanical theories for calculating reaction observables. 

Low-energy nuclear theory has been invigorated by the introduction and promulgation of novel, rigorous theoretical methods for many-body bound states, allowing for broadly successful
$A$-body 
calculations
(we present an illustrative, albeit not exhaustive, list  in Secs. \ref{manybody} and \ref{connections}). The low-energy community has also become aware of the need for similarly improved calculations for low-energy scattering and reaction theory, particularly for interpreting experiments at 
rare isotope beam facilities around the world.  There are only a handful of dedicated low-energy reaction theorists at work today, however, with  more effort going to structure calculations using bound-state frameworks. In addition, current few-body reaction frameworks often approximate the microscopic many-body structure of the target/projectile, while current many-body structure methods typically limit or entirely neglect continuum degrees of freedom. In between are methods to connect structure calculations to reaction observables.

\begin{figure}[h]
\centering
\includegraphics[width=0.5\textwidth]{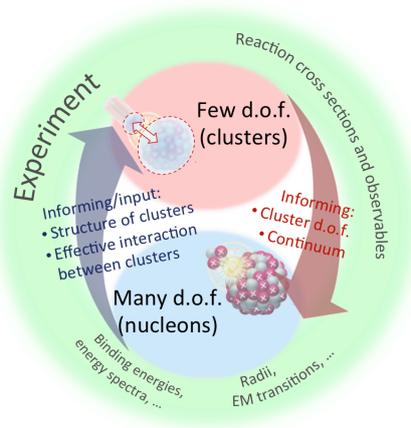}
\caption{\label{fig:workshop_overview} The goal of the FRIB-TA topical program was to help develop tools to calculate reaction observables that are directly measured in experiments, primarily by expanding existing state-of-the-art few-body and many-body theories but also by motivating the development of innovative approaches that can build upon connections between these theories.}
\end{figure}

This paper reports on the presentations and discussions of the 2018 Facility for Rare Isotope Beams Theory Alliance (FRIB-TA) workshop ``From bound states to the continuum: Connecting bound state calculations with scattering and reaction theory''. The goal of the workshop was to 
discuss current and future 
tools to calculate reaction observables that are directly measured in experiments, primarily by expanding existing state-of-the-art few-body and many-body theories but also by motivating the development of innovative approaches that can build upon connections between these theories (Fig. \ref{fig:workshop_overview}). In particular, the main themes were:
(1) to bridge from many-body theories based  on bound state formalisms to continuum degrees of freedom, and (2) to expand few-body techniques to include microscopic degrees of freedom.  This paper discusses scattering and reaction theory, bound state structure calculations, and especially work on the interface, with the aim to identify and lower the technical barriers to bridge between bound state calculations and the continuum.

The program discussed the accomplishments and limitations of various methods, and how well these methods or combinations of the methods can address important experimental questions. Among the issues covered,
which by no means exhausts all those of importance, were:
\begin{itemize}
\item The commonplace use of localized basis in many bound-state calculations and the need to
use and advance hybrid degrees of freedom;
\item The need to address collectivity, clustering, and non-resonant continuum; 

\item The critical need for reliable effective inter-cluster interactions, often called ``optical potentials,''  that can be employed in many currently available reaction codes used by theorists and experimentalists;
\item The key role of thresholds (energy differences) and asymptotic normalization coefficients (ANC) in describing reactions;
\item The importance of correlations in the nuclear wave functions, such as clustering, collectivity, and coupling to the continuum;
\item The complementary roles of 
approaches that work at different scales and resolution,
and experimental data;
\item The important support of experiment that can provide measurements on a grid of the nuclear chart allowing for theoretical interpolations, with a focus on masses. In particular, theory may lead to smaller uncertainties if it interpolates between experimental data, instead of extrapolating to very neutron-rich nuclei.
\end{itemize}

\red{In Sec.~\ref{sec.anaoqs} we give a brief overview of nuclei as open quantum systems, a theme underlying much 
of this work.  Then in Sec.~\ref{expt} we review the experimental context, especially the astrophysical 
and nuclear structure drivers in Sec.~\ref{sec.nucl_astro}.  A specific example is charge-exchange reactions,
discussed in Sec.~\ref{CEX}, that provide important input to neutrino physics, but the interpretation of those 
experiments is constrained by the quality of understanding the reactions.  

In the rest of this paper, we review theoretical approaches, beginning with a broad overview of  few-body methods in Sec.~\ref{sec.fewbody}, which typically emphasize continuum degrees of freedom, and then in Sec.~\ref{manybody} 
discuss several current many-body methods, which primarily, though not exclusively, are built from 
bound single-particle states.  Finally, in Sec.~\ref{connections} we discuss a number of approaches to 
connecting bound state methods with continuum degrees of freedom. } Some specific technical details for getting started are included in the appendices.

\subsection{Atomic nuclei as open quantum systems}\label{sec.anaoqs}

\begin{figure}[htb]
	\centering
	\includegraphics[width=0.45\textwidth]{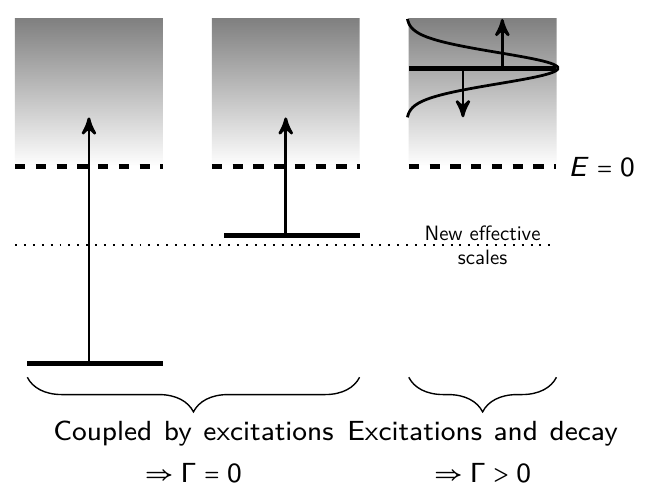}
    \caption{Three typical cases in quantum systems: a well bound state \red{(left-hand panel)}, a weakly bound state \red{(middle panel)}, and a decaying resonance state \red{(right-hand panel)}. In the first two cases, couplings between the discrete state and continuum states are only possible through excitations, 
    while in the third case both excitations and decays are possible ways to couple to continuum states. Because of the strong continuum couplings in the last case, many continuum states have a structure very similar to the discrete state, which gives the characteristic energy dispersion or width observed experimentally.}
    \label{fig_cont_coupl}
\end{figure}

One of the major goals of nuclear physics is the exploration of the drip lines, defined as the limit of nuclear stability with respect to the emission of one neutron or proton, and provide information on how many neutrons and protons can stick together.
The main theoretical difficulty in the description of nuclei close to the drip lines and beyond comes from the emergence of new effective scales associated with the increasing importance of couplings to continuum states, represented in Fig.~\ref{fig_cont_coupl}. 
Indeed, couplings to continuum states are often neglected when describing well bound states, as considerable energy would be required to break such states apart \red{(left-hand panel of Fig.~\ref{fig_cont_coupl})}, leading to well-localized wave functions. Weakly bound systems \red{(middle panel of Fig.~\ref{fig_cont_coupl})}, however, do not require much energy to break apart; this translates into localized but extended wave functions, as if weakly bound systems were ready for the emission of one or more particles.

Above the particle-emission threshold \red{(right-hand panel of Fig.~\ref{fig_cont_coupl})}, the situation is more complex. One particle or more can leave the system and consequently the process is intrinsically time-dependent. However, it is known experimentally that when scattering a neutron or a proton on a nucleus at low (positive) energies, there are some specific energies at which the absorption increases significantly, which give peaks of various widths in the cross section.  These are \textit{resonances}, and in 
a time-dependent picture  the scattered nucleon spends a significant amount of time around the target before departing.

Coupling to the environment of scattering states and decay channels cannot be taken into account by an appropriate modification of the Hermitian Hamiltonian of a closed quantum system \cite{Michel_2010}. The matrix problem involving discrete and continuum states is complex-symmetric and leads to new phenomena such as, e.g.,  resonance trapping \cite{ref1a,ref1b,ref1c,ref2} and  super-radiance \cite{ref3a,ref3b},  multichannel coupling effects in reaction cross sections \cite{ref4a,ref4b} and shell occupancies \cite{ref5a,ref5b}, the modification of spectral fluctuations \cite{ref6}, and  deviations from Porter-Thomas resonance widths distribution \cite{ref2,ref7a,ref7b}.
The appearance of collective states and clustering close to the corresponding cluster emission threshold is yet another consequence of the continuum coupling \cite{Barker_1964,clusterA,clusterB}.
Ikeda et al. \cite{ikeda} observed that $\alpha$-cluster states can be found in the proximity of $\alpha$-particle decay thresholds. 
It has been conjectured \cite{clusterA,clusterB,2,lee} that the interplay between internal configuration mixing by interactions and external configuration mixing via decay channels leads to 
this near-threshold collectivity.  
This may explain
why many states, both on and off the nucleosynthesis path, exist ``fortuitously'' close to open channels.  Given their clear importance, cluster states have been extensively explored within various cluster models (for a review, see \cite{FreerHKLM18}), but in many cases, remain a challenge to many-body theory, especially as they do not arise easily out of a spherical shell model.

\subsection{Experimental context}

\label{expt}

While the driver for this paper is experiment, it is useful for us to  
identify two broad classes of measurements. The first we call \textit{direct measurements},
where the quantity being measured is directly of physical interest. These include, for example, cross sections of interest to astrophysics, scattering phase shifts for constraining interactions, and excitation energies.  The second are \textit{interpreted} measurements, where one uses  reactions or scattering to get at some other quantity of physical interest (Table \ref{tab:reactiontypes}). These include charge-exchange, transfer, breakup, and knockout reactions, Coulomb excitations, and so on. The interpretation of these experiments depend significantly upon the  reaction theory used, and so robust interpretations require robust theory. 

\begin{table}
\begin{center}
    \begin{tabular}{ |l|c|p{6cm}| } 
    \hline
    Reaction & Example & Nuclear information \\
    \hline
    \multirow{3}{*}{Elastic Scattering} & A(a,a)A & \multirow{3}{6cm}{Extracts effective interactions (optical potentials), interaction radii,
    density distributions}\\
    & \nuc{208}{Pb}(n,n)\nuc{208}{Pb} &  \\
    & & \\
    \hline
    \multirow{2}{*}{Inelastic Scattering} & A(a,a$'$)A$^{*}$ &  \multirow{2}{6cm}{Extracts electromagnetic transitions or
    nuclear deformation}\\ 
    & \nuc{90}{Zr}($\alpha,\alpha'$)\nuc{90}{Zr}$^{*}$ &  \\ 
    \hline
    \multirow{2}{*}{Charge exchange} & A(a,c)C & \multirow{2}{6cm}{Studies the isovector response and extracts weak-interaction strengths} \\ 
    & \nuc{14}{C}(p,n)\nuc{14}{N} &  \\ 
    \hline
    \multirow{2}{*}{Capture} & A(a,$\gamma$)C & \multirow{2}{6cm}{Determines resonance energies and widths;
     relevant for reaction networks}\\ 
    & \nuc{16}{O}$(\alpha,\gamma)$\nuc{20}{Ne} & \\ 
    \hline
    \multirow{2}{*}{Breakup} & \red{A(a,bc)A$^*$}  & \multirow{2}{6cm}{Extracts properties of loosely bound states} \\ 
    & \red{\nuc{90}{Zr}(d,pn)$\nuc{90}{Zr}^{*}$ } &  \\ 
    \hline
    \multirow{3}{*}{Knockout} & A(a,a$'$b)B &  \multirow{3}{6cm}{Probes structure of weakly bound nuclei}\\ 
    & $\nuc{16}{O}(\alpha,2\alpha)\nuc{12}{C}$  &  \\ 
    & \red{ $\nuc{48}Ca(e,e^\prime p)\nuc{47}{K}$ } &  \\ 
    \hline
    \multirow{2}{*}{Stripping} & A(a,c)C & \multirow{2}{6cm}{Extracts spin, parity and orbital occupancy  }\\ 
    & \nuc{90}{Zr}(d,p)\nuc{91}{Zr}$^{*}$ & \\ 
    \hline
    \multirow{2}{*}{Pickup} & A(a,c)C & \multirow{2}{6cm}{Extracts spin, parity and orbital occupancy }\\ 
    & $\nuc{157}{Gd}(\nuc{3}{He},\alpha)\nuc{156}{Gd}^{*}$ &  \\ 
    \hline
    \end{tabular}
    \caption{Examples of direct reactions and their role in extracting nuclear information.
    A,B, and C are generic targets and products, while a,b, and c are generic projectiles. \label{tab:reactiontypes}}
\end{center}
\end{table}

\begin{figure}[th]
\centering
\includegraphics[width=1\textwidth]{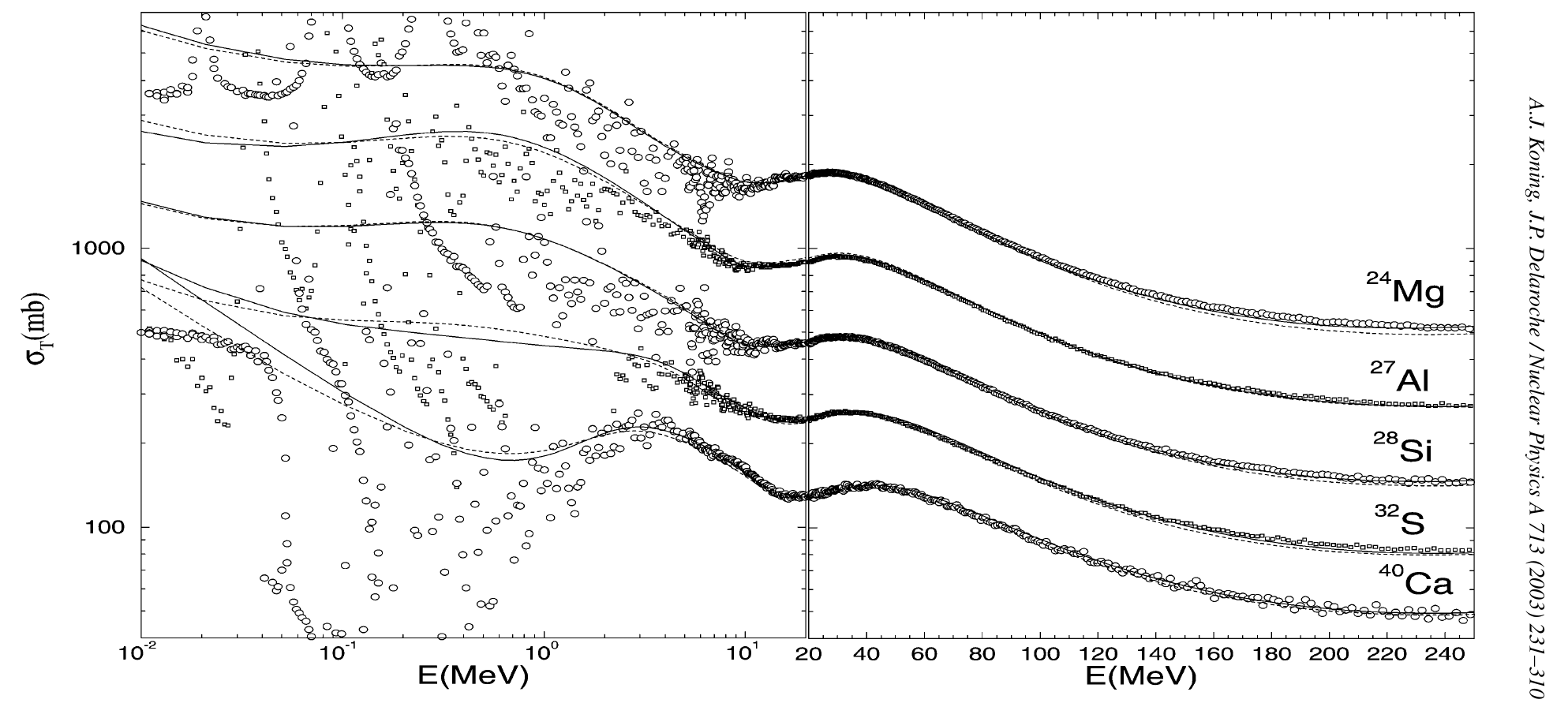}
\caption{\label{fig:phenomOptPot}Successful phenomenological optical potentials cannot address the low-energy regime where isolated resonances become important. Comparison of predicted neutron total cross sections (solid/dashed) and experimental data (circles). 
\newline
\red{
{\it Source:} Figure from A.J. Koning and J.P. Delaroche (2003) Nucl. Phys. A 713, 231 $\copyright$~Elsevier. Reproduced with permission. \url{http://dx.doi.org/10.1016/S0375-9474(02)01321-0}.
}
}
\end{figure}

The nuclear physics community currently utilizes a diverse set of radioactive and stable beam facilities, and is eagerly anticipating the new opportunities for studying unstable nuclei which will be provided by FRIB and other forefront rare-isotope beam facilities.
The types of reactions will include \red{scattering}, 
radiative capture, knockout, transfer, charge-exchange and others, both on neutron-rich and proton-rich sides. Many current reaction models rely on phenomenology including the phenomenological optical potential, valence shell-model calculations, $R$-matrix methods, and Glauber theory; fusion reactions in heavier nuclei use the Hauser-Feshbach model. An 
some of the challenges is illustrated in Fig.~\ref{fig:phenomOptPot}: while phenomenological optical potentials do well at comparatively high projectile energy, at low energies they fail to account appropriately for isolated resonances. 
To analyze and interpret data, theory with controlled approximations is needed, with uncertainties at least about 10\%.

Major progress in this area is expected by labs with current and planned radioactive beam facilities, such as FRIB (USA), GSI
(Germany), TRIUMF (Canada), RIKEN
(Japan), GANIL
(France), CERN (Europe), and RAON (South Korea).  The facilities available to the U.S. nuclear science community have been recently summarized in two White Papers; see in particular Sec.~3 of Ref.~\cite{Arc17} and Sec.~3.6 of Ref.~\cite{Car17}. The capabilities include radioactive and stable beams that vary in size from large user facilities 
to small university laboratories. In addition, there are facilities which provide neutron and photon beams. These facilities are both specialized and complementary, and provide researchers with a broad suite of tools for addressing questions in nuclear physics. The synergy obtained by facilities that can efficiently produce rare-isotope beams and facilities that can perform high-precision studies at or near the valley of stability is very important for better understanding decay and scattering processes,  nuclear reactions key to unfolding information about the nuclear forces that bind nucleons into nuclei, \red{and applications to nuclear astrophysics}.

It is not our goal to propose specific experiments. For example discussions on 
the interplay between theory and proposed experiment for unstable nuclei -- worth reading by both experimentalists 
and theorists -- see  Refs. \cite{scienceFRIB,REAupgrade,HRS,ISLA,GRETA,FRIB400,FRIBdecaystation}.

\subsubsection{Experimental nuclear astrophysics and structure}
\label{sec.nucl_astro}

A big science driver for reactions is 
nuclear astrophysics, a field concerned with the origin of the elements and energy generation in the big bang, quiescent stellar burning, and cataclysmic events such as novae, x-ray burst, supernovae, and neutron star mergers. The state of the field and future U.S. plans for research in this area are summarized in a 2017 White Paper~\cite{Arc17}.

One priority for future work is to determine the nuclear physics which defines the $r$~process, the source of many heavy nuclei, including gold and uranium~\cite{Mum16}. The most important work will be mass measurements, which define  separation energies and other reaction/decay thresholds. Measurements of decay properties (half lives, $\beta$-delayed particles) by means of decay spectroscopy and resonance-decay spectroscopy is also very important. Rare isotope beam facilities will be key for this work, as they can produce many of the neutron-rich nuclei involved in the $r$~process. These observables are related to the nuclear shell structure in the region, which will be the primary tool for interpreting measurements. 

Another focus area is radiative capture, which plays a particularly important role in nuclear astrophysics. The capture of charged particles~\cite{Bru15} and neutrons~\cite{Kap11} followed by the emission of one or more photons is important in the big bang, quiescent burning (e.g., solar fusion reactions and the $s$ process) as well as explosive burning. 
The capture of charged particles by radioactive nuclei is studied using recoil separators, such as the Separator for Capture Reactions (SECAR) at FRIB. For capture on stable nuclei, one may use separators or intense beams of protons or $\alpha$~particles in normal kinematics. Several facilities for the latter exist around the world, with some located underground for the purpose of reducing backgrounds. For neutron capture, there are several dedicated neutron facilities~\cite[Sec.~3.4.1]{Arc17}. 
These may be also addressed by extracting level density and $\gamma$-ray strength, e.g., by using the $\beta$-Oslo method when counts are low \cite{Rekstad_1983,PhysRevC.83.034315}, or by measurements of (d,p) to deduce the properties of isolated levels, as well as (d,p$\gamma$) for continuum integrated studies. 

Many other types of reactions, including (p,$\alpha$), ($\alpha$,p), and ($\alpha$,n), are important in nuclear astrophysics, and measurements of these are being vigorously pursued using stable and radioactive beams. These reactions, as well as radiative capture, may be further classified according to whether they are dominated by non-resonant (direct) processes, isolated resonances, or many resonances (statistical or Hauser-Feshbach regime). These regimes depend upon the mass, relevant excitation energy range, and nuclear structure of the compound nucleus.

Many critical reaction and decay rates in nuclear astrophysics cannot be measured directly, due to practical considerations such as very small cross sections, unavailability of beams, or the infeasibility of measuring neutron-induced reactions on radioactive isotopes. 
Even in the case of direct measurements, it is often the case that experiments alone do \red{not} provide all of the needed information. A good example is the ${}^7{\rm Be}({\rm p},\gamma){}^8{\rm B}$ reaction. 
 Experiments are unable to perform measurements at energies corresponding to the core of the sun, due to the small cross section. It is thus necessary to use experiment {\em and} theory to arrive at the best estimate for the stellar reaction rate.

A critical tool for nuclear astrophysics is thus indirect methods, which may be able to determine or constrain cross sections or decay rates by other measurements. 
Much of  future experimental work will involve indirect methods, which are critically reliant upon reaction theory for interpreting measurements and propagating uncertainties.
One example is transfer reactions, which in favorable cases can determine excitation energies, spins, parities, and partial widths of resonant states. 
Indeed, the methods for analyzing these reactions to unbound states (the typical nuclear astrophysics scenario) are quite rudimentary.
Another example is charge-exchange reactions, \red{discussed in section \ref{CEX}}, from which weak interaction strengths can be extracted that can be used to benchmark and guide the development of theoretical models for estimating weak reaction rates in stellar environments. 

There are also many cases where experiments are primarily motivated by nuclear structure questions, and accurate reaction theory is needed to extract the nuclear structure information. Transfer reactions, which probe the single-particle or cluster structure of nuclei, are again a major area of application. A recent example is a study of the ${}^{86}{\rm Kr}($d,p$){}^{87}{\rm Kr}$ reaction~\cite{Wal19}, where measurements at different bombarding energies are analyzed simultaneously to extract neutron spectroscopic factors in ${}^{87}{\rm Kr}$. Another example is provided by analyses using the dispersive optical model, which are discussed below in Sec.~\ref{sec:DOM}. In one case, a large body of experimental data, including n$+{}^{48}{\rm Ca}$ scattering and total cross section measurements over a wide range of energies, have been analyzed to yield the neutron skin thickness of ${}^{48}{\rm Ca}$~\cite{Mahzoon:2017}.

\subsubsection{Neutrino physics and charge-exchange reactions}
\label{CEX}
Charge-exchange reactions at intermediate energies ($E\gtrsim 100$ MeV/$u$) are important for detailed studies of the isovector response of nuclei \cite{RevModPhys.64.491,HAR01,Ichimura:2006,Fujita2011549,FRE18}, with important applications in astro- and neutrino physics. A variety of probes are used, ranging from light ion probes, such as (n,p) and (p,n), ($^{3}$He,t) and (t,$^{3}$He), and (d,$^{2}$He), to heavy-ion probes such as ($^{7}$Li,$^{7}$Be) and ($^{12}$C,$^{12}$B/$^{12}$N). In the past, $\pi$-induced charge-exchange experiments have also been performed. In addition, with the advent of rare-isotope beam facilities, novel unstable beam probes have been developed, such as ($^{10}$Be,$^{10}$B) \cite{Scott2015a}, ($^{10}$C,$^{10}$B)  \cite{Sasamoto2011,Sasamoto2012} and ($^{12}$N,$^{12}$C) \cite{PhysRevLett.120.172501}.  The different probes have different sensitivities and advantages for studying specific features of the isovector response. Over the past decade, significant progress has been made in charge-exchange experiments, primarily by using the (p,n) \cite{Satou:2011,SAS11,PhysRevLett.121.132501,LIP18} and  ($^{7}$Li,$^{7}$Be) \cite{PhysRevLett.104.212504, PhysRevLett.108.122501} reactions in inverse kinematics. 

A particularly useful feature of charge-exchange probes at intermediate energies is that the differential cross section at small linear momentum transfer ($q$) for transitions associated with \red{angular momentum transfer}  $\Delta L=0$ 
is proportional to the Fermi B(F) or Gamow-Teller B(GT) strengths 
for those transitions \cite{Taddeucci1987125,Sasano:2009,Zegers:2006,Perdikakis:2011}. As this proportionality can be calibrated for transitions for which these strengths are known from $\beta$-decay experiments, it enables the near model-independent extraction of these strengths at any excitation energy, including and especially beyond the $Q$-value window for $\beta$ decay.  Theoretical reaction calculations (typically in distorted-wave Born or impulse approximation) are only necessary to extrapolate the measured cross sections from finite $q$ to $q=0$ and to decompose \cite{Moinester:1989} the angular momentum contributions to the differential cross section in order to isolate the $\Delta L=0$ contributions to the excitation-energy spectra from higher angular-momentum transfers. Even with phenomenological and not necessarily well constrained optical potentials the uncertainties in the extraction of Gamow-Teller strengths are relatively small, as the absolute scale of the reaction calculations is irrelevant. Remaining uncertainties are mostly due to the tensor-$\tau$ component of the nucleon-nucleon (NN) interaction that interferes with the $\sigma\tau$ component, causing interference between $\Delta L=0$ and $\Delta L=2$ components \cite{Zegers:2006} that depend on the structure of the initial and final states involved. The uncertainty due to this interference becomes stronger for transitions with smaller B(GT). 

Unfortunately, the proportionality between strength and differential cross section described above is not established for transitions associated with higher angular momentum transfer. And although there is a strong interest to extract transition strengths for dipole and higher angular momentum transfers (for example, to test theoretical models used in the estimation of matrix elements of relevance for neutrinoless double $\beta$ decay \cite{10.3389/fphy.2017.00055} and for estimating neutrino-induced reaction rates in astrophysical environments \cite{RevModPhys.75.819}), the extraction of those strengths is much more uncertain and reliant on high-quality optical potential parameters. In addition, transitions and giant resonances at high excitation energies (beyond the thresholds for particle emission) are important \cite{HAR01}, even though most reaction codes employed require that the single-particle wave functions are bound to achieve convergence. Such uncertainties also have an impact on the extraction of Gamow-Teller strengths at high excitation energies, where the $\Delta L=0$ contributions are relatively small, and the contributions from higher-order monopole excitations, primarily the isovector (spin) giant monopole resonance \cite{HAR01}, must be estimated and subtracted to estimate the total Gamow-Teller strength \cite{SAK04,YAK05}. 

Further, only about 50-60\% of the GT sum-rule strength is observed when including strengths up to excitation energies that include the giant-resonance region \cite{GAA81,GAA85}. While significant advances have been made by experiment and theory, see e.g. Refs. \cite{br88,RAY90,WAK97,SAK04,YAK05,GYS19}, experiments with neutron-rich (proton-rich) unstable isotopes with zero summed GT strength in the $\beta -$ (or $\beta +$) direction and dominant GT strength compared to other contributions, are potentially good candidates for making further progress \cite{LIP18}. The ability to make good estimates for the optical potentials for such unstable systems and for properly accounting for continuum effects will be important.

These efforts are important for better understanding the long-standing problem of the quenching of Gamow-Teller strength and the axial-vector coupling constant, which also has important implications for the search for neutrinoless double $\beta$ decay (0$\nu\beta\beta$) \cite{10.3389/fphy.2017.00055}, as further discussed below.

Similarly, double charge-exchange (DCX) processes are a promising tool to explore nuclear structure and in particular the study of two-body correlations in nuclei \cite{Cappuzzello:2018wek,Zheng:1990vg,Vogel:1988ve}. In the 1980s, the DCX reactions using pion beams produced in the three meson factories at LAMPF \cite{lind1987charge}, TRIUMF \cite{helmer1987triumf}, and SIN \cite{blaser1966progress} were performed successfully providing interesting nuclear structure information.
At present, there is a renewed interest in DCX reactions, to a large extent due to the extensive studies of double beta ($\beta\beta$)-decay, both the decay in which two neutrinos are emitted 
and neutrinoless double beta decay. 
In DCX and $\beta\beta$-decay, two nucleons are involved. The pion, however, interacts weakly with states involving the spin and the pion DCX reactions do not excite the states involving the spin, such as the double Gamow-Teller (DGT) state; 
instead one turns to light-ion DCX reactions to probe the DGT state. The ($^{18}${O},$^{18}${Ne}) reaction has been used \cite{BLOMGREN199534,MAT13} in the past and is now under investigation at NUMEN in Catania \cite{cappuzzello2016nuclear}. The ($^{12}$C,$^{12}$Be) DCX reaction is being investigated to study DGT resonances as well. 
The ($^{8}$He,$^{8}$Be) reaction was used to find a candidate resonant tetraneutron state \cite{PhysRevLett.116.052501}. 
Experimental studies of the DGT transition, and in particular the giant DGT resonance, may 
further our understanding of
the quenching mechanism and its role in 0$\nu\beta\beta$ decay. 

\begin{figure}[th]
\centering
\includegraphics[width=0.55\textwidth]{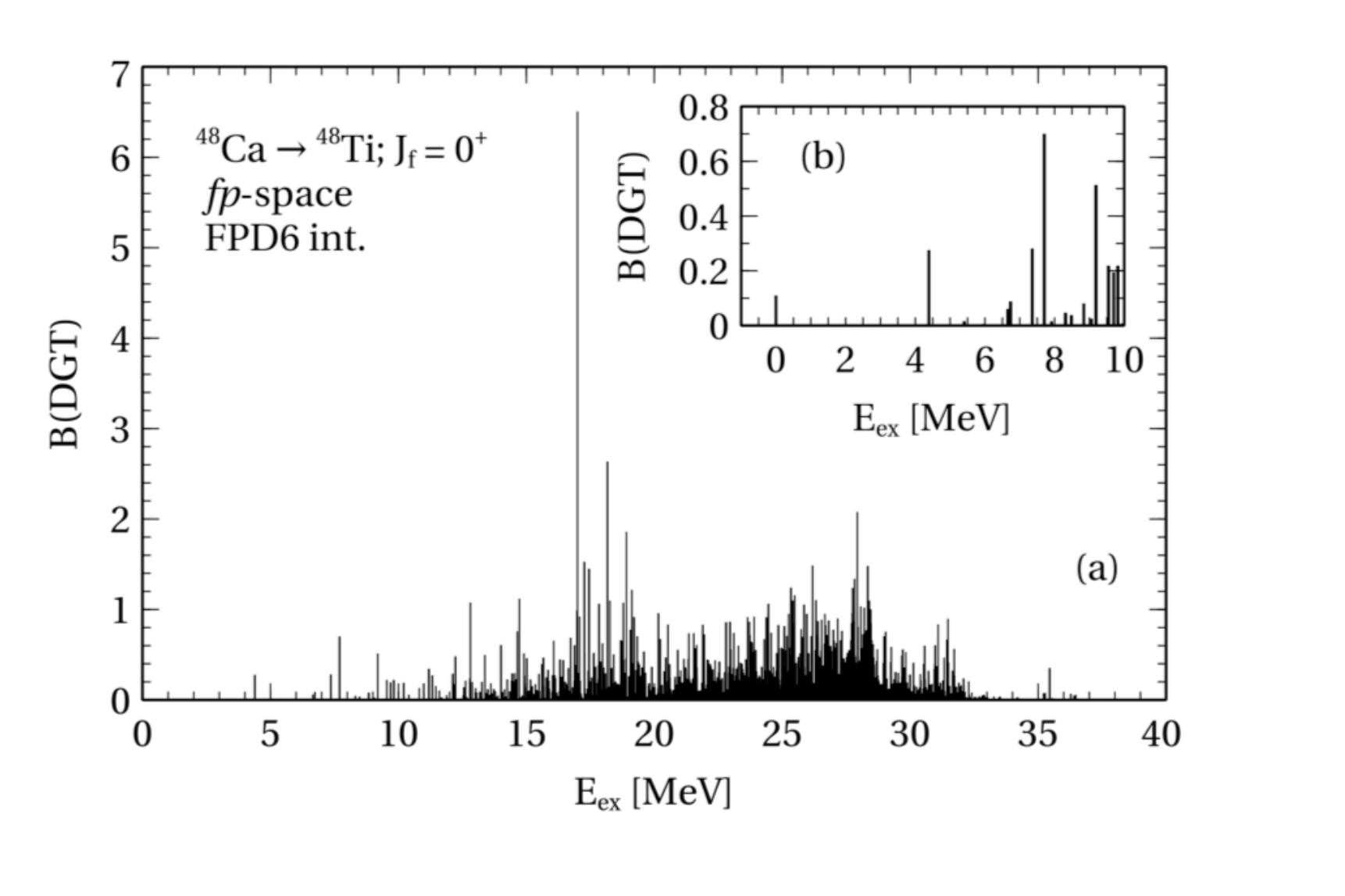}
\caption{\label{fig:DGT} (a) $B(\rm{DGT}; 0^{+} → 0^{+})$ in $^{48}$Ca. (b)  The \red{double Gamow-Teller} (DGT) transitions to low-lying states.
\newline
\red{
{\it Source:} Figure from N. Auerbach and Bui Minh Loc (2018) Phys. Rev. C 98, 064301 $\copyright$~APS. Reproduced with permission. \url{http://dx.doi.org/10.1103/PhysRevC.98.064301}.
}
}
\end{figure}

These experimental efforts are accompanied by new theoretical efforts, related to the structure of double giant resonances \cite{AuerbachM18} and  the reaction theoretical aspects \cite{LENSKE2019103716,BEL20}.
Recent calculations of the DGT transition strength distributions in even-$A$ calcium isotopes were calculated in the full fp-model space, 
by applying the single Gamow-Teller operator twice
on the parent ground state (Fig. \ref{fig:DGT})~\cite{Shimizu:2017qcy,AuerbachM18}. 
Of particular interest were the limiting cases 
when the SU(4) symmetry holds or when the spin orbit-orbit \red{interaction} is put to zero.

\section{Overview of few-body methods}
\label{sec.fewbody}

\begin{figure}[h]
\centering
\includegraphics[width=0.45\textwidth]{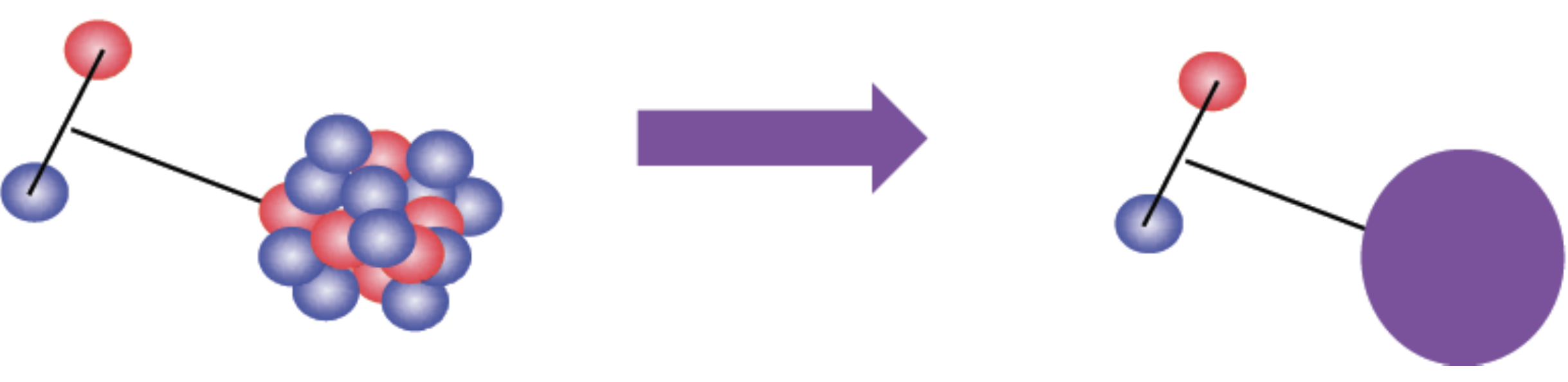}
\caption{\label{fig:fb}  Few-body techniques rely on the reduction of the scattering problem,  inherently a many-body problem, to a problem involving only a few relevant degrees of freedom.}
\end{figure}

\red{A tool widely used in understanding and interpreting  experiments, especially scattering and 
reaction experiments, are few-body methods.}
For complex systems and/or heavy composites, it is understood that a full microscopic description of the reaction is not feasible. One then relies on the reduction of the scattering problem,  inherently a many-body problem, to a problem involving only a few relevant degrees of freedom~\cite{nunes-hites2012,elster-jpg2012} (Fig. \ref{fig:fb}). Once these degrees of freedom have been identified, specific formulations can be developed. Here we discuss the main methods currently in use in the field.

Since our objective is ultimately to connect to experiment, the desired observables are cross sections. The cross sections can be constructed from the $T$-matrix, which can generally be written in post form as \cite{nrbook}:
\begin{equation}
T^{exact} = \langle \chi_f^{(-)} \; |  V  | \; \Psi_i^{(+)}\rangle ,
\end{equation}
\noindent
where $|\Psi_i^{(+)}\rangle$ represents the exact incoming wave, solution of the scattering equation:
\begin{equation}
H \Psi^{(+)}  = E \Psi^{(+)} 
\label{eq:hexact}
\end{equation}
with scattering boundary conditions $\Psi^{(+)} (R \rightarrow \infty) \equiv  (F + T H^+) $.  Here $F$ represents the regular Coulomb wave and $H^+$ is the outgoing \red{Coulomb} Hankel function \cite{nrbook}. The transition operator
$V$ is responsible for the reaction channel under study and $\chi_f^{(-)}$ is the outgoing distorted wave. If the many-body problem can be mapped onto a two-body problem,  Eq. (\ref{eq:hexact}) is a trivial two-body scattering equation.  However, most often one needs to consider more degrees of freedom and the solution of Eq. (\ref{eq:hexact}) rapidly becomes very challenging.

Most nuclear probes are peripheral, and therefore a correct asymptotic treatment of the problem is critical. This can be a challenge for two reasons: a) standard single-particle bases used in many-body methods are bound-state bases, which perform poorly in expanding the asymptotics of scattering states, and b) the long-range Coulomb force. As the projectile and/or target charge increases, the Coulomb effect will become dominant and the asymptotic properties of the system may not be known analytically. Furthermore, as more degrees of freedom are included, there are several relevant thresholds (Q-values) that need to be considered. Observables are extremely sensitive to these thresholds, requiring a precise match to the experimental values. When nuclei breakup into the continuum, the non-resonant continuum can be just as important at the resonant continuum, and final state interactions can have an important effect on phase-space. We briefly summarize below the few-body methods that have been developed to address these challenges.

One should keep in mind a number of important inputs needed for a few-body reaction theory. As mentioned above, Q-values are critical inputs. As opposed to many-body methods that rely on the NN force, few-body methods rely on optical potentials, effective interactions between the relevant fragments, the dynamics of which have been determined to be important to describe the process. In addition, for some processes one needs structure quantities such as overlap functions, transition densities, etc. Because it is hard to obtain the level of accuracy needed from many-body theoretical predictions, these quantities are either directly taken from data or strongly constrained by data. 

\subsection{The standard toolkit}

Widely used for scattering and reaction theory are the distorted-wave Born approximation, 
the Faddeev equations and their generalizations, the continuum discretized coupled channels 
method, and the eikonal approximation. 

Traditionally in reaction theory, perturbation theory is used to expand the exact $T$-matrix into the well known distorted-wave Born (DWBA) series \cite{nrbook}. By retaining only the first and/or second terms of the expansion, a method referred to as DWBA, one avoids the complications of solving Eq. (\ref{eq:hexact}) exactly. While DWBA may still be preferred in the analysis of data, the theory community has long advocated for better methods, because DWBA is not reliable particularly when there are strong clustering effects and/or the proximity of breakup channels. A larger array of non-perturbative methods have been taking over the field in the last few decades. For the purposes of illustration, we focus here on three-body methods, although analogous theories are applicable to four-body scattering problems. Currently, there are no efforts to expand few-body methods in nuclear reactions to more than four-bodies.

If the reaction A+a  can be cast as a three-body problem, A+b+x, where  to a good approximation the projectile is well represented by a$=$b+x, the exact treatment of the scattering problem is provided by the Faddeev equations, which couple all rearrangement channels to all orders \cite{alt:67a}. In the Faddeev method, an overcomplete basis spanning the three rearrangement channels is used and this ensures their separation  in the asymptotic region. This is an important aspect of the method, since then we know how to impose the correct boundary conditions. In nuclear physics, the Faddeev equations are often solved in the $T$-matrix form in momentum space \cite{alt:67a,deltuva2009}. Examples of applications include the analysis of scattering and transfer of halo nuclei as well as other reaction channels \cite{deltuva2009,deltuva2009b,cravo2010}. Note that the solution of the Faddeev equations for systems with the Coulomb interaction becomes very challenging because the equations become non-compact \cite{nunes-cgs2018}. Work to address this challenge using separable interactions is underway \cite{mukhamedzhanov2012,upadhyay2014,hlophe2014}.

The continuum discretized coupled channels (CDCC) method for describing the reaction A+a in terms of the three-body problem A+b+x relies on the expansion of the three-body wavefunction in a couple set of eigenstates of the system b+x \cite{austern1987,yahiro2012continuum}. These states include bound and scattering states. For practical reasons, the continuum is discretized, usually into energy bins and represented in terms of square-integrable {\it wave-packets} so that the resulting coupled-channel equations can be solved. As opposed to the Faddeev method, CDCC does not couple to all orders the rearrangement channels and therefore in some cases it cannot provide a complete picture of the reactions. However, for many cases in which rearrangement channels are not important, it offers the best alternative to the Faddeev method \cite{upadhyay2012,ogata-cdcc2016}. Particularly for heavier systems and for reactions in which larger clusters are involved, when the current Faddeev methods fail, CDCC is the best alternative. Recently, CDCC has been applied to a wide variety of cases, including the study of $\alpha$ yield in $^6$Li induced heavy-ion reactions \cite{lei2017}, the role of resonant states in fusion reactions \cite{gomez-camacho2018} and the sensitivity of the NN force in (d,p) reactions \cite{gomez-ramos2018}.

(Few-body coupled channel calculations can also be cast in the 
Gamow shell model formalism \cite{PRC_GSM_CC_Yannen2,COSM_ref,MercenneMP19}
which we discuss in more depth in Sections \ref{berggren} and in  \ref{sec.berggren}.)

Both Faddeev and CDCC are computationally intensive.
When the energy scales involved in the b+x and the A+a systems can be well separated, some non-perturbative approximations are used to reduce the problem without sacrificing accuracy. The adiabatic approximation consists of writing the problem as in CDCC, but neglecting the excitation energy in the b+x system \cite{ogata-ad2017}. By making all the eigenstates  of the b+x system degenerate with the ground state, the CDCC equations reduce to a simplified form which then can be solved parametrically in one of the variables \cite{nrbook}. One can also use the adiabatic approximation in particular reaction channels with enormous success \cite{johnson-soper1970,johnson-tandy1974,nunes-ad2011}. Although primarily valid at higher energies, the adiabatic method for (d,p) performs well even at $E \approx 10$ MeV/A. Applications include the analysis of the comprehensive  $^{10}$Be(d,p)$^{11}$Be data \cite{schmitt2012,yang2018} and the extraction of $^{30}$P(n,$\gamma)^{31}$S from the corresponding (d,n) reaction \cite{kankainen2017}. 

If the energy is large enough, then the eikonal approximation may become appropriate \cite{nrbook}. The standard eikonal model assumes that deviations from a straightline trajectory for the projectile can be neglected. Then, the effects of the interaction with the target are encapsulated into the so-called eikonal phase. This phase $\phi(b)$ is readily computed per impact parameter $b$ from the integral over the projectile's path of the contributions of the interaction.  Few-body eikonal theories for reactions are popular in the field due to their simplicity. Nuclear knockout reaction experiments have consistently over the years been interpreted with eikonal theory \cite{tostevin2014,aumann2017}. In order to stretch the high-energy approximation, there also have been many studies to improve on the eikonal descriptions (e.g., \cite{capel2008,hebborn2018}).   

To add to the large body of work on three-body methods for reactions, the community has now been paying attention to the need to include excitations of one or more of the clusters in the scattering problem of A+b+x. Examples of these new developments are studies of breakup reactions including core excitation in CDCC \cite{summers2006,summers2006b,moro2012,diego2014,gomez2015}, \red{the eikonal-CDCC} \cite{PhysRevC.68.064609}, 
\red{and the dynamic eikonal approximation \cite{PhysRevLett.95.082502}}, 
and Faddeev \cite{crespo2011,deltuva2015}. 
Four-body extensions in reaction theory have also been pursued, particularly when considering two-nucleon halo projectiles (inherently three-body in structure) \cite{matsumoto2006,rodriguez2009} and reactions where both target and projectile are loosely bound \cite{descouvemont2018}. 

The few-body reaction community has focused on developing increasingly sophisticated few-body theories that aim to solve the few-body problem very accurately. In contrast, uncertainties in the inputs to these theories are still ambiguously quantified. Recently, there has been some effort to use rigorous statistical methods to quantify the uncertainties in the reaction observables, associated with the optical potential, an essential input to any of these reaction theories. Examples focusing on the uncertainties in (d,p) reactions \cite{lovell2018,king2018} show that more work is necessary to fully understand, not only how to quantify the uncertainties, but also how to reduce them to the level needed so that, when combined with reaction data, one can extract the desired information. 




\subsection{The dispersive optical model}
\label{sec:DOM}

\red{An important input to many reaction calculations is the effective inter-cluster potential, also called 
the optical potential or optical model.  While optical potentials applied to reactions are often 
phenomenological, in Sec.~\ref{Sect:fromMB} we discuss efforts to derive them directly from $A$-body calculations. 
Here we discuss an intermediate approach, the dispersive optical model (DOM). }

Mahaux and Sartor \cite{Mahaux:91} developed the dispersive optical model  using  the dispersion relation, Eq.~(\ref{eq:sdisprel}) below,  
to link  the energy domain of elastic scattering to the binding potential at negative energy which generates the levels of the nuclear shell model. 
Their approach has been later expanded to applications
  to unstable nuclei.
In addition, more information related to experimental properties of the ground state has been included, in particular  the charge density, by allowing fully nonlocal potentials~\cite{Mahzoon:14,Dickhoff:17,Dickhoff:2019}.
This approach has  been used to make predictions of the neutron skin of ${}^{48}$Ca~\cite{Mahzoon:2017}.

The dispersion relation for the self-energy is employed in its subtracted form
\begin{eqnarray} 
\!\!\!\!\! \mbox{Re}\ \Sigma_{\ell j}(r,r';E)\! &=& \!  \mbox{Re}\ \Sigma_{\ell j} (r,r';\varepsilon_F) \hspace{2.0cm}  \label{eq:sdisprel} \\
&-& \! {\cal P} \!\!
\int_{\varepsilon_T^+}^{\infty} \!\! \frac{dE'}{\pi} \mbox{Im}\ \Sigma_{\ell j}(r,r';E') \left[ \frac{1}{E-E'}  - \frac{1}{\varepsilon_F -E'} \right]  \nonumber  \\
&+& {\cal P} \!\!
\int_{-\infty}^{\varepsilon_T^-} \!\! \frac{dE'}{\pi} \mbox{Im}\ \Sigma_{\ell j}(r,r';E') \left[ \frac{1}{E-E'}
-\frac{1}{\varepsilon_F -E'} \right],
\nonumber
\end{eqnarray}
where $\mathcal{P}$ is the principal value.  The self-energy $\Sigma$ is an effective one-body interaction between the particle (or hole) \red{ with orbital angular momentum $\ell$ and total angular momentum $j$}, and the $A$-particle system, and is an exact representation of the Feshbach optical potential \cite{fesh1} generalized for both bound and continuum states \cite{CapMah:00,escher02} (cf. Sec. \ref{Sect:Green}).

The representation (\ref{eq:sdisprel}) 
allows for a link with empirical information both at the level of the real part of the non-local self-energy at the Fermi energy (probed by a multitude of Hartree-Fock calculations) and also through empirical knowledge of the imaginary part of the optical potential (constrained by experimental data) that consequently yields a dynamic contribution to the real part by means of Eq.~(\ref{eq:sdisprel}).
In addition, the subtracted form of the dispersion relation emphasizes contributions to the integrals from the energy domain nearest to the Fermi energy on account of the $E'$-dependence of the integrands of Eq.~(\ref{eq:sdisprel}). 
Recent DOM applications include experimental data up to 200 MeV of scattering energy and are therefore capable of determining the nucleon propagator in a wide energy domain as all negative energies are included as well.

\begin{figure}[h]
\centering
\includegraphics[width=1\textwidth]{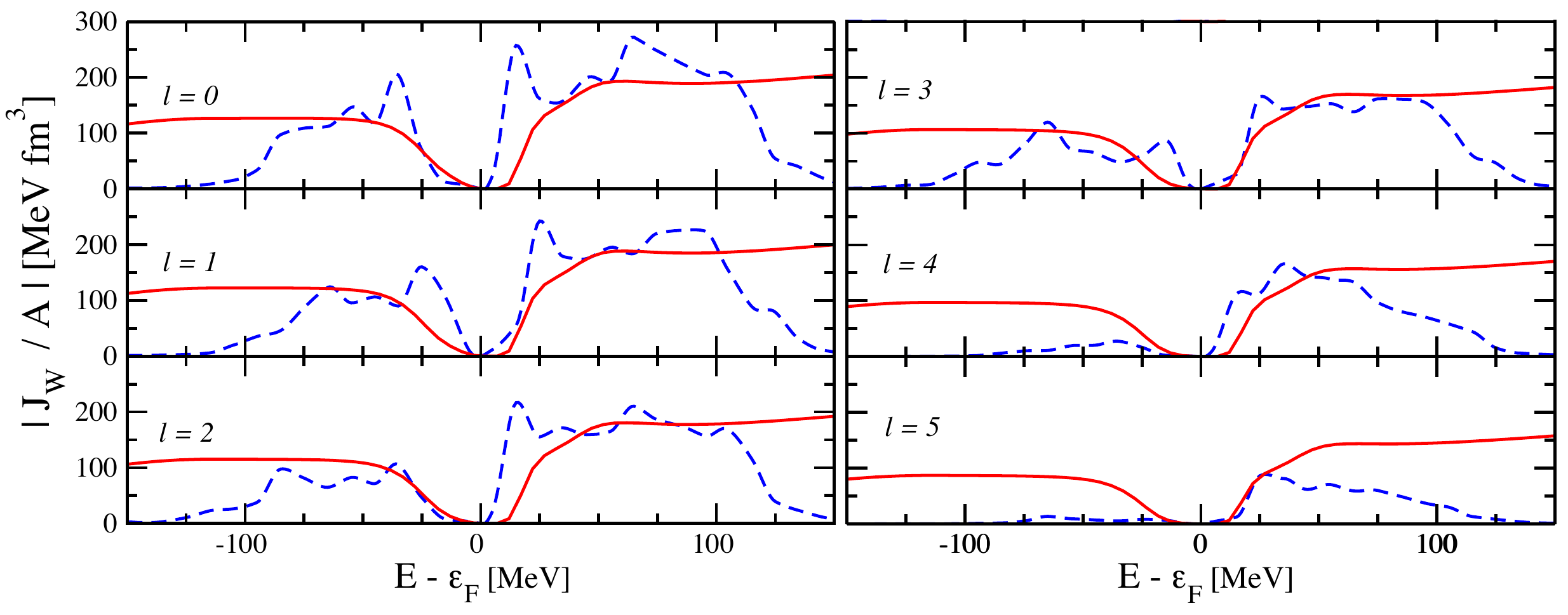}
\caption{\label{fig:vol_int}Comparison of volume integrals of the imaginary part of the nonlocal 
DOM (red solid \red{lines}) for ${}^{40}$Ca with 
FRPA calculations (blue dashed \red{lines}) with the AV18 interaction \cite{wiringa1995}.
\red{Data from \cite{Waldecker:11}.}
}
\end{figure}

The importance of this formulation of an empirical representation of the nucleon self-energy is contained in its translation of experimental data into a theoretically accessible quantity.
While the DOM assumes standard functional forms of the potentials, its parameters are constrained by data.
The resulting information can therefore be used as an interface with 
\textit{ab initio} methods
 for the nucleon self-energy.
An example is provided in Fig.~\ref{fig:vol_int} where the imaginary part of the central DOM self-energy of ${}^{40}$Ca is compared for different $\ell$-values with calculations based on the Faddeev random phase approximation (FRPA)~\cite{Waldecker:11,Dickhoff:17}, one of the incarnations of the self-consistent Green's function approach, discussed
 in Section \ref{Sect:Green}. \red{(Other \textit{ab initio} approaches to the effective inter-cluster interaction 
 are discussed in Sec.~\ref{Sect:fromMB}.)}

This comparison clearly demonstrates that a reasonable correspondence can be generated with theory and empirical results but also identifies the limitations that 
\textit{ab initio} 
methods face.
First it should be noted that  FRPA  generates a self-energy that is a collection of many discrete poles. 
A phenomenological procedure is therefore required which assigns these poles a width that increases with distance to the Fermi energy.
This will always be the case in all such calculations unless a continuum basis is employed.
The inevitable finite configuration space also does not provide a good description beyond roughly 50 MeV away from the Fermi energy.
Furthermore, the agreement for higher \red{single-particle} $\ell$-values deteriorates rapidly.
Such comparisons clearly demonstrate the usefulness of the DOM as a vehicle to assess the quality of theoretical approaches.
A comparison with volume integrals obtained from multiple scattering approaches may provide better agreement at higher energy but will most likely fail at lower energy.

Finally, the DOM provides a new approach to the analysis of nuclear reactions, as it can provide distorted waves for \red{incoming projectile} protons and neutrons \red{(and  bound single-particle states in the target)}, as well as relevant overlap functions with their normalization (spectroscopic factors).
Early applications already found such ingredients to provide a good description of transfer reactions~\cite{Nguyen:2011}.
More recently applications of the $(d,p)$ reaction have been investigated with DOM potentials for ${}^{40,48,60}$Ca in Ref.~\cite{Potel2017} generating physically more reasonable results as compared to those obtained with the global Koning-Delaroche potential because DOM potentials provide also a good description at negative energy.
A very recent analysis, including never-before published data of the ${}^{40}$Ca$(e,e'p)^{39}$K reaction,  demonstrates that DOM ingredients provide all the necessary ingredients to accurately account for these data using the distorted-wave impulse approximation~\cite{Atkinson:2018} at energies around 100 MeV for the outgoing proton.
The spectroscopic factors provided for this analysis were constrained by other data and therefore provided a consistency check of their interpretation.
The results furthermore demonstrate that the Nikhef analysis with separate phenomenological local potentials for bound and scattering states~\cite{Lapikas:1993} slightly underestimated the values for spectroscopic factors by about 0.05.
A DOM analysis for the $^{48}$Ca$(e,e'p)^{47}$K was recently published~\cite{Atkinson:2019} that pointed to the importance of proton reaction cross sections in a wide energy domain in constraining the corresponding spectroscopic factors for valence protons.
The combined $^{40-48}$Ca results demonstrate a non-negligible decrease of the proton spectroscopic factors when 8 neutrons are added.

\subsection{Halo effective field theory}
\label{sec:haloEFT}

Halo nuclei are weakly bound systems where the valence nucleons (usually one or two neutrons) are spatially decoupled from a tightly bound core
~\cite{TanihataA:1985,TanihataB:1985,tanihata1996neutron,jonson2004light}. 
$^{11}$Li nuclei has just 11 nucleons, but its valence neutron orbitals have a matter radius 
$\sim 3.3$ fm  comparable to that of a lead nucleus with 208 nucleons, resulting from  two weakly bound valence neutrons forming a halo around a tightly bound 
$^9$Li core~\cite{Hansen:1987}. \red{As such, halo nuclei are treated as few-body systems. }

\red{A recent approach to halo nuclei is through effective field theory (EFT).}  
A central idea in an EFT formulation is the separation of the physics at a given energy scale 
of interest from the physics at the higher energy scale.  At the energy scale of interest, 
once the degrees of freedom are identified no attempt at modeling the high energy scale physics is made. All the interactions allowed by the relevant symmetries are included. 
Observables are expressed as an expansion in the ratio of low energy over the higher energy scale. This allows a systematic estimation of theory error from the higher order terms of the expansion that were not included in the 
calculation~\cite{Manohar:2018aog,Kaplan:2005es,Polchinski:1992ed}. 
Weinberg's pioneering work~\cite{Weinberg:1990rz,Weinberg:1991um} led to the construction of nucleon-nucleon interactions from an EFT of pion-nucleon interaction.  

There are two aspects of Quantum Chromodynamics (QCD) that dominate low-energy
 physics, especially in light nuclei:   
(1) Chiral symmetry that dictates the lightness of pions, kaons, etas, and their interaction with nucleons; and  (2) large scattering lengths and weakly bound systems that proliferate \red{in} nuclear physics. Both of these aspects are consistently treated in the continuum and lattice formulations of EFT (\red{see Section \ref{sec:latticeEFT}}). 

\begin{figure}[th]
\begin{center}
  \includegraphics[width=0.6\textwidth,clip=true]{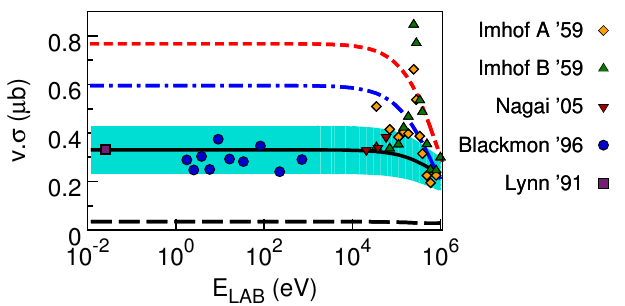}
\end{center}
\caption{$^7\mathrm{Li}(n,\gamma)^8$Li capture cross section~\cite{Fernando:2011ts}. Black long-dashed and solid \red{lines} are the EFT results for the E1 capture to the excited state and the total E1 capture, respectively. The shaded area shows the estimated $30\%$ EFT errors in the latter. The results of two traditional potential model calculations, \cite{PhysRevC.68.045802} and~\cite{Tombrello:1965}, are given respectively by the blue dot-dashed and red dashed \red{lines}. The references to the experimental data represented as colored dots are in  Ref.~\cite{Fernando:2011ts}.
\newline
\red{{\it Source:} Figure from \cite{Fernando:2011ts}, \url{http://dx.doi.org/10.1140/epja/i2012-12024-7}.}
}
\label{fig:Li8}
\end{figure}


%

Halo nuclei are characterized by small separation energies of the valence nucleons compared to the core excitation or breakup energy. This separation in energy scale is used to construct halo EFT \cite{Bertulani:2002sz,Bedaque:2003wa}.  Later calculations describe elastic scattering, electromagnetic captures and electromagnetic form factors \cite{Rupak:2016mmz,Hammer:2017tjm}. This formalism is also applicable to nuclear systems that have a cluster description at low energy such as the reaction $^3$He$(\alpha,\gamma)^7$Be in terms of point-like $^3$He and $\alpha$ clusters~\cite{Higa:2016igc,Zhang:2018qhm,Zhang:2019odg}. In halo EFT, both nuclear clusters and valence nucleons are treated as point-like particles interacting  through short-range forces, making halo EFT similar to, and sharing many calculational tools with,  pionless 
EFT~\cite{vanKolck:1998bw,Chen:1999tn}.

 Halo EFT can provide important insight into capture reactions at astrophysical energies. Typically experimental measurements have to be extrapolated to low energies for big bang nucleosynthesis (BBN) and stellar burning calculations. Theoretical error estimates are crucial for these extrapolations. Halo EFT provides a systematic expansion where the cross sections are related to universal parameters that can be constrained by observables. The
$^7\mathrm{Li}(n,\gamma)^8$Li reaction provides an example of this~\cite{Fernando:2011ts,Rupak:2011nk}.

\red{In Fig.~\ref{fig:Li8}, we compare an EFT calculation to two model calculations that used a Woods-Saxon potential. The parameters (potential strength $v_0$,  radius $R$, diffusiveness $a$) of the Woods-Saxon potentials in Refs.~\cite{PhysRevC.68.045802} and ~\cite{Tombrello:1965} were fitted to reproduce the $^8$Li ground state binding energy. The potential model results, Fig.~\ref{fig:Li8},  differ from each other and the data though they used the same physical input, the binding energy. The reason for this discrepancy has a simple physical origin. The $^8$Li ground (and excited) state is constructed as a $p$-wave bound state of neutron and $^7$Li in these calculations. The asymptotic normalization constant (ANC) of $p$-wave bound state is sensitive to the effective range. The two potential model calculations picked two different sets of $v_0$, $R$ and $a$ to reproduce the binding energy but this resulted in generating two different $p$-wave effective ranges. In contrast, the early work in halo EFT~\cite{Bertulani:2002sz,Bedaque:2003wa} identified that two operators are necessary at leading order (LO) to describe a $p$-wave bound state. Thus two parameters are constrained in halo EFT at LO by the binding energy and the effective (or the correct asymptotic  normalization), reproducing the data.  The halo EFT construction is more efficient in that only two parameters were needed as opposed to three for the Woods-Saxon potential. However, the real insight was that the $p$-wave effective range constitutes an irreducible source of error in the theory calculations at LO. Varying the effective range over a small region interpolates between the potential model calculations~\cite{Rupak:2011nk,Fernando:2011ts}.  Reactions involving $p$-wave bound states $^7\mathrm{Be}(p,\gamma)^8$B,$^3$He$(\alpha,\gamma)^7$Be, $^3$H$(\alpha,\gamma)^7$Li are impacted in a similar manner.} 
The electric charge form factor of bound states can also be related to effective range corrections~\cite{Chen:1999tn,Ryberg:2015lea}. Thus halo EFT can provide key physical insight into connections between physical observables in terms of universal parameters that are model-independent.

In two-body systems, an important distinction between traditional potential models and halo EFT is the inclusion of  electro-weak currents not constrained by the Siegert theorem in the latter. A well-known example from pionless EFT is the two-body axial current with coupling $L_{1,A}$, important in constraining solar neutrino-deuteron scattering cross section~\cite{Butler:2000zp}. In the recent $^3$He$(\alpha,\gamma)^7$Be cross section calculation~\cite{Higa:2016igc}, two-body currents not constrained by elastic scattering are found to contribute at LO. At the same time, due to the sensitivity of the ANC to the effective range $\rho$, a 10\% change in the value of $\rho$ can
accommodate vanishing two-body current contribution in the energy region where data is available. However, at solar energies. 
the cross section with and without two-body corrections differ~\cite{Higa:2016igc}. Halo EFT can provide meaningful error estimates in solar burning and BBN energy regime where data is lacking.

\section{Overview of many-body methods}
\label{manybody}

\red{Unlike few-body methods, which regularly incorporate continuum degrees of freedom, most ``many-body'' 
methods (the boundary between the two is not rigidly defined and is, furthermore, porous) are typically built using 
bound single-particle states.  In this section we review several widely-used approaches, while in the 
next section we discuss how practitioners can connect these many-body methods to the continuum.}

There are a bewildering variety of many-body methods, though most of them draw from 
the same pool of theoretical tropes. To introduce them, we begin with the interacting 
shell model or just the shell model (SM), often called the configuration-interaction method in fields outside of 
nuclear physics. There are more powerful methods than the SM, several of which will 
be discussed below, but the SM is a conceptually straightforward paradigm, and many of its
weaknesses with regards to the continuum are shared by other many-body methods. 

\subsection{The shell model}

The SM wave function is just an expansion in a basis $\{ | \phi_\alpha \rangle \}$: 
\begin{equation}
| \mathrm{SM} \rangle = \sum_\alpha c_\alpha | \phi_\alpha \rangle.
\label{CIexpand}
\end{equation}
Unlike few-body methods, which typically work in 
relative (or Jacobi) coordinates, the SM wave function is usually in single-particle coordinates, 
that is the $A$-body wave functions are of the form $\phi_\alpha(\vec{r}_1, \vec{r}_2, \ldots, \vec{r}_A)$, where $\vec{r}_i$ is the coordinate of the $i$th nucleon in laboratory frame. The primary driver for this is the antisymmetry of fermions.  Thus the basic 
element of SM wave functions are, either explicitly or implicitly, Slater determinants (or, more properly, their occupation-number representations), antisymmetrized products of 
single-particle wave functions, which is trivial in second quantization representation 
using fermion single-particle creation and annihilation operators.
In Jacobi coordinates, one must explicitly antisymmetrize by taking all possible permutations.  In additional to its conceptual simplicity, 
the SM  can compute low-lying excited states
nearly as easily as the ground state. As illustrated in Eq.~(\ref{CIexpand}), one conceptualizes the SM, as well as some other many-body methods, through the lens of the wave function; this is one of the conceptual gulfs between many-body and few-body methods, which frequently focus on the scattering or $T$-matrix. Equally foundational is the idea that one finds the coefficients $c_\alpha$ by a variational principle, minimizing the energy.

The coefficients also lead us to the primary disadvantage of the SM model:  the lack of 
correlations in any given Slater determinant, so one must include many configurations, 
up to 24 billion in the largest calculations to date.  Most variants on the SM, such 
as symmetry-adapted configuration-interaction SM and cluster-based SM, as well as alternatives such as 
coupled cluster (CC) (Sec. \ref{sect:CC}) and Green's function Monte Carlo (GFMC) \cite{WiringaS98,WiringaP02,PastoreBCGPSW18}, build in correlations. This in turn has 
a cost, of course, of additional complexity or loss of flexibility (CC and GFMC must work harder to get excited states). 

Single-particle coordinates
have disadvantages, of which the primary one is that separation  of intrinsic or relative motion from center-of-mass motion is not trivial. Because of this, it is difficult to 
truly identify relative cluster motion, for example the asymptotic behavior of a single 
particle (or alpha or other cluster) at a long distance from the remainder. This in turn leads to the simplifying assumption of the boundary condition that  wave functions must vanish at infinity. These problems drive several of the current methodologies for connecting to the continuum,
such as the resonating group method (RGM) and related methods, Sec.~\ref{RGM}, and complex-momentum bases, such as the 
Berggren basis, which impose outgoing boundary conditions, Sec.~\ref{berggren} and \ref{sec.berggren}.

The community has amassed a great deal of phenomenological/empirical knowledge 
about the shell model, especially those carried out in valences spaces \cite{BG77,br88,caurier2005}.
Although methods starting from realistic interactions, 
such as the no-core shell model (Section \ref{sect:ncsm}), have become more widespread in recent years,
the phenomenological (or empirical) shell model remains important in interpreting many experimental 
results, not least of which because phenomenology still can reach further than, 
say, the no-core shell model.  In light nuclei, where both approaches can be applied, 
the empirical and no-core shell models have strong overlaps, even in areas not 
directly constrained by experiment, 
such as group-theoretical decompositions \cite{PhysRevC.91.034313}. 
Recently there are efforts to build more rigorous models that look like the empirical 
shell model but which arise out of realistic interactions \cite{stroberg2019nonempirical}.
And, independent of the rise of realistic interactions, the empirical shell model 
underpinned efforts to connect to the continuum, such as the continuum shell model \cite{mahwei,barz} 
and the shell model embedded in the continuum \cite{benn00,rot05a,rot05b}, as well as the 
the Gamow shell model using the Berggren basis, which is discussed in Section \ref{berggren}.

\subsubsection{No-core shell model}
\label{sect:ncsm}

The basic idea of the no-core shell model (NCSM) is simply to treat all $A$ nucleons in a nucleus as active, i.e., to write down the Schr\"odinger
equation for $A$ nucleons and then to solve it numerically. This approach avoids problems related to excitations of nucleons from the core, such as core-polarization effects, because there is no core, and being a non-perturbative approach, there are
no difficulties related to convergence of a series expansion. It may also be formulated in terms of an intrinsic Hamiltonian,
so as to avoid spurious center-of-mass (COM) motion.

The starting Hamiltonian 
\begin{equation}
H = T_{\rm rel} + V= \frac{1}{A}\sum_{i<j}\frac{(\vec p_i - \vec p_j)^2}{2m}+\sum_{i<j}^A (V_{\rm NN})_{ij}  + \sum_{i<j<k}^A (V_{\rm 3N})_{ijk} + \ldots + V_{\rm Coulomb}, 
\label{ham}
\end{equation}
where $m$ is the nucleon mass, $V_{\rm NN}$ is the nucleon-nucleon (NN) interaction, $V_{\rm 3N}$ is the three-nucleon interaction, and $V_{\rm Coulomb}$ is the Coulomb interaction between the protons. There is no restriction on the $V_{\rm NN}$ and $V_{\rm 3N}$ interactions used, typically derived in the chiral effective field theory \cite{BedaqueVKolck02,EpelbaumNGKMW02,EntemM03,Epelbaum06} as mentioned in Sec.~\ref{sec:free_srg}, meson-exchange theory \cite{Machleidt01}, or  inverse scattering {harmonic oscillator representation of scattering equations} (HORSE) formalism \cite{InvAMS1,JISP16a,JISP6JPG,ShirokovMZVW07} (see Sec. \ref{HORSE}). The NCSM uses a harmonic oscillator (HO) single-particle basis that allows preservation of translational symmetry of the nuclear self-bound
system, even if single-particle coordinates are utilized.  This is possible as long as the basis is truncated by a maximal total HO energy of the $A$-nucleon system [or selected according to SU(3) symmetry, as discussed in Sec. \ref{SA}]. The NCSM employs a large but finite harmonic oscillator (HO) basis. 
%

The NCSM is best matched to light nuclei due to its computational efficacy, but is limited by the explosive growth in computational resource demands with increasing number of particles and size of the spaces in which they reside. To address this, the NCSM framework has been extended to heavier nuclei by using truncating schemes as in the Importance Truncation NCSM \cite{RothN07}, the Monte Carlo NCSM \cite{AbeMOSUV12}, symmetry-adapted bases (see Sec. \ref{SA}). Other 
approaches  re-introduce the core and derive effective interactions for a valence shell from NCSM \cite{LisetskiyBKNSV08,Dikmen,Nadya2019}, coupled cluster method (CC)~\cite{CC-EffectiveInteractions}, and the in-medium similarity renormalization group (IMSRG)~\cite{CC-EffectiveInteractions,IMSRG-EffectiveInteractions,IMSRG-EffInt2016,IMSRG-EffInt2017} (see also Sec. \ref{IMSRG_CC}).

\subsubsection{The symmetry-adapted no-core shell model}
\label{SA}

\begin{figure}[th]
 \begin{minipage}[t]{0.5\textwidth}
\begin{tabular}{cc}
  \includegraphics[width=115pt]{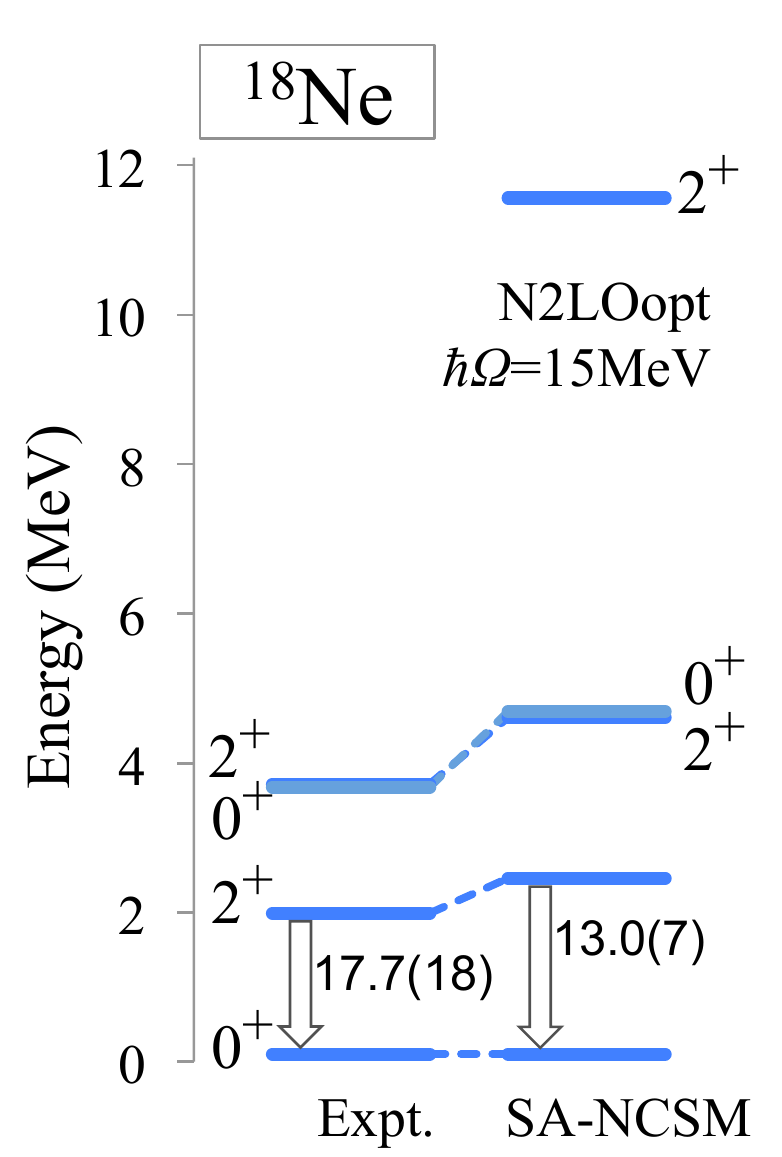} &
   \includegraphics[width=95pt]{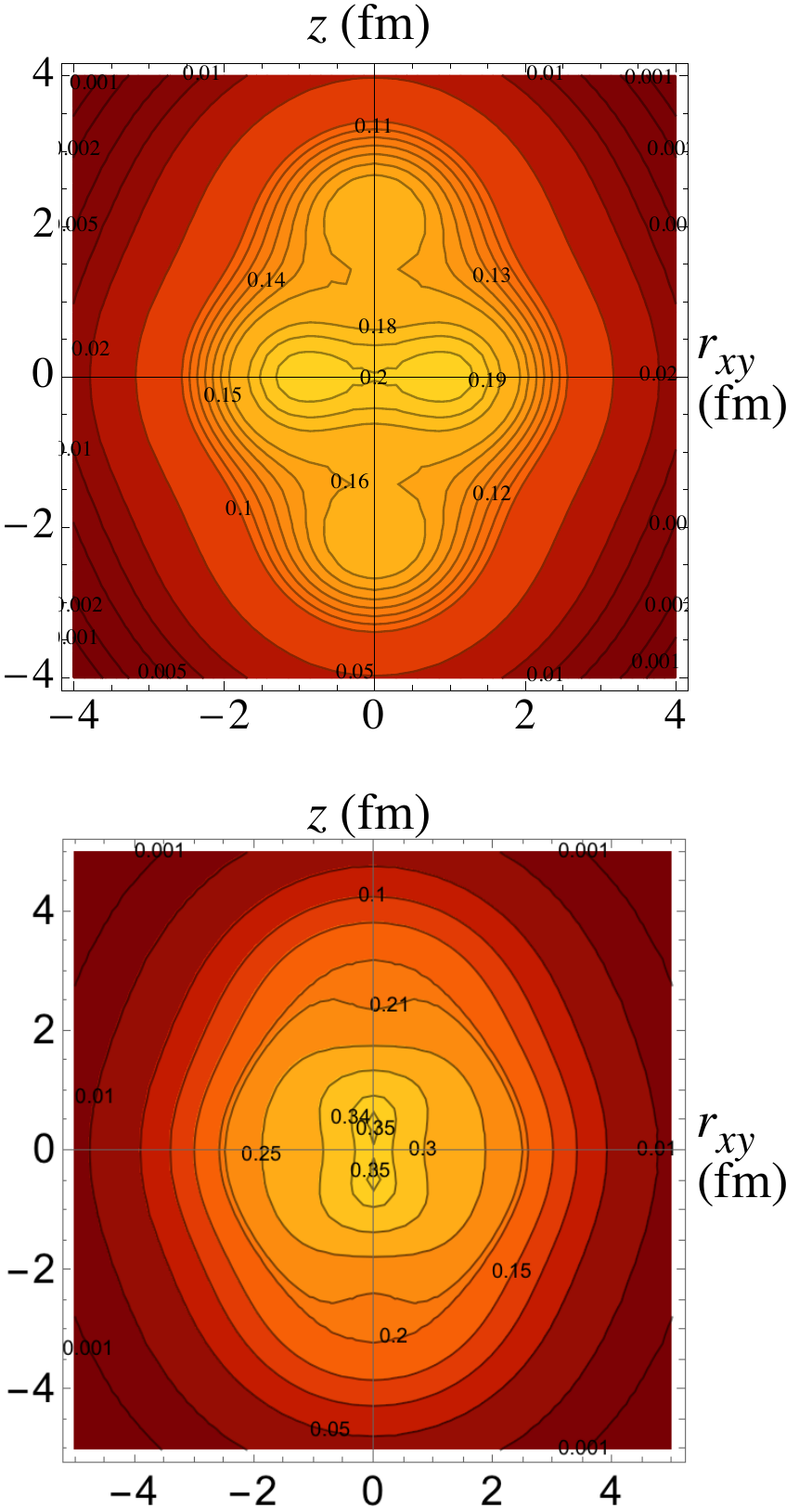}\\
   (a) & (b)
\end{tabular}
\end{minipage}
\begin{minipage}[*]{0.49\textwidth}
  \caption{ \label{sd_pf}
{\it Ab initio} SA-NCSM calculations using the chiral NNLO$_{\rm opt}$  \cite{Ekstrom13} NN interaction. (a) Energy spectrum of   $^{18}$Ne  in 9 HO major shells, along with the $B(E2;2^+ \rightarrow 0^+)$ strength in W.u. reported for 33 shells.  (b) Density profile of the ground state of $^{20}$Ne (top) and  $^{48}$Ti (bottom) \cite{LauneySOTANCP42018}. 
\newline
\red{
{\it Source:} Figure from K. D. Launey et al. (2018) AIP Conf. Proc. 2038, 020004 $\copyright$~AIP. Reproduced with permission. \url{http://dx.doi.org/10.1063/1.5078823}.
}
}
  \end{minipage}
\end{figure}

Built upon the \textit{ab initio} NCSM framework, the symmetry-adapted framework exploits exact and approximate symmetries of the nuclear many-body dynamics  (reviewed in Refs. \cite{DytrychSDBV08_review,LauneyDD16}). The symmetry is utilized to construct the basis states, that is, the model space is reorganized to a symmetry-adapted (SA) basis that respects the  symmetry. Hence,  calculations are not limited {\it a priori} by any symmetry and employ a large set of basis states that can,  if the nuclear Hamiltonian demands, describe a significant symmetry breaking. In particular, the symmetry-adapted no-core shell model (SA-NCSM) \cite{LauneyDD16,DytrychLDRWRBB20}, has achieved significantly reduced  model spaces without compromising the accuracy for various observables, and has accommodated nuclei beyond the light species, as well as modes of enhanced deformation and spatially extended clustering (Fig. \ref{sd_pf}) \cite{DytrychLMCDVL_PRL12,DytrychMLDVCLCS11,DytrychHLDMVLO14,LauneySOTANCP42018}. 

Some hallmarks of nuclear structure are deformation, vibrations and rotations, which, 
following the pioneering work of Bohr \& Mottelson \cite{BohrMottelson69},  Elliott  \cite{Elliott58,Elliott58b,ElliottH62}  and the microscopic no-core formulation by Rowe \& Rosensteel \cite{RosensteelR77,Rowe85}, lead to choosing SU(3) as a basis symmetry. 
Earlier applications, which have been typically limited to just a few basis states and symmetry-preserving interactions, have provided successful descriptions of dominant collective features of nuclei -- from the lightest systems  \cite{RoweTW06,DreyfussLTDB13}, through intermediate-mass nuclei \cite{DraayerWR84,TobinFLDDB14,LauneyDD16}, up  to strongly deformed nuclei of the rare-earth and actinide regions \cite{Rowe85,CastanosHDR91,JarrioWR91,BahriR00}. An important result is that, even when one starts from first-principle considerations, these dominant features of nuclei naturally emerge and are found to track with SU(3) and Sp(3,R)  symmetries  \cite{LauneyDD16, DytrychLDRWRBB20}. A major advantage of the SA-NCSM is that  the SA model space can be down-selected to a subset of SA basis states that describe equilibrium and dynamical deformation, and within this selected model space the spurious center-of-mass motion can be factored out exactly~\cite{Verhaar60,Hecht71}. 
The  many-nucleon SU(3)-scheme basis states are constructed using efficient group-theoretical algorithms \cite{AkiyamaD73,DraayerLPL89}. 


The symplectic \SpR{3} symmetry provides a  further organization of the nuclear model space, and underpins the many-nucleon \SpR{3}-scheme basis states  (reviewed in Refs. \cite{DytrychSDBV08_review,LauneyDDSD15,LauneyDD16}). The main feature is that, within a symplectic configuration, particle-hole excitations with \red{ 
combined orbital angular momentum $L=0$ and $L=2$ } are driven by the total kinetic energy operator (or equally, the nuclear monopole moment that describes the ``size'' of the nucleus) and quadrupole moment (that describes the deformation of the nucleus). Indeed, operators that preserve the symplectic symmetry (do not mix symplectic configurations) include the monopole  and quadrupole moment operators, the many-body kinetic energy, generators of rigid and irrotational flow rotations, and the total orbital momentum $L$. Using \SpR{3}-scheme basis is not as straightforward as the SU(3)-scheme basis, as there are no known \SpR{3} coupling/recoupling coefficients. The SA-NCSM with \SpR{3}-scheme basis resolves this by diagonalizing an \SpR{3} symmetry-preserving operator calculated in the SU(3) basis \cite{DytrychLDRWRBB20}.  The resulting Hamiltonian matrix is drastically small in size and its eigensolutions, the nuclear energies and states, can be  calculated without the need for supercomputers.  Alternative methods build the symplectic basis recursively, as outlined in Refs. \cite{reske1984:diss,SuzukiH86,Suzuki86}, and later generalized to a no-core shell-model framework in Refs. \cite{EscherD98,DytrychSBDV_PRL07,DytrychSDBV08_review} and in the {\it ab initio} symplectic no-core configuration interaction model (SpNCCI) \cite{mccoy2018:diss,mccoy2018:spncci-busteni17}. 
{\it Ab initio} 
SA-NCSM calculations reveal the predominance of only a few symplectic configurations in low-lying nuclear states in isotopes up through the calcium region -- this implies that these states are typically made of only one or two equilibrium nuclear shapes (deformed or not) with associated vibrations and rotations  \cite{DytrychSBDV_PRL07, LauneyDD16,DytrychLDRWRBB20}.

In order to converge long-range properties of the nucleus, that is, those
which depend on the ``tails'' of the nuclear wave function, it is necessary,
in the context of an oscillator basis, to include orbitals with large numbers of
radial nodes and thus many oscillator quanta.  Indeed, in a symplectic basis,
once the dominant symplectic configurations contributing to the nuclear wave function
are identified, the calculation can be extended, within these configurations, to include
basis states with many more oscillator quanta than would be possible in a
traditional NCSM calculation. In this way, one can accommodate collective correlations that are essential to account for deformation, as well as include small but critical configurations from the continuum.
Such a pattern is encouraging for possible use of the SA framework to
provide the structure component to reaction models that properly account for the continuum \cite{Burrows:2018ggt,MercenneLEDP19}  (e.g., see Secs. \ref{RGM} and \ref{sec:MultiScatt}).


\subsection{Effective field theory and interactions}
\label{sec:free_srg}

Much of the progress in nuclear structure theory in the past decade has been enabled by the use of Effective Field Theory (EFT) and Renormalization Group (RG) methods. Two- and three-nucleon interactions from chiral EFT have become the standard input for \emph{ab initio} structure theory -- see Refs. \cite{Epelbaum:2009ve,Machleidt:2011bh,Machleidt:2016yo,Meisner:2016fk} for reviews and discussions of open issues. 

\begin{figure}[th]
\begin{center}
\includegraphics[width=0.38\textwidth]{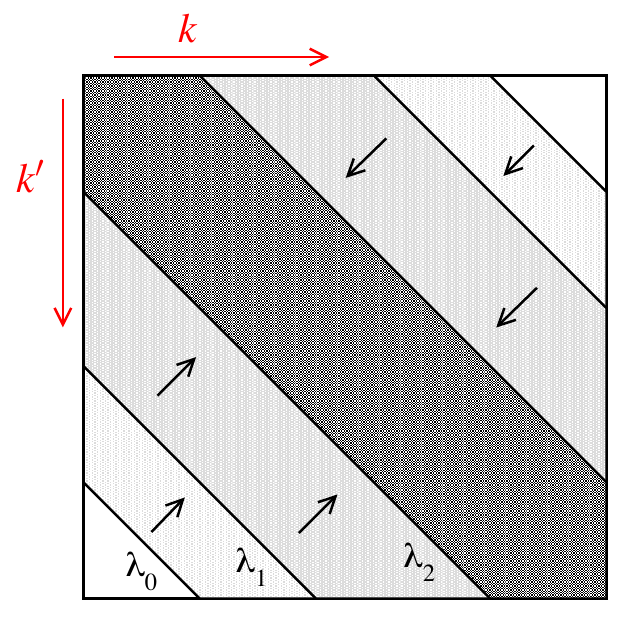}\hspace{12pt}
 \includegraphics[width=0.36 \textwidth]{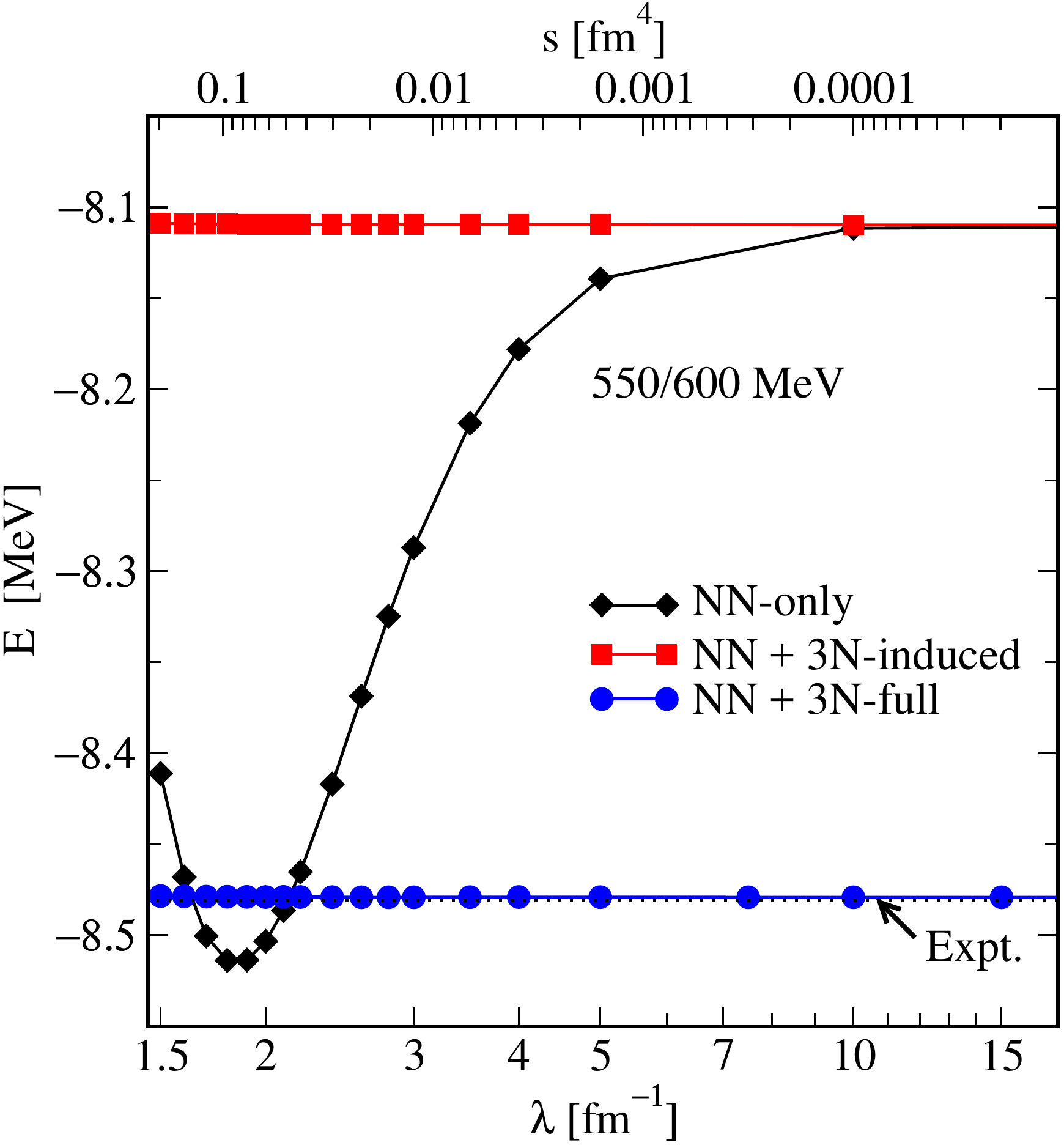}\\
 \hspace{0.28in} (a) \hspace{2.2in}(b)
    \caption{\label{fig:srg}SRG evolution of nuclear Hamiltonians. (a) Schematic view of the SRG evolution in momentum space. (b) Ground state energy of $\nuc{3}{H}$ as a function of the flow parameter $\lambda$ for chiral \NNLO{} NN and NN$\!+\!$3N interactions (see Ref. \cite{Hebeler:2012ly} for details). NN-only \red{(black diamonds)} means initial and induced 3N interactions are discarded, NN$\!+\!$3N-induced \red{(red squares)}, takes only induced 3N interactions into account, and 3N-full \red{(blue circles)} contains initial 3N interactions as well. The black dotted line shows the experimental binding energy \cite{Wang:2012uq}. Data for the figure courtesy of K.~Hebeler.  \red{Adapted from H. Hergert Phys. Scr. \textbf{92} 023002 (2017) 
    \cite{Hergert:2017kx}, 
    doi.org/10.1088/1402-4896/92/2/023002}   }
    \end{center}
\end{figure}

Renormalization group (RG) methods are a natural companion to EFT models, because they allow one to smoothly dial the {resolution scale or cutoff} of a theory. In principle, they could be used to 
rigorously connect various EFTs of the strong interaction by systematically integrating out degrees of freedom (high-momentum modes, composite particles, etc.), possibly starting from QCD. In practice, applications are typically simpler but nevertheless extremely useful. By lowering the resolution scale of input chiral two- and three-nucleon forces, we decouple their low- and high-momentum modes and greatly accelerate the convergence of few- and many-body methods that rely on Hilbert space expansions. This decoupling is achieved by means of a continuous unitary transformation, which we implement via the operator flow equation
\begin{equation}\label{eq:flow}
  \frac{d}{ds}H(s) = \comm{\eta(s)}{H(s)}\,.
\end{equation}
To decouple momenta, we construct the generator of the transformation using the relative kinetic energy,
\begin{equation}\label{eq:def_eta}
  \eta(s) \equiv \left [ \frac{\vec{k}^2}{2\mu},{H(s)} \right ]
\end{equation}
Clearly, $\eta$ would vanish -- and the SRG evolution would stop
-- if the Hamiltonian were diagonal in momentum space. In applications, it is convenient to parameterize the flow by $\lambda\equiv s^{-1/4}$. From Eqs.~(\ref{eq:flow}) and (\ref{eq:def_eta}), it is clear that $\lambda$ has the dimensions of momentum. Its meaning is illustrated in  Fig.~\ref{fig:srg}~(a): the figure shows the SRG evolution of a two-nucleon Hamiltonian matrix in momentum space, and $\lambda$ measures the width of \red{the} diagonal band. In other words, it limits the momentum that the interaction can transfer in an NN scattering process to $|\vec{k}-\vec{k}'|\lesssim\lambda$. Thus, $\lambda$ can be identified with the resolution scale of the evolved Hamiltonian.

The benefits of using the SRG to decouple low- and high-momentum physics in nuclei come at a cost: the evolved nuclear Hamiltonian will contain induced many-body operators, even if we start from a two-body interaction. Numerically, the effect of induced interactions is demonstrated in Fig.~\ref{fig:srg}~(b), which shows the evolution of $\nuc{3}{H}$ ground-state energies that have been calculated with a family of SRG-evolved chiral NN and NN+3N interactions (see Ref. \cite{Hebeler:2012ly} for details). If we neglect the induced three-body terms in the evolved interaction and the SRG generator, the energy varies by 5--6\% as we evolve the Hamiltonian over a typical range of $\lambda$ values (NN-only, \red{black diamonds}). If we include induced 3N interactions (\red{red squares}), the unitarity of the transformation is restored and the energy no longer varies with $\lambda$. Note that the NN+3N-induced results now match the ground-state energy we would have obtained with the unevolved NN interaction, while the NN+3N-full results \red{(blue circles)} are obtained from a consistently evolved NN+3N starting Hamiltonian that was fit to reproduce experimental triton data \cite{Epelbaum:2009ve,Machleidt:2011bh,Gazit:2009qf,Wang:2012uq}).

Given that this example shows the importance of tracking induced interactions, one may ask whether the added complexity of dealing with 3N (and higher) many-body forces makes RG evolving the Hamiltonian worthwhile at all. The answer is affirmative, because the induced interactions will still be of low-momentum/low-resolution character; their improved convergence behavior in many-body calculations (far) outweighs the increased complexity of the Hamiltonian. Moreover, a hierarchy of many-nucleon forces naturally appears in chiral EFT anyway, because its degrees of freedom are composite objects like nucleons and pions rather than quarks.

Another important consequence of working with SRG and chiral EFT is that {all operators of interest} must be constructed and RG-evolved consistently to ensure that observables like expectation values or cross sections remain invariant under specific choices of calculation schemes and resolution scales (see, e.g., Refs. \cite{Duguet:2015lq,Furnstahl:2010zr,Furnstahl:2012fn,More:2015bx,More:2017eu}). This will be especially important when we work to connect nuclear structure and reaction theories. The SRG provides us with a useful diagnostic in these efforts: As discussed for Fig.~\ref{fig:srg}, truncations of the SRG flow can lead to a violation of unitarity that manifests as a $\lambda$-dependence of calculated observables. We can use this dependence as a tool to assess the size of missing contributions, although they have to be interpreted with care \cite{Duguet:2015lq}.


\subsection{Decoupling the model space}


While configuration interaction methods are straightforward to understand, they have 
limitations, most notably the exponential explosion in the basis dimension.  While one 
approach is to choose a smart truncation, as in symmetry-adapted methods (Sec.~\ref{SA}), another 
approach is to transform the space so as to fully or approximately decouple the 
model space from a much larger Hilbert space.   


\begin{figure}[t]
  \setlength{\unitlength}{\textwidth}
  \begin{center}
  	\includegraphics[width=0.75\unitlength]{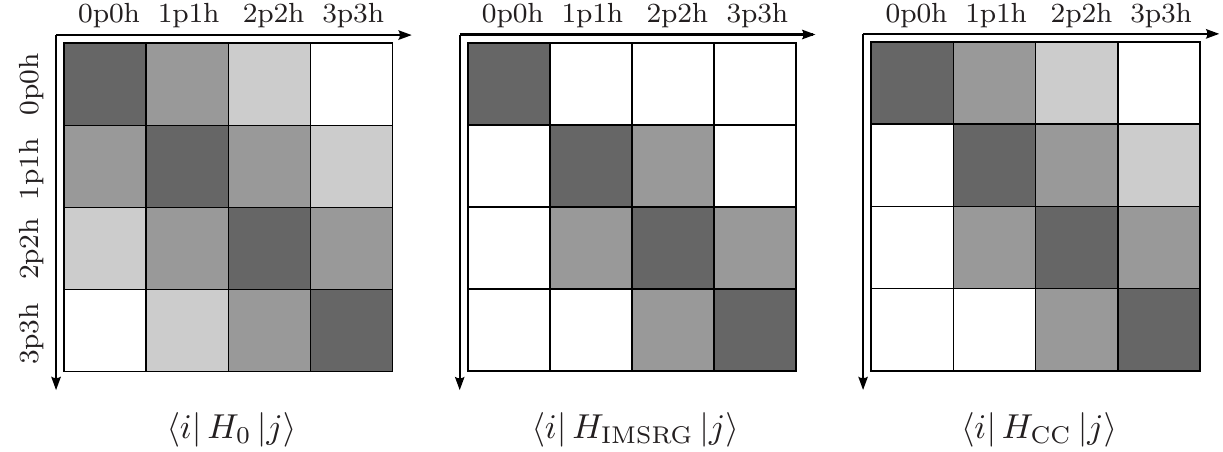}\\
	\hspace{0.3in}(a) \hspace{1.5in} (b) \hspace{1.5in} (c)
  \end{center}
  \caption{\label{fig:imsrg_cc}Decoupling of particle-hole excitations from a 0p0h reference state:  the schematic matrix representation of the initial Hamiltonian $H_0$ (a) and the transformed Hamiltonians obtained from IMSRG  (b) and CC (c), respectively. (See text for details.)}
\end{figure}


The Coupled Cluster (CC) method and the In-Medium SRG (IMSRG) avoid the basis size explosion by avoiding the construction of the Hamiltonian matrix altogether, and hence can reach heavy nuclei. 
Instead, they use similarity transformations that act on the operator directly and implicitly decouple specific states or groups of states from the rest of the Hamiltonian matrix. 


\subsubsection{In-medium renormalization}

The original application of the SRG was to the nuclear interaction in relative 
coordinates. One consequence is the induction of three-body and higher-rank interactions.
As the community grew to appreciate the power of decoupling through SRG, 
a new approach was introduced, the in-medium similarity renormalization group or IMSRG.
Here one works in single-particle coordinates (laboratory frame),  approximating
 the many-body forces through normal-ordering against the ``medium'' (here a reference 
state).

We start with a general one- plus two-body Hamiltonian, which can be written in second quantization as
\begin{equation}
  H = E + \sum_{ij}f_{ij} \{\aaO_i\aO_j\} + \frac{1}{4}\sum_{ijkl}\Gamma_{ijkl} \{\aaO_i\aaO_j\aO_l\aO_k\}\,.
\end{equation}
The \red{braces} indicate that the strings of creation and annihilation operators have been \emph{normal ordered} with respect to a reference state,
which is typically a Slater determinant constructed from harmonic oscillator or Hartree-Fock orbitals (see Refs. \cite{Hergert:2016jk,Hergert:2017kx,Hjorth-Jensen:2017it} for details). After normal ordering, $E$ is the energy expectation value of the reference state, while $f$ and $\Gamma$ are the \emph{in-medium} mean-field Hamiltonian and residual two-nucleon interaction, respectively. 

Figure \ref{fig:imsrg_cc}~(a) shows the matrix representation of the Hamiltonian in the basis consisting of our reference Slater determinant and its particle-hole excitations. Note that $H$ is band diagonal because it can at most couple $n$p$n$h states to $(n\pm2)$p$(n\pm2)$h states.
The goal of the IMSRG is to decouple the one-dimensional block in the Hamiltonian matrix that is spanned by our reference state Slater determinant (labeled 0p0h in Fig.~\ref{fig:imsrg_cc}) from all excitations, using the flow equation (\ref{eq:flow}). 
 In principle, we can use a suitably chosen reference to target different eigenstates, e.g., by taking references which are expected to have the largest overlap with the target (see Chapter 10.3 of Ref. \cite{Hjorth-Jensen:2017it}). In practice, we usually target the ground state by using a Hartree-Fock Slater determinant as our reference.

To achieve the desired decoupling, we write $H\equiv H_d + H_{od}$, where $H_{od}$ denotes the part  we want to suppress, and $H_d$ is the desired Hamiltonian  at the end of the IMSRG flow. In the free-space SRG discussed in Sec.~\ref{sec:free_srg}, one decouples momentum scales by choosing $H_d=\vec{k}^2/2\mu$ and driving interaction matrix elements that couple states with $|\vec{k}-\vec{k'}|\gtrsim \lambda$ to zero. Here, we need to suppress
\begin{equation}\label{eq:def_Hod}
  H_{od} \equiv \sum_{ph} H_{ph} \{\aaO_p\aO_h\} + \frac{1}{4}\sum_{pp'hh'} H_{pp'hh'} \{\aaO_p\aaO_{p'}\aO_{h'}\aO_{h}\} + \mathrm{H. c.}\,,
\end{equation}
which are the terms of $H$ that couple the reference Slater determinant to 1p1h and 2p2h excitations, respectively. In analogy to the free-space SRG, we define the generator
\begin{equation}
  \eta(s) \equiv \comm{H_d(s)}{H_{od}(s)} = \comm{H_d(s)}{H(s)}
\end{equation}
to evolve the Hamiltonian operator and implicitly transform it to the matrix representation shown in Fig.~\ref{fig:imsrg_cc}~(b). We note that 
this will not only decouple the ground state from excitations, but also eliminate the outermost band in the Hamiltonian matrix, which makes the evolved Hamiltonian an attractive input for subsequent configuration interaction or equation-of-motion approaches (see, e.g., Refs. \cite{Parzuchowski:2017yq,IMSRG-EffInt2017,Gebrerufael:2017fk}).

In most IMSRG applications to date, we truncate all operators at the two-body level, which defines the so-called IMSRG(2) scheme. Since we work with normal-ordered operators, the omission of induced three-body terms causes much smaller issues than in the free-space SRG: We are only truncating  residual 3N interactions, while in-medium contributions to the zero-, one-, and two-nucleon parts of the Hamiltonian are accounted for. For more details about the method, we refer our readers to Refs. \cite{Hergert:2016jk,Hergert:2017kx,Hjorth-Jensen:2017it}.

\subsubsection{Coupled cluster}
\label{sect:CC}

In contrast to the IMSRG, the CC method decouples sectors of the Hamiltonian matrix through a non-unitary similarity transformation. Traditionally, the discussion of CC focuses on the correlated wave function, for which the ansatz
\begin{equation}
  \ket{\Psi_{CC}}= e^T\ket{\Phi}
  \label{CC_ansatz}
\end{equation}
is introduced. Here, $T$ is the so-called cluster operator, which is defined as
\begin{equation}\label{eq:def_tcc}
  T=\sum_{ph} t_{ph}\{\aaO_p\aO_h\} + \frac{1}{4}\sum_{pp'hh'}t_{pp'hh'}
  \{\aaO_p\aaO_{p'}\aO_{h'}\aO_h\} + \ldots\,,
\end{equation}
and $t_{ph}, t_{pp'hh'}, \ldots$ are the cluster amplitudes (see, e.g., Refs.~\cite{Shavitt:2009,coupled_cluster,Hjorth-Jensen:2017it}). In practical applications, the cluster operator is usually truncated to include up to 2p2h (CC with Singles and Doubles, or CCSD) or some of the 3p3h terms (CCSDT, including Triples). Acting on a Slater determinant reference state \red{$| \Phi \rangle$}, $e^T$ admixes arbitrary powers of such correlated few-particle, few-hole excitations. Note, however, that the cluster operator $T$ is not anti-Hermitian because it lacks de-excitation operators, and therefore $e^T$ is not unitary. 

The cluster amplitudes are determined by demanding that the transformed Hamiltonian,
\begin{equation}\label{eq:def_HCC}
  H_{CC} \equiv e^{-T}He^T,
\end{equation}
does not couple the reference \red{state  $| \Phi \rangle $} to 1p1h and 2p2h states. Defining $\ket{\Phi^{p\ldots}_{h\ldots}}=\{\aaO_p\ldots\aO_h\ldots\}\ket{\Phi}$, the decoupling conditions lead to the following system of non-linear equations:
\begin{eqnarray}\label{eq:cc_eqs_E}
  \bra{\Phi}e^{-T}He^T\ket{\Phi}&=E_{CC}\,,\\
  \bra{\Phi^{p}_{h}}{e^{-T}He^T}\ket{\Phi} &=0\,,\label{eq:cc_eqs_1p1h}\\
  \bra{\Phi^{pp'}_{hh'}}{e^{-T}He^T}\ket{\Phi} &=0 \label{eq:cc_eqs_2p2h}\,.
\end{eqnarray}
Here, $E_{CC}$ is the CC approximation to the ground-state energy, which corresponds to the upper left entry in $H_{CC}$'s matrix representation, as shown in Fig.~\ref{fig:imsrg_cc}~(c). The other blocks in the first column of the matrix vanish because of the CC equations. 

As a consequence of the non-unitarity of the CC transformation, care must be taken when one evaluates observables using the CC wave function, or uses the  non-Hermitian $H_{CC}$ (cf.~Fig.~\ref{fig:imsrg_cc}) as input for subsequent diagonalization. In this regard, the CC methods are less convenient than unitary transformation methods like the IMSRG. An advantage of CC over a unitary method is that the Baker-Campbell-Hausdorff series appearing in Eqs.~(\ref{eq:cc_eqs_E}--\ref{eq:cc_eqs_2p2h}) terminates at finite order because of the properties of the cluster operator, while additional truncations must be used in unitary approaches. 

\subsection{Lattice EFT}
\label{sec:latticeEFT}

Lattice EFT is a numerical method for calculating nuclear properties exactly in a periodic box from an EFT defined on a space-time lattice. Lattice methods for field theory \red{were} introduced by Wilson in the context of 
QCD~\cite{Wilson:1974}.  Lattice methods for nuclear \textit{ab initio} calculations from an EFT were introduced in Ref.~\cite{Muller:1999cp}. In lattice EFT, one starts with an initial state wave function $|\Phi\rangle$ (a Slater determinant) and evolves it in Euclidean time $\tau$ as $\exp(-\tau H)  |\Phi\rangle$ with the microscopic Hamiltonian $H$ derived from EFT interactions.  The exponential behavior of the partition function 
$\mathcal Z = \langle\Phi|\exp(-\tau H)  |\Phi\rangle$ at large Euclidean times allows one to extract information about ground and excited state energies. Expectation values of observable $\mathcal{O}$, including higher order energy corrections, can be calculated as 
$\langle\Phi|\exp(-\tau H/2) \mathcal{O} \exp(-\tau H/2)  |\Phi\rangle/\mathcal Z$. The review article~\cite{Lee:2008fa} describes  the implementation of lattice EFT for few- and many-body calculations.  Both the pionless EFT and chiral formulation in the so called Weinberg power counting on a space-time lattice are described in detail there. 
The lattice EFT calculations are performed by Monte Carlo simulations over possible field configurations between the initial and final states. 

The lattice EFT methods have been applied to a wide range of systems from few nucleons to $A\sim 30$~\cite{Borasoy:2005yc,Lahde:2013kma}. The first accurate \textit{ab initio} calculation of the Hoyle state energy was performed in lattice EFT~\cite{Epelbaum:2011md}. Beyond static properties, reaction cross sections can be calculated using lattice methods. At low energy one usually considers the reactions A(a,c)C, A(a,$\gamma$)C where A, C, a, and c are nuclear clusters.
Two recent algorithmic developments -- adiabatic projection method and pinhole algorithm -- allow for \textit{ab initio} reaction calculations in lattice EFT. 

\begin{figure}[tbh]
\begin{center}
  \includegraphics[width=0.6\textwidth,clip=true]{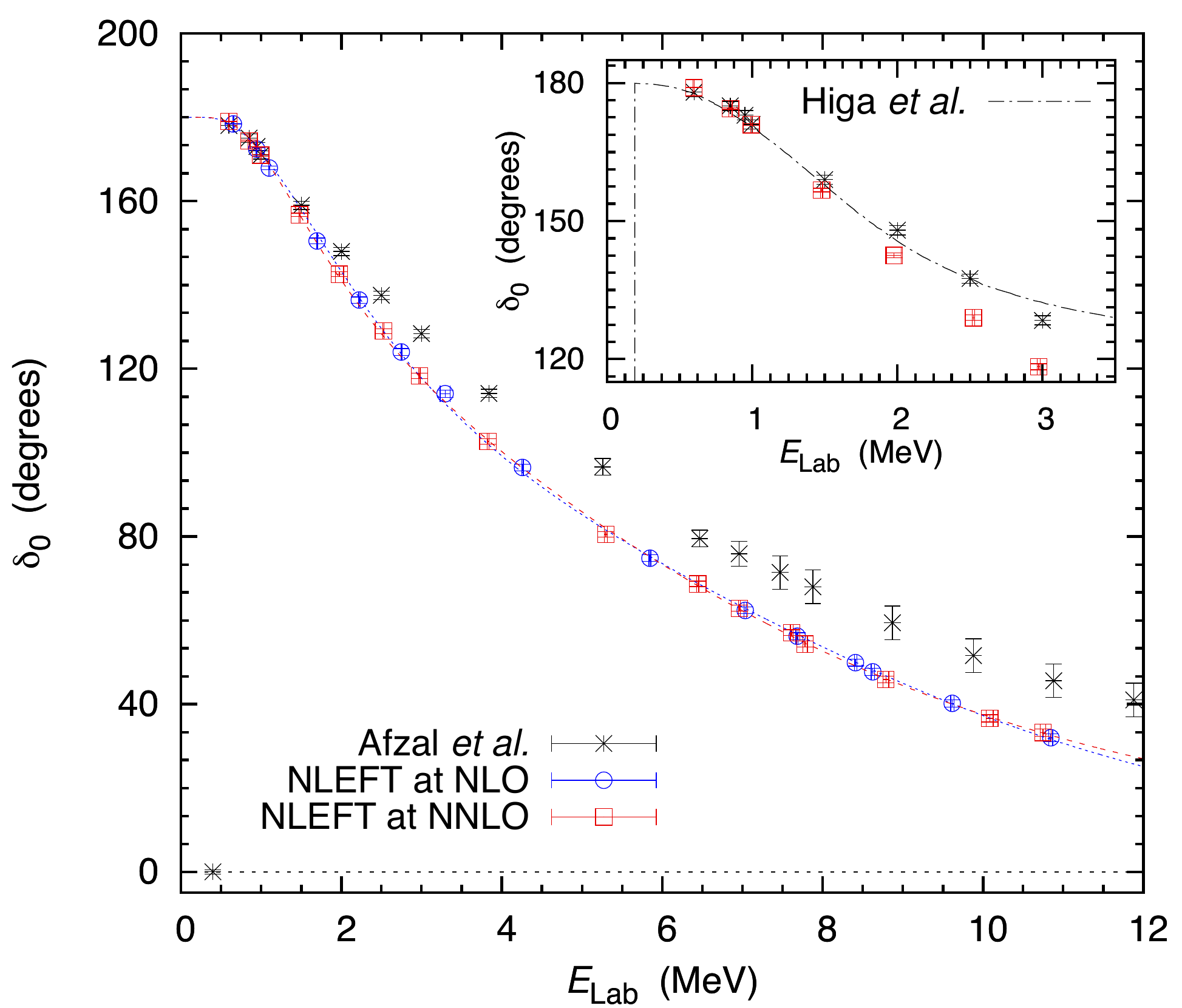}
\end{center}
\caption{$\alpha$-$\alpha$ $s$-wave elastic scattering phase shift from an \textit{ab initio}lattice EFT calculation~\cite{Elhatisari:2015iga}. \red{See text for discussion}.
\newline
\red{{\it Source:} Figure from S. Elhatisari et al. (2015) Nature 528, 111 $\copyright$~Springer Nature. Reproduced with permission. \url{http://dx.doi.org/10.1038/nature16067}.
}
}
\label{fig:AlphaAlpha}
\end{figure}

In the adiabatic projection method~\cite{Pine:2013zja}, an initial  trial state $|\vec{\bm R}_0\rangle$ 
approximating two nuclear clusters with separation $\vec{\bm R}_0$ is evolved with the microscopic Hamiltonian as 
$|\vec{\bm R}_\tau\rangle=\exp(-\tau H)|\vec{\bm R}_0\rangle$. Energy measurements with normalized wave functions determines the adiabatic Hamiltonian as 
$H_a=( \langle \vec{\bm R}'_\tau|\vec{\bm R}_\tau\rangle)^{-1} \langle \vec{\bm R}'_\tau|H|\vec{\bm R}_\tau\rangle$. 
The Hamiltonian $H_a$ defined in the cluster coordinates has matrix dimensions $L^3\times L^3$ whereas the microscopic Hamiltonian is $L^{3(A-1)}\times L^{3(A-1)}$ for an $A$-body system. The cluster Hamiltonian $H_a$ is applicable at energies below the breakup of the nuclear clusters. It includes all deformation and polarizations of the clusters in the presence of other clusters from an \textit{ab initio} 
calculation, without any modeling.  As an example, $\alpha$-$\alpha$
 $s$-wave scattering phase shifts calculated from the $8$-body Hamiltonian using the adiabatic projection method 
 is shown in Fig.~\ref{fig:AlphaAlpha}. Details of the calculations are in Ref.~\cite{Elhatisari:2015iga}.
 
 In the adiabatic projection method~\cite{Pine:2013zja}, an initial  trial state $|\vec{\bm R}_0\rangle$
approximating two nuclear clusters with separation $\vec{\bm R}_0$ is evolved with the microscopic Hamiltonian as
$|\vec{\bm R}_\tau\rangle=\exp(-\tau H)|\vec{\bm R}_0\rangle$. Energy measurements with normalized wave functions determines the adiabatic Hamiltonian as
$H_a=( \langle \vec{\bm R}'_\tau|\vec{\bm R}_\tau\rangle)^{-1} \langle \vec{\bm R}'_\tau|H|\vec{\bm R}_\tau\rangle$.
The Hamiltonian $H_a$ defined in the cluster coordinates has matrix dimensions $L^3\times L^3$ whereas the microscopic Hamiltonian is $L^{3(A-1)}\times L^{3(A-1)}$ for an $A$-body system. The cluster Hamiltonian $H_a$ is applicable at energies below the breakup of the nuclear clusters. It includes all deformation and polarizations of the clusters in the presence of other clusters  from an \textit{ab initio} calculation, without any modeling.  
The $\alpha$-$\alpha$ $s$-wave scattering phase shifts calculated from the $8$-body Hamiltonian using the adiabatic projection method is shown in Fig.~\ref{fig:AlphaAlpha}.  
 \red{The results are significant for a couple of reasons. It is the first ab initio calculation of  nuclear collision, involving more than a few nucleons, starting from a microscopic description without any modeling. Success of the lattice EFT result is judged by how well it reproduces the low energy phase shifts. Further, it also demonstrates the effectiveness of the adiabatic projection method to calculate Minkowski space correlation from Euclidean time simulation in non-relativistic systems involving more than a few nucleons.} 
 Details of the calculations that also include the $d$-wave phase shift results are in Ref.~\cite{Elhatisari:2015iga}.
 
In  lattice EFT calculations, the particle locations in an atomic nuclei are not accessible.  This deficiency has been remedied now with the introduction of the pinhole algorithm~\cite{Elhatisari:2017eno}. An opaque screen with $A$ pinholes is inserted in the middle of the Euclidean time steps. The location, and spin-isospin labels of the pinholes are selected by the nuclear distribution in the simulation as determined by the microscopic interaction. This is implemented by inserting in the middle of the time step the $A$-body density operator $:\rho_{\alpha_1, a_1}(\bm{r}_1)\cdots  \rho_{\alpha_A, a_A}(\bm{r}_A):$ 
with spin label $\alpha_i$ and isospin label $a_i$, respectively. Once the coordinates of the nucleons ($\vec{\bm r}_i$) are calculated from the pinholes, the center-of-mass location $\vec{\bm R}_\mathrm{c.m.}$ is determined by minimizing
 $\sum_{i=1}^A|\vec{\bm R}_\mathrm{c.m.}-\vec{\bm r}_i|^2$. The nucleon distributions were calculated for carbon isotopes $^{12}$C, $^{14}$C, $^{16}$C; and accurately reproduced experimental measurements where available~\cite{Elhatisari:2017eno}. For example, the lattice EFT calculation for the $^{14}$C charge radius $2.43\pm0.07$ fm is compatible with data $2.497\pm0.017$ fm~\cite{Schaller:1982} within the error bars. 


\subsection{Self-Consistent Green's function}
\label{Sect:Green}

Green's functions, a representation of the solutions to a time-independent Hamiltonian $H$ \cite{dickhoff05_b63},
\red{allow one to calculate one-particle direct scattering processes, that is, a generalized Feshbach optical potential \cite{fesh1} for both bound and continuum states \cite{CapMah:00,escher02}. }

Green's functions are directly used to solve the many-body problem following a diagrammatic expansion scheme \cite{DICKHOFF2004, soma14, barbieri17, idini12, idini15, broglia16}. The expansion will determine the many-body approximation used and the physical processes included in the self-energy, hence also in the propagator. Starting from a general, time-independent,  two and three-body Hamiltonian, as the one in Eq. (\ref{ham}), the application of diagrammatic rules reduces it to a one-body effective interaction, referred to as the irreducible self-energy $\Sigma^\star$.  The propagator represents the probability amplitude related to a particle in a given state subjected to this effective interaction generated by the nucleus. The self energy $\Sigma^\star$ represents the effective nucleus-particle interaction, also called the optical potential.  For this reason, the Green's function representation can be used in different contexts (cf. Secs. \ref{sec:DOM}, \ref{optPot}). Furthermore, the Green's function can, in principle, be constructed with densities and eigenstates from any many-body method (cf. Sec. \ref{Sect:fromMB} and  \ref{App:Green}).

Different methods can be employed to construct the self energy using Green's functions and related methods, e.g. direct diagrammatic expansion \cite{broglia16}, equation of motion \cite{idini12}, and Algebraic Diagrammatic Construction (ADC) \cite{Schirmer82,barbieri17}. Particular schemes to sum infinite-order contributions have been implemented for infinite systems with full self-consistency, often the means of \red{the} Dyson equation,
\begin{equation}
G = G^0 + G^0 \Sigma^\star G,
\label{eq:Dyson}
\end{equation}
that iterates the calculation of the propagators to consider infinite order contributions from a given set of diagrams. The Dyson equation is solved iteratively to yield the dressed propagator, $G$, from the bare propagator, $G^0$, over the single-particle vacuum. This makes this method self-consistent and non-perturbative.  In the case of the ring-diagram approximation for the polarization propagator, one recovers the so-called $GW$ approximation~\cite{hedin65, Holm1998} which has found ample applications in condensed matter physics.

Ladder diagram summation is particularly relevant for nuclear systems and have been successfully implemented with full self-consistency at finite temperature for different realistic interactions (see e.g. Ref.~\cite{Rios2014}).
Finite temperature calculations avoid consideration of pairing solution although these can be incorporated as well~\cite{Ding2016}.

In finite nuclei, the main method used today to calculate self-consistent \textit{ab initio} Green's functions is the Algebraic Diagrammatic Construction (ADC) ~\cite{Schirmer82,soma14, barbieri17}. This method systematically includes orders of diagrams into the definition of the irreducible self-energy $\Sigma^\star$, obtaining two components,
\begin{equation}
 \Sigma^\star (E) = \Sigma^{\infty} + \widetilde{\Sigma}(E),
\end{equation}
where $\Sigma^{\infty}$ is the static, energy independent part arising from mean-field-like contributions, including an eventual external static potential; and $\widetilde{\Sigma}(E)$ is the energy-dependent part arising from correlations. ADC defines a hierarchy of possible many-body truncations, denoted ADC$(n)$. State-of-the-art ADC$(3)$ calculations in finite nuclei using two and three-body interactions correspond to considering all possible 2 particle-1 hole, 2 holes-1 particle and some 3 particle-2 holes, 3 holes-2 particles configurations in the three-body sector \cite{raimondi18}. This expansion yields  many-body correlation similar to Coupled Cluster singles and doubles with perturbative triples (cf. Sec. \ref{sect:CC}).

Self-consistent Green's function in ADC$(3)$ using NNLO$_{\textrm{sat}}$ interaction has been recently used to compute the optical potential and the corresponding neutron elastic scattering absolute cross section of Ca and O isotopes in \cite{idini19}. The results were also compared to no-core shell model with continuum calculations (cf. Sec.~\ref{RGM}). As expected, including more particle-hole configurations increases the absorption and improves the reproduction of experimental data. Therefore, future efforts should be focused in this direction.


\subsection{Nuclear density functional theory} 

The nuclear energy density functional (EDF) formalism of nuclear density functional theory (DFT) is a viable
microscopic approach 
for heavy isotopes~\cite{Bender03}. 
The nuclear energy density functional representing the effective (in-medium) nuclear interaction is constructed from local nucleonic densities and currents. Its parameters are adjusted 
to reproduce a collection of nuclear structure properties. The main advantage of the 
EDF approach is that the resulting framework scales well with the number of particles, 
making EDF a reliable tool to study  systems such as neutron-rich nuclei close to the drip-line.
Of particular interest is the formulation of nuclear DFT in  configuration space,  
convenient for the description of continuum coupling in the presence of pairing correlations.  For more discussion, see \cite{dobaczewski2013hartree} and references therein.
 In particular, the structure of the quasi-particle continuum, important in the context of many applications, has been discussed in Refs.~\cite{belyaev1987pairing,PhysRevC.84.024311}.

In Hartree-Fock (HF) theory, the contribution of the two-body interaction $V_{\rm NN}(\boldsymbol{r})$ to the total energy can be obtained by contracting the two-body potential matrix elements with the density matrix. This can be expressed  in relative ($\boldsymbol{r}$) and center of mass ($\boldsymbol{R}$) coordinates as
\begin{eqnarray}
E^{\rm NN} & = \frac{1}{2} {\rm Tr_1 Tr_2} \int d \boldsymbol{R} \int d \boldsymbol{r} \langle \boldsymbol{r} \sigma_1 \tau_1 \sigma_2 \tau_2 | V_{\rm NN}(\boldsymbol{r}) | \boldsymbol{r} \sigma_3 \tau_3 \sigma_4 \tau_4 \rangle \\ 
& \times \left[ \rho_1 \left( \boldsymbol{R} + \frac{\boldsymbol{r}}{2} \right) \rho_2 \left( \boldsymbol{R} - \frac{\boldsymbol{r}}{2} \right) - \rho_1 \left( \boldsymbol{R} - \frac{\boldsymbol{r}}{2}, \boldsymbol{R} + \frac{\boldsymbol{r}}{2} \right) \rho_2 \left( \boldsymbol{R} + \frac{\boldsymbol{r}}{2}, \boldsymbol{R} -\frac{\boldsymbol{r}}{2} \right) P_{12}^{\sigma \tau} \right],
\label{eq:HF_energy}
\end{eqnarray}
where the traces indicate summation over the spin $\sigma$ and isospin $\tau$ quantum numbers and $P_{12}^{\sigma \tau}$ is an exchange operator. While in principle a realistic NN interaction can be used in Eq. (\ref{eq:HF_energy}) to calculate the HF energy, the non-local nature of the densities in the exchange term makes such calculations rather involved. In practice, interactions with specific symmetries are developed to deal with non-localities, such as the contact terms of a Skyrme functional \cite{erler2011self} or the Gaussian functions of a Gogny interaction \cite{berger1991time}. In these cases parameters are adjusted to reproduce nuclear structure data.

The phenomenological nature of this approach makes the implementation of systematic improvements, both in the description of experimental data and in predictive power, rather complicated. 
Additional terms can be added to the EDF to reduce \red{the} difference between theory and experiment. This is not necessarily done following a clear hierarchy among different terms. However, several efforts follow the prescriptions of an order-by-order expansion either as terms in an expansion series \cite{skyrme1958effective,davesne2013skyrme} 
or in a regularised EFT \cite{bennaceur2017nonlocal}, or drawing a correspondence with pionless EFT \cite{PhysRevC.95.054325}.


In order to implement the systematic order by order improvements of EFT into the EDF framework, and use a chiral interaction in Eq. (\ref{eq:HF_energy}), the non-local densities can be treated with a density matrix expansion (DME). The DME approach can be considered analogous to a Taylor expansion in the sense that it expands a non-local density in terms of a local density and its derivatives
\begin{equation}
\rho \left( \boldsymbol{R}-\frac{\boldsymbol{r}}{2}, \boldsymbol{R}+\frac{\boldsymbol{r}}{2}\right) \approx \sum_{n=0}^{n_{\rm max}} \Pi_n(kr) \mathcal{P}_n(\boldsymbol{R})
\end{equation}
where the $\Pi_n$ functions are determined by the DME variant, $\mathcal{P}_n$ denote derivatives of the local density and the arbitrary momentum scale $k$ is usually chosen to be the Fermi momentum $k_F$.

This DME approach has recently been fully implemented in \cite{Navarro18} with a chiral two-pion exchange potential including $\Delta$ resonances \cite{Piarulli14}. Although this EDF is microscopically constrained, a phenomenological contribution to the EDF is still necessary to recover the many-body correlations. 




\section{Connecting \red{few- and many-body methods} }
\label{connections}

Because of the clear importance of the continuum to nuclear physics, theorists using 
bound-state methods
have not ignored it. \red{Building upon the methods reviewed in Sec.~\ref{sec.fewbody} and \ref{manybody},} 
 we \red{now} describe several approaches \red{to connect many-body methods, which are 
primarily albeit not exclusively built from bound degrees of freedoms, to few-body methods and 
the continuum}:
the resonating group method,
which builds integro-differential 
scattering equations from fully microscopic calculations;  $J$-matrix methods, 
which are discrete but exact alternatives to the resonating group method, using bound single-particle states as a basis such as harmonic oscillator states; and the continuum shell model, the 
shell model embedded in the continuum or using the 
Berggren basis, which is a single-particle basis with outgoing boundary conditions.

From there we outline approaches for computing effective inter-cluster interactions, 
more popularly known as optical potentials. The rest of this section describes a number 
of additional methods relevant to reaction theory.

\subsection{The resonating group method}

\label{RGM}


One way to address the long-distance behavior of nuclear wave function and to introduce coupling to the continuum is by using a basis that explicitly considers cluster degrees of freedom. This can be done through a combination of the NCSM with the Resonating Group Method (RGM) \cite{navratil08,QuaglioniN09,navratil11,MercenneLEDP19}.
In the RGM \cite{WildermuthT77}, nucleons are organized within different groups, or clusters, ``resonating'' through the intercluster exchange of nucleons. 
This antisymmetrization between the different clusters safeguards the Pauli exclusion principle, which along with the consideration of internal structure for the clusters, is one the most important features of the approach.
In the case of two clusters $(A-a)$ and $a$, the cluster states for a channel $\nu$ are defined as ${ \ket{ { \Phi }_{ \nu r } }= {{ \ket{ A-a } \otimes \ket{ a  } } Y_{ \ell } ({ \hat{ r } }_{ A-a,a })  } \frac{ \delta(r - { r }_{ A-a,a }) }{ r { r }_{ A-a,a } } }$ for a relative distance between the clusters ${ r }_{ A-a,a }$. The nuclear wave function is given in terms of the cluster states 
  \begin{equation}
    \ket{ { \Psi } } = \sum_{\nu} \int_{r} dr { r }^{ 2 } \frac{ { g }_{ \nu }(r) }{ r } \hat{ \mathcal{A} } \ket{ { \Phi }_{ \nu r } } \;,
    \label{RGM_ansatz}
  \end{equation}
with unknown amplitudes ${ { g }_{ \nu }(r) }$ that are determined  by solving the integral 
Hill-Wheeler equations (that follow from the Schr\"odinger equation): 
  \begin{equation}
    \sum_{\nu} \int dr { r }^{ 2 } \left[ { H }_{ \nu' \nu } (r',r) - E { N }_{ \nu'\nu }(r',r) \right] \frac{ { g }_{ \nu }(r) }{ r } = 0.
    \label{RGM_equations}
  \end{equation}
  Here,  $H _{ \nu'\nu }(r',r) ={ \bra{ { \Phi }_{ \nu' r' } } \hat{ \mathcal{A} } H \hat{ \mathcal{A} } \ket{ { \Phi }_{ \nu r } } }$ is the Hamiltonian kernel and $N_{ \nu' \nu }(r',r)={ \bra{ { \Phi }_{ \nu' r' } } \hat{ \mathcal{A} } \hat{ \mathcal{A} } \ket{ { \Phi }_{ \nu r } } }$ is the norm kernel, where ${ \hat{ \mathcal{A} } }$ is the antisymmetrizer. The kernels are computed using the microscopic wave functions of the clusters that can be obtained, e.g., in the NSCM. Once the kernels are computed, Eq.(\ref{RGM_equations}) can then be solved using a microscopic ${ R }$-matrix approach \cite{BayeB00,DescouvemontB10}, the code for which is publicly available \cite{Descouvemont16}.
 
 \subsubsection{No-core shell model with continuum}
 \label{sec:ncsmc}
 
 A hybrid basis approach, the no-core shell model with continuum (NCSMC) \cite{baroni13a,baroni13b,navratil16_1956}, uses mixed shell-model and RGM basis to achieve a faster convergence \cite{PRL_Hupin}. 
Another way to include the continuum  in the  NCSM framework  is to start with a continuum single-particle basis, such as the Berggren basis (see Sec. \ref{berggren}).

\subsubsection{Scattering in a bound-state basis}
\label{HORSE}

Somewhat counter-intuitively, one can carry out scattering calculations in a 
basis of bound single-particle states. 
The $J$-matrix  formalism in scattering theory was originally developed
in atomic physics~\cite{HeYa74} utilizing the so-called
Laguerre basis which is a Sturmian-type basis for the Coulomb problem. A generalization of this formalism
utilizing either the Laguerre \red{basis} or the harmonic oscillator basis was suggested in Ref.~\cite{YaFi-m}.
Later the harmonic-oscillator version of the $J$-matrix method was independently rediscovered in
nuclear physics~\cite{filippov1980,filippovGF1981,okhrimenko1984allowance,NeSm82a,NeSm82b}. The $J$-matrix with oscillator basis is sometimes also referred to as an { Algebraic
Version of RGM}~\cite{filippov1980,filippovGF1981,okhrimenko1984allowance} or as  a { harmonic oscillator representation of scattering equations}
(HORSE)~\cite{Bang-m}. 

Within the HORSE approach, the model space is split into internal and external regions. 
In the internal region which includes the basis states with oscillator  quanta $N\leq N_{\max}$,
the Hamiltonian completely accounts for the kinetic and potential energies. The internal region
can be naturally associated with the shell-model space. In the external
region, the Hamiltonian accounts for the relative kinetic energy of the colliding clusters only (and for 
their internal Hamiltonians if needed) and its matrix takes a form of an infinite tridiagonal
matrix of the kinetic-energy operator (plus the sum of eigenenergies of the  colliding clusters at
the diagonal if 
they have an internal structure). The external region  
represents the scattering channels under consideration. If the eigenenergies~$E_{\nu}$,
$\nu=0$, 1,~...\,, and the
respective eigenvectors of the Hamiltonian matrix in the internal region are known, one can
easily calculate the $S$-matrix,  phase shifts and other parameters characterizing the
scattering process (see, e.g., Refs.~\cite{YaFi-m,Bang-m,Svesh-mem,zaitsev1998true}).

The HORSE formalism was successfully utilized in numerous studies of the nuclear continuum
with two- and three-body open channels in cluster models either with phenomenological inter-cluster
interactions (see Ref.~\cite{LurAnn} and references therein) or within the RGM framework
(see Refs.~\cite{Kiev1,Kiev2,Kiev3,lashko2017microscopic,SolIg}
and references therein). The inverse scattering HORSE formalism was also developed~\cite{Ztmf1a,Ztmf1b,Ztmf1c,InvAMS1,InvAMS2,InvAMS3}
and utilized for designing high-quality NN interactions JISP6~\cite{JISP16a,JISP6JPG} and JISP16~\cite{ShirokovMZVW07} describing
the NN scattering data and deuteron properties together with observables in light nuclei.
However, an extension of the NCSM to the states in the continuum by
a direct implementation of the HORSE formalism seems to be unpractical since it requires 
a calculation of 
all spurious-free NCSM eigenstates of a given spin-parity which is impossible 
in  modern large-scale NCSM studies. This drawback can be overcome by the use of a specific
Lanczos-type reformulation of the HORSE method~\cite{Jerry}, which has not been implemented yet.
Another option is to use the so-called Single-State HORSE (SS-HORSE) 
approach~\cite{SSHORSE1,SSHORSE2,blokh1,blokh2,SSHORSE-Coulomb} which makes it possible to calculate the low-energy $S$-matrix and 
resonance energies and widths by a simple analysis of the dependence of the lowest NCSM
eigenstate of a given spin-parity on  parameters defining the many-body NCSM model space, the
oscillator frequency~$\hbar\Omega$ and the basis truncation boundary~$N_{\max}$. The
NCSM-SS-HORSE approach was successfully applied to the studies of the N$\alpha$ 
scattering~\cite{SSHORSE1,SSHORSE2,SSHORSE-Coulomb}
and the resonance in a system of four neutrons (tetranetron)~\cite{PRL4n}.
The Lanczos-type HORSE formalism~\cite{Jerry}, the SS-HORSE~\cite{SSHORSE1,SSHORSE2,blokh1,SSHORSE-Coulomb}
extensions of the NCSM, as well as HORSE-RGM applications \cite{Kiev1,Kiev2,Kiev3,lashko2017microscopic,SolIg, KravvarisV19} are  prospective for future
studies of nuclear reactions.

\red{The advantage of the SS-HORSE approach is its simplicity. As a result, it can be easily implemented in any approach utilizing multi-shell many-body oscillator basis. However the SS-HORSE approach has strong limitations. In particular, only the single-channel version of SS-HORSE has been developed; the multi-channel SS-HORSE version is possible but will be much more complicated, requiring much more computational efforts.  SS-HORSE  can calculate scattering phase shifts and resonance energies and widths but not the scattering wave functions, hence this method cannot be used for calculations of reactions like radiative capture, etc. 

Recently a new technique for calculating reactions using bound-state bases has 
been suggested \cite{PhysRevC.99.034620} based on a reformulation of the Hulth\'en-Kohn variational method  \cite{Zhukov1972}. This approach simplifies calculations of scattering wave functions and cross sections and seems to be applicable to nuclear systems with many scattering channels, but has been tested only on model problems.}


\subsection{Berggren basis}
\label{berggren}


To address the challenges of describing nuclei as open quantum mechanical systems (Sec. \ref{sec.anaoqs}), the quasi-stationary formalism has been developed,
where the state of a many-body system in the continuum is described as a stationary wave with outgoing boundary conditions. While this formalism has been first introduced in 1884 \cite{thomson1884}, it appeared in nuclear physics with the work of G. Gamow in 1928 \cite{gamow28_500}. The same idea was introduced in atomic physics by A. F. J. Siegert in 1939 \cite{siegert39_132}.

Using the bound states, resonances, and scattering states of a given potential, T. Berggren \cite{berggren1968,berggren93_481} demonstrated that a complex-energy basis can be formulated (detailed in  \ref{sec.berggren}), which is especially suitable for the description of loosely bound nuclei. This last attribute makes the Berggren representation easy to implement in codes that use localized basis states, such as harmonic oscillator or Gaussian basis functions. Existing published codes provide the building blocks, i.e. bound states, resonant states and complex energy scattering states, which are eigenfunction of a given mean-field \cite{1982Vertse,1995Ixaru,2018Baran} and can be used to replace the localized basis. 
While dealing with divergent radial wave function or complex energy scattering state requires some extra skill, in addition to the increased computational complexity, the gain of using the Berggren basis is to have correct single-nucleon asymptotics for weakly bound and unbound states. 

\begin{figure}[th]
 \begin{minipage}[]{0.55\textwidth}
  \includegraphics[width=0.9\textwidth]{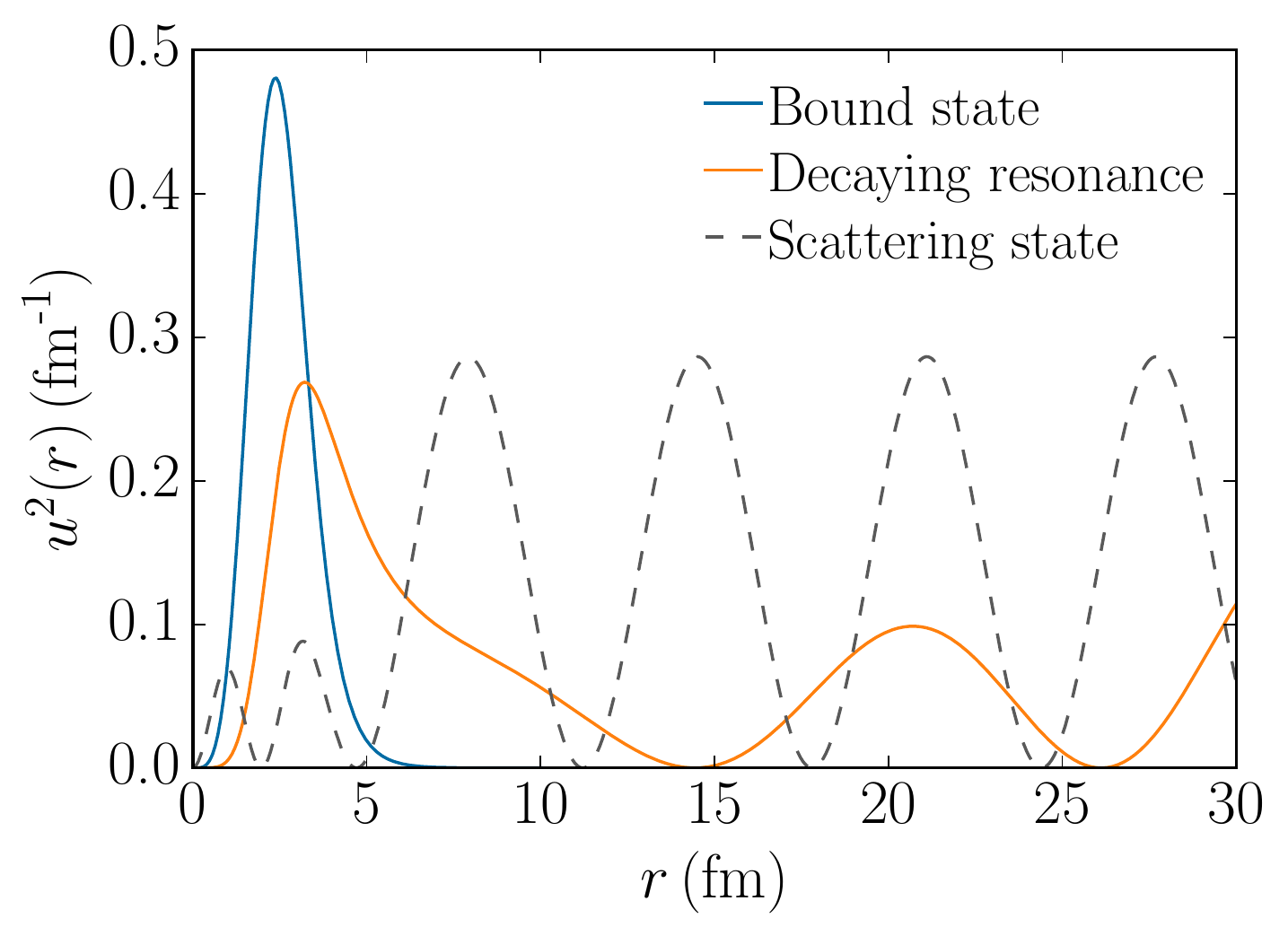}
\end{minipage} \hspace{12pt}
\begin{minipage}[]{0.35\textwidth}
    \caption{    \label{fig_wfs} Typical bound state, real part of the decaying resonance, and scattering radial wave functions. The scattering and decaying resonance wave functions are not normalizable in the usual sense (Hilbert space norm).}
\end{minipage}
\end{figure}

In the quasi-stationary formalism, the energy of a decaying resonance is complex and can be written as 
\begin{equation}
	E = {E}_{0} - i \frac{\Gamma}{2} \quad {\rm with} \quad {T}_{1/2} = \frac{\hbar}{\Gamma} \ln(2),
    \label{eq_Ecx}
\end{equation}
where the real part corresponds to the energy position of the resonance (the peak in the scattering cross section) and the imaginary part is associated to the energy dispersion or width of the resonance (width of the peak). The width can then be used to get the lifetime of the resonance. In short, the quasi-stationary formalism gives an access to the structure of quantum systems in the continuum without using a time-dependent approach.
The price to pay for describing an intrinsically time-dependent process in a time-independent approach is that some of the solutions are not square-integrable anymore, as shown in Fig.~\ref{fig_wfs}.
While the bound state  in Fig.~\ref{fig_wfs} has a localized wave function that falls off at large distances, the scattering state wave function has a plane-wave shape, and the real part of the decaying resonance has a localized inner part but also an \red{oscillating} external part that grows as  the decay width \red{grows}. In fact, resonant and complex energy scattering states are not part of the Hilbert space, but find their place in quantum mechanics when defined in a rigged Hilbert space \cite{gelfand61_b1,gelfand68_b2,maurin68_b4}.

\begin{figure}[htb]
	\centering
	\includegraphics[width=0.75\textwidth]{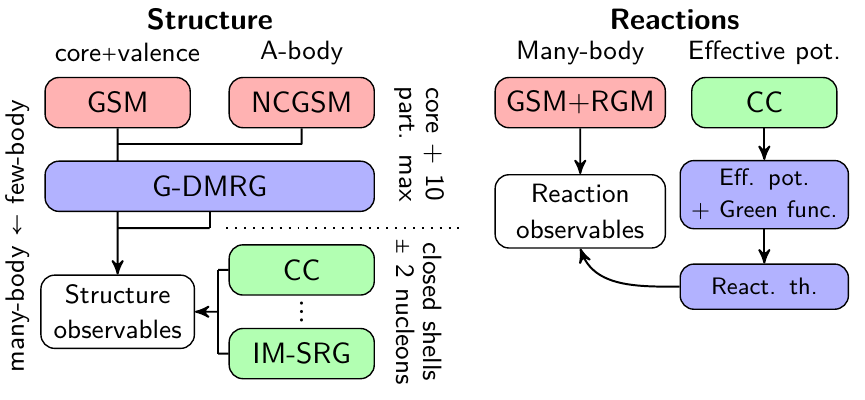}
	\caption{Snapshots of the methods using the Berggren basis for structure (left) and reaction (right) calculations. On the left, several methods are missing such as the particle-plus-rotor model or the Gamow coupled channel in Jacobi coordinates. On the right, the current strategies to obtain reactions observables can be divided in two groups: the use of RGM cluster basis as an extension of structure calculations and the effective potentials methods.
	}
	\label{fig_BB_strcut_react}
\end{figure}

The Berggren basis is a versatile tool that has been used in several nuclear many-body approaches (Fig.~\ref{fig_BB_strcut_react}), starting with the Gamow shell model (GSM) \cite{1996Liotta,2,3}, the density matrix renormalization group method for open quantum systems or Gamow-DMRG \cite{rotureau06_15,rotureau09_140}, as well as in the coupled cluster (CC) method \cite{coupled_cluster}. It has also been used in the particle-plus-rotor model for atomic and nuclear physics problems \cite{fossez16_1335,PhysRevA.94.032511,PhysRevA.98.062515}, and was recently used in the Gamow coupled-channel approach in Jacobi coordinates \cite{COSM_ref,PhysRevLett.120.212502,PhysRevLett.122.122501} to solve the three-body problem. 
The configuration interaction approaches that use the Berggren basis are currently limited to about ten particles (with or without a core), while the CC or IMSRG approaches can reach a hundred active particles, but only around (sub-)closed shell nuclei plus or minus one or two particles. The main issue with all these structure approaches, is that while in principle they take into account channel couplings through the configuration mixing thanks to the Berggren basis, it is not clear how to distinguish individual decay channels. As a consequence, these approaches cannot provide partial decay widths or any reaction observable. In order to circumvent this problem and access reaction observables in Berggren-based approaches, two possibilities are available (Fig.~\ref{fig_BB_strcut_react}, right panel). One possible way is the extension of the many-body structure formalism to the resonating group method (RGM) where reaction channels are defined explicitly, similarly to what was done in the no-core shell model \cite{navratil16_1956}. This strategy was adopted in the GSM+RGM (also referred to as coupled-channel Gamow shell model, GSM-CC) and allows to compute reactions observables directly from the many-body calculations \cite{PRC_GSM_CC_Yannen1,PRC_GSM_CC_Yannen2,Mercenne16,PRC_GSM_CC_Kevin,Dong2017}. The second way is to generate an effective potential for the ``target'' from many-body structure calculations \cite{kumar17_1965,idini19,idini17_1983,rotureau17,Burrows:2018ggt,gennari18_1984} using the many-body Green's function formalism \cite{dickhoff05_b63} (cf. Sec. \ref{optPot}).

\subsubsection{Gamow shell model}

\begin{figure}[th]
\centering{
\includegraphics[width=0.7\textwidth]{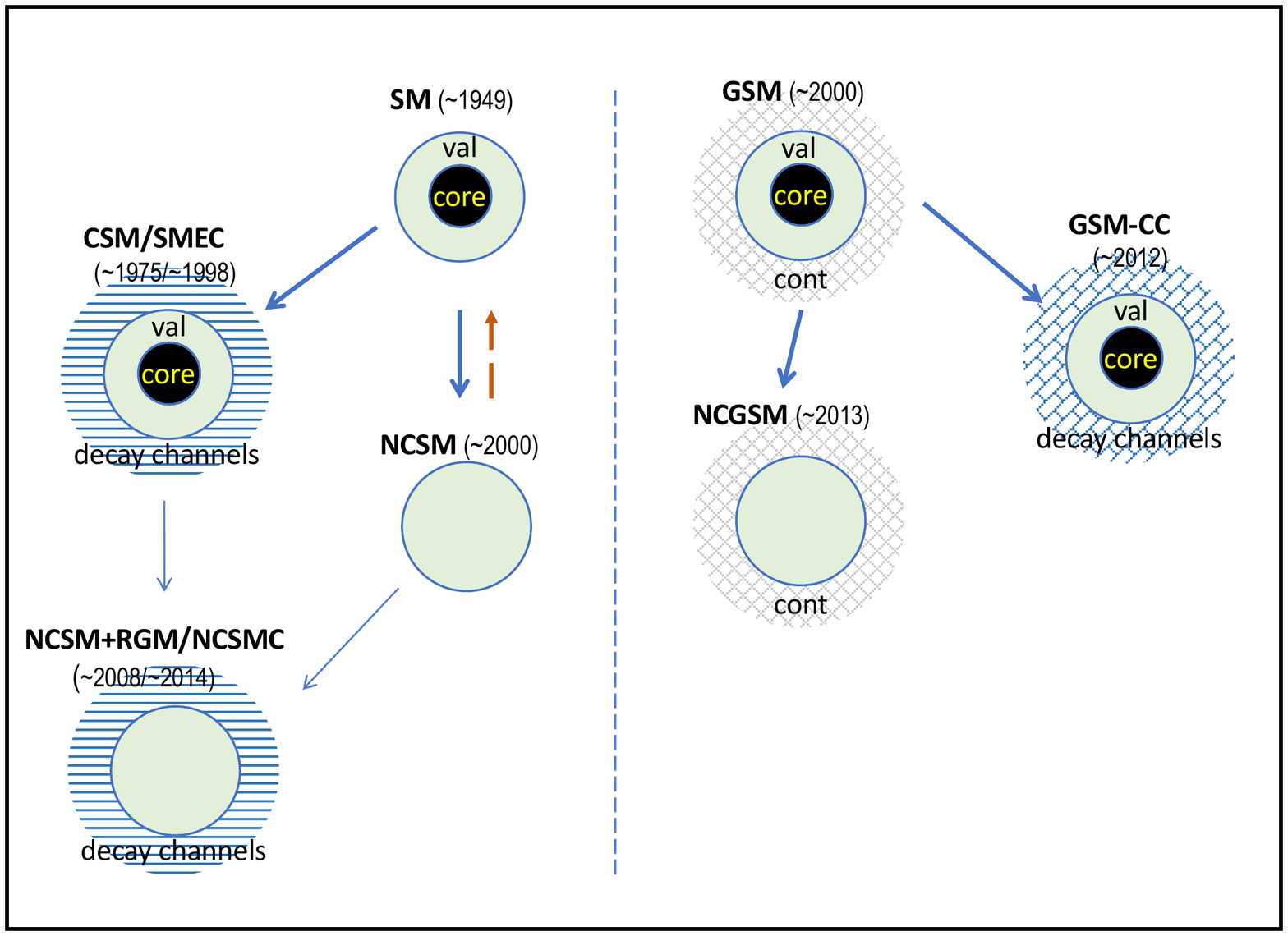}
}
\caption{Recent evolution of paradigms in the low-energy nuclear structure and reaction theory from closed to open quantum systems, \red{from the shell model (SM) and Gamow shell model (GSM) to the continuum shell (CSM)/ shell
model in the continuum (SMEC), the no-core shell model (NCSM), and the no-core-shell model plus continuum (NCSMC); 
and to the GCM with coupled clusters (GSM-CC) and the no-core Gamow shell model (NCGSM).}}
\label{Fig1}
\end{figure}

Early attempts to reconcile discrete and continuum aspects of nuclear many-body problems have been based on the projection formalism \cite{fesh1,fesh2} and lead to the development of the continuum shell model  (CSM) \cite{mahwei,barz} and, more recently, the shell-model embedded in the continuum (SMEC) \cite{benn00,rot05a,rot05b} which provides a unified approach to low-energy nuclear structure and reactions. In the SMEC, one couples eigenstates of the phenomenological SM Hamiltonian with relevant reaction channels to describe the level spectroscopy and the reaction cross sections in the same many-body formalism \cite{Oko03} (Fig. \ref{Fig1}).

The first open quantum system formulation of the nuclear SM, respecting unitarity at the particle emission threshold(s), has been achieved in the Gamow shell model (GSM) \cite{1996Liotta,2,3,mic03,JPG_GSM_review}. The many-body states in GSM are given by the linear combination of Slater determinants defined in the Berggren ensemble of single-particle states.
Reaction channels are not explicitly identified, so the GSM with this Slater determinant representation is the tool for spectroscopic studies of bound and unbound states and their decays. Most numerical applications of the GSM have been done by separating an inert core and using the cluster orbital shell model \cite{10} relative variables in the valence space. In this way, spurious center-of-mass excitations are removed. 

The solution of an eigenvalue problem involving the continuum states is an acute numerical problem. 
The dimension of the many-body valence space increases catastrophically with the number of valence nucleons and the size of the single-particle basis. In GSM, each single-particle state of the discretized scattering contour becomes a new shell in the many-body calculation. Moreover, the use of Berggren ensemble implies complex-symmetric matrices for the representation of the Hermitian Hamilton operator. 

For the description of scattering properties and reactions, the entrance and exit reaction channels have to be identified. This can be achieved in GSM by expressing wave functions in the complete basis of the reaction channels (in a procedure that is based on the RGM described in Sec. \ref{RGM}). This coupled-channel representation of the Gamow shell model (GSM-CC) has been recently applied for various observables involving one-nucleon reaction channels, such as the excitation function and the proton/neutron elastic/inelastic differential cross sections \cite{PRC_GSM_CC_Yannen1,PRC_GSM_CC_Yannen2,Mercenne16}, or low-energy proton/neutron radiative capture reactions \cite{PRC_GSM_CC_Kevin,Dong2017}. Channels in these reactions are given by the initial/final GSM eigenvectors of $(A-1)$-body system coupled to proton/neutron in the continuum states. Resulting $A$-body wave functions are antisymmetrized and the separation of core and valence particles allows the GSM-CC to be applied in medium-heavy and heavy nuclei. One should stress that channels in GSM-CC are built from GSM wave functions which respect the unitarity at the decay thresholds of each cluster subsystem. The extension of the GSM-CC approach to reactions involving cluster reaction channels, such as deuteron or $\alpha$-particle reaction channels, has been recently  developed as well \cite{MercenneMP19}. In particular, the calculated center-of-mass differential cross sections of the $^4$He(d,d) elastic scattering in the GSM-CC \cite{MercenneMP19} reproduces both the resonance spectrum of $^6$Li, the phase shifts and the elastic scattering cross sections with a similar precision as the NCSMC  \cite{PRL_Hupin}. However, GSM-CC can be applied to study the elastic scattering and transfer reactions involving many-nucleon projectiles whose composites cannot be reached in the NCSMC approach. 

%
For 
light nuclei, the no-core Gamow shell model (NCGSM) formulation has been recently successfully applied to investigate the resonant states of helium and hydrogen isotopes and the existence of tetraneutron \cite{5,PRL_Fossez}.

\begin{figure}[th]
\includegraphics[scale=0.3]{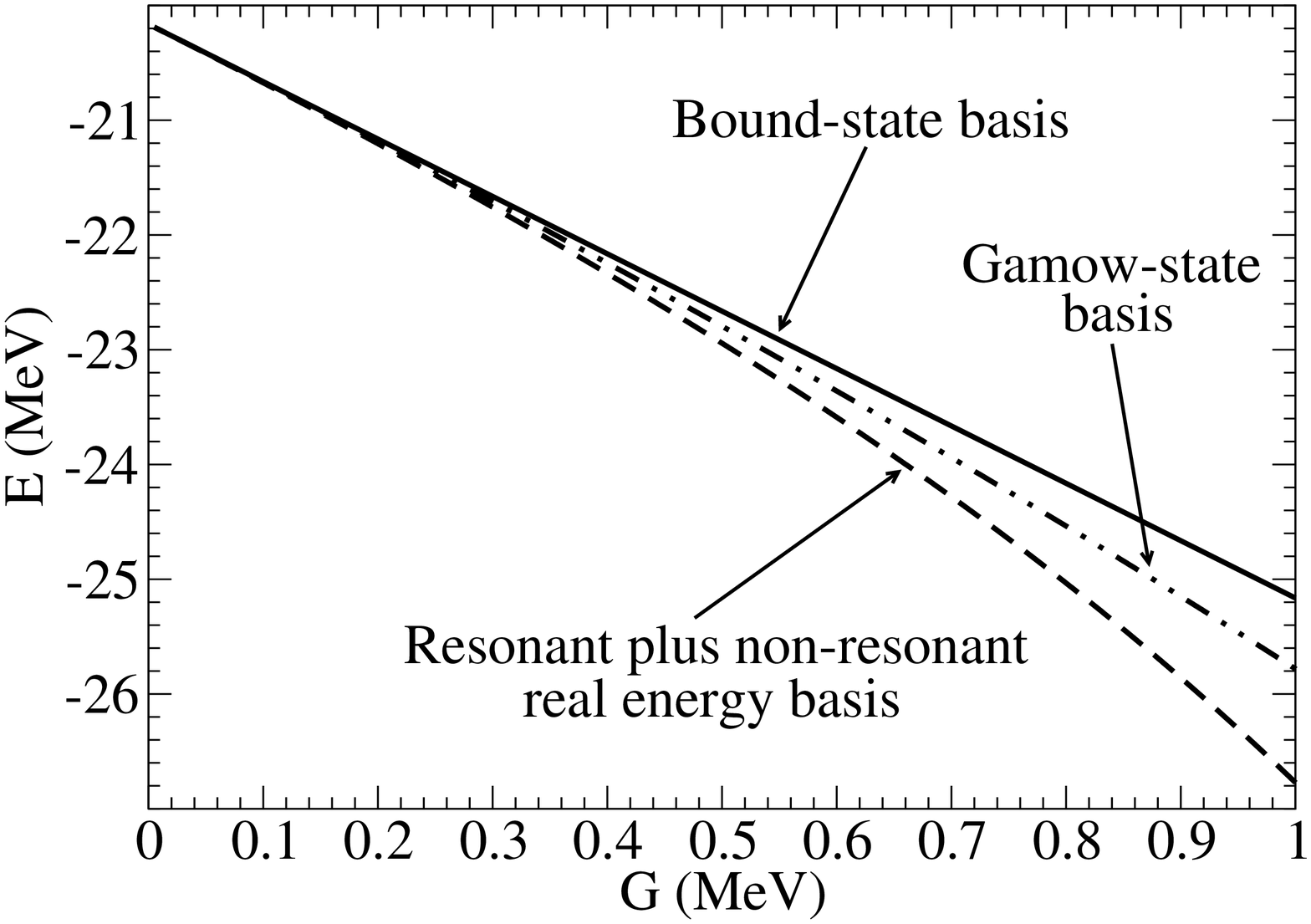}
\includegraphics[scale=0.3]{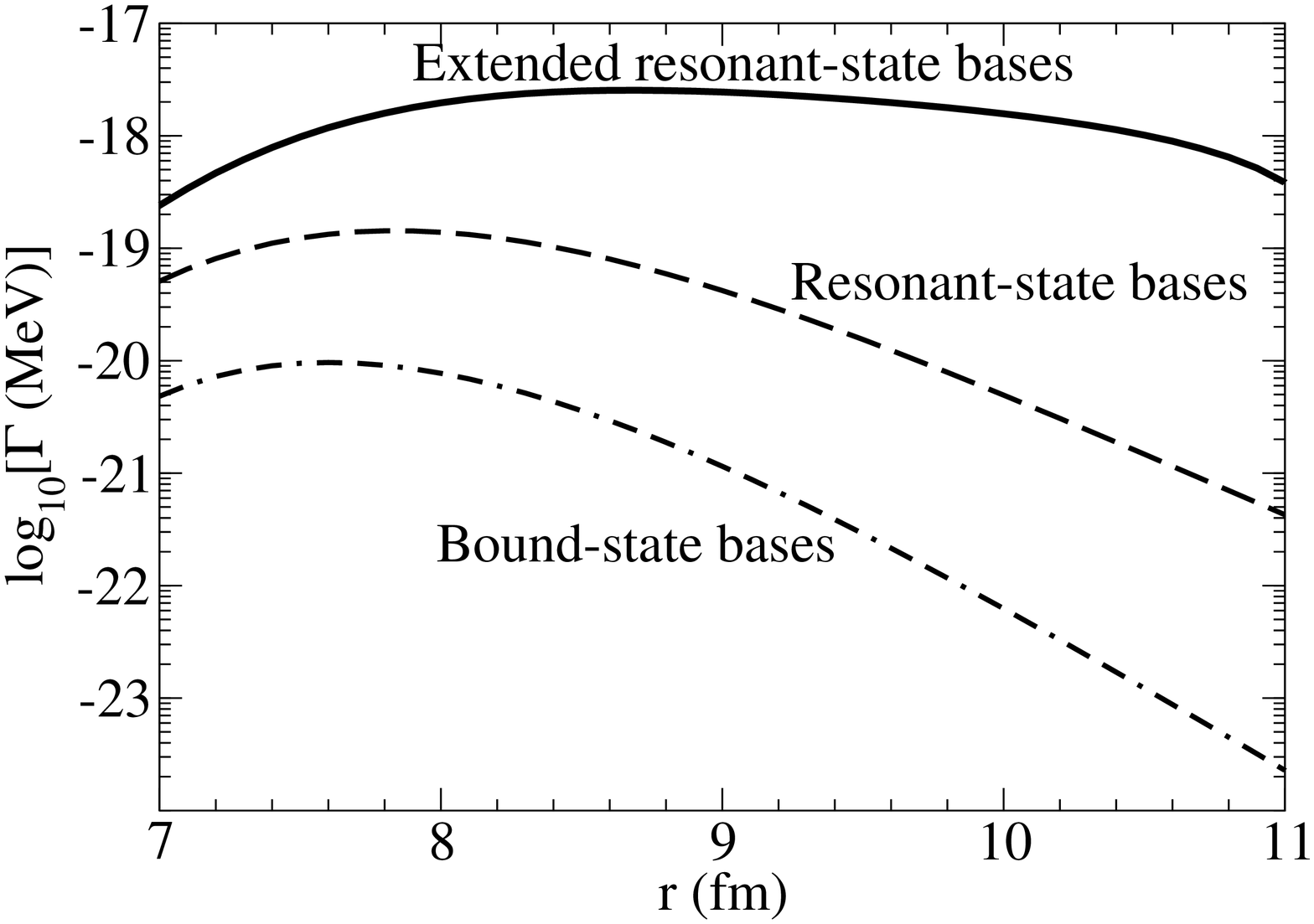}
\caption{\label{fig.c22} (a) Assessment of the bound states and the continuum in the ground state energy of the drip-line nucleus $^{22}$C as a function of the pairing strength $G$. (b) Assessment of the continuum in the alpha decay calculation in $^{212}$Po.
\newline
\red{
{\it Source:} Figure 15(a) from \cite{2012IdBetan} $\copyright$~APS. Reproduced with permission. \url{http://dx.doi.org/10.1103/PhysRevC.85.064309}; 
Figure 15(b) from \cite{2012IdBetanWitek} $\copyright$~APS. Reproduced with permission. \url{http://dx.doi.org/10.1103/PhysRevC.86.034338}
}
}
\end{figure}

One can explore the coupling to the continuum through exactly solvable models, for example,   pairing \cite{1957Bardeen}  which is 
important in nuclear systems \cite{1958Bohr,1959Belyaev} and has a key role in 
weakly bound systems \cite{PhysRevC.53.2809}.
Richardson introduced a method to solve the system with constant pairing but with 
non-degenerate single-particle energies \cite{1963Richardson} which has since become popular \cite{2000Sierra,2005Brink,Guan12,2013Dukelsky}. The application 
to continuum states via the Gamow states \cite{2003Hasegawa,2012IdBetan} must include the correlations with the resonant and non-resonant parts of the continuum spectrum of energy.
Figure \ref{fig.c22} (a) shows the relative contribution of the bound, resonant and non-resonant continuum in the drip-line nucleus $^{22}$C.
 The analytic extension to the complex energy plane shows that the Berggren basis emerges naturally in the conserving particle number solution of the pairing Hamiltonian \cite{2016Mercenne,2017IdBetan}. This can be applied to studies of both loosely bound  and  unbound nuclei \cite{2012IdBetan}.

One of the biggest challenges in the microscopic description of nuclear drip-line physics is the  asymptotic behavior of the many-body wave function. This is particularly so for  alpha decay. In the $R$-matrix formalism the $\alpha$-decay absolute width is the product of two functions, i.e. penetrability and reduced width \cite{1954Thomas}, but when 
using a harmonic oscillator basis \cite{1957Mang,1988Dodig,2000Delion},  the
incorrect asymptotic behavior leads to an absolute width strongly depends on the radial coordinate. Since the Berggren basis  includes both the continuum and the proper asymptotics, one expects that the  product of the two functions becomes independent of the radial coordinate. 
This is demonstrated in 
Fig.~\ref{fig.c22} (b), where  the behavior of the absolute width $\Gamma$ as the single-particle model space increases from a representation which includes only bound states, resonant states, and finally an extended resonant-state basis, when a schematic separable force is used \cite{2012IdBetanWitek}. 
A proper treatment of the microscopic alpha decay in the vicinity of the drip-line nuclei  should take into account all nucleon-nucleon correlations, e.g.,  with a suitable effective interaction in the continuum \cite{2017Yannen,2018Auranen}.

\subsubsection{Decoupling methods with Berggren basis}
\label{IMSRG_CC}

The complex-energy Gamow-Berggren framework has been recently merged with the IMSRG approach that uses a chiral EFT force \cite{8}. The Gamow IMSRG has been successfully applied to neuron-rich carbon isotopes with significant results, e.g., pointing to a halo structure for the dripline nucleus $^{22}$C. Some interesting resonant excited states are predicted, which would be valuable for future experiments. 
However, there are still some issues needed to be solved in the Berggren calculations. The complex resonant Berggren wave functions are not square integrable, which results in problems in the direct calculations of nuclear radii and electromagnetic transitions. The long-range Coulomb interaction is also difficult to  handle, due to the non-integrability of resonant wave functions. Though one can write an intrinsic Hamiltonian with the center-of-mass motion removed, the spurious center-of-motion excitation in the Berggren wave functions is not easily treated in the cases where a Woods-Saxon or Hartree-Fock single-particle basis is used; for comparison, for the harmonic oscillator basis, the spurious center-of-motion excitation can be straightforwardly removed by using the Lawson method \cite{9}. One may use the so-called cluster-orbital shell-model framework to avoid the spurious center-of-mass excitation in the Berggren many-body wave functions \cite{10, 11}. However, in the \textit{ab initio} types of the Gamow calculations, the transformations of realistic nucleon-nucleon interactions to the cluster-orbital shell-model scheme are very difficult to handle. 
In addition, the Gamow numerical computations are performed with complex numbers and complex functions, therefore one needs to pay special attentions on the convergences of complex-number calculations. Due to the existence of scattering channels, the model space of a Gamow calculation can be huge, requiring substantial computer resources. Finally, the stability of the Gamow numerical computations might be an issue, particularly for the imaginary parts of the resonance or continuum results. Nevertheless, the Gamow \red{framework} provide a powerful tool to calculate the structure and reactions of bound, weakly-bound and unbound systems.


\subsection{Effective inter-cluster interactions (optical potentials)}
\label{Sect:fromMB}

Exact solutions for the scattering
problem have only been formulated and carried out for few-nucleon systems. Todays most
advances in exact scattering  calculations have been carried out for the four-nucleon
system \cite{Fonseca:2017koi,Deltuva:2017bia,fad2b,Viviani:2016cww}. 
Applying this 
approach to nuclear reactions has been and still is isolating
important degrees of freedom, thus reducing the many-body problem to a few-body problem,
and solving the few-body problem exactly~\cite{deltuva2009}.

Isolating important degrees of freedom means projecting onto a reduced Hilbert space and
thus creating effective interactions between the degrees of freedom that are treated
either exactly or with ``controlled'' approximations. Since the 1960's (or earlier) such
effective interactions have been constructed by fitting relevant experimental data with
usually complex functions, leading to the well known phenomenological optical model
potentials (see e.g.~\cite{Furumoto:2019anr,Weppner:2009qy,koning:2003zz,Varner:1991zz}),
 which are local and energy-dependent, and largely still the only method available for 
 practical applications.  
Nonetheless, 
an overarching goal is to construct such effective interactions
from the same first principles that govern recent advances in many-body approaches to
nuclear structure. 

The effective interaction between a nucleon and a nucleus is one of the most important
ingredients for reaction theories. Theoretical formulations have been introduced early
on by Feshbach, leading to the Green's function formulation \cite{CapMah:00}. 
The theoretical approach to elastic scattering of a nucleon from a nucleus, pioneered by
Watson~\cite{Watson1953a,Watson1953b}, made familiar by Kerman, McManus, and
Thaler~\cite{KMT} and being refined further as the spectator
expansion~\cite{Siciliano:1977zz} 
leads to a  multiple scattering expansion that  can employ structure and
reaction contents on equal footing in an order by order fashion.
Both approaches start from a many-body 
Hamiltonian employing two- and three-nucleon forces.  



\subsubsection{Low energies: Green's function plus coupled cluster method \label{optPot}}
The computation of a many-body propagator can be used to generate an effective interaction between the few sub-systems (clusters) that participate in a
reaction process 
(e.g., nucleon-nucleus optical potential).
An ideal framework for constructing the nucleon-nucleus optical potential is the Green's function propagator. This is the propagator related to the $A+1$ and $A-1$ system with respect to the $A$ system, and the self-energy 
arising from the Dyson equation, see Eq. (\ref{eq:Dyson}),
is the desired effective potential \cite{escher02}.
The Green's function can be calculated self-consistently from first principles \cite{barbieri17,idini19} or constructed from a phenomenological approach \cite{idini15} (cf. Sec. \ref{Sect:Green}). It can eventually be built also using solutions from other many-body methods \cite{rotureau17,gfccpap2} (cf. \ref{App:Green}).

For the CC method (Sec. \ref{sect:CC}), the matrix elements of the corresponding \red{one-particle} Green's function,
\red{evolving from initial single-particle state $\alpha$ to final state $\beta$, are}
\begin{eqnarray}
\lefteqn{G^{CC}(\alpha,\beta,E) \equiv }\nonumber\\
&& \langle \Phi_L|\overline{a_{\alpha}}\frac{1}{E-(\overline{H}-E^{A}_{gs})+i\eta}\overline{a^{\dagger}_{\beta}}|\Phi_{}\rangle \nonumber \\
&+&\langle \Phi_{L}|\overline{a^{\dagger}_{\beta}}\frac{1}{E-(E^{A}_{gs}-\overline{H})-i\eta}\overline{a_{\alpha}}|\Phi_{}\rangle ,
\label{gfcc}
\end{eqnarray}
where $\overline{a_{\alpha}}=e^{-T}a_{\alpha}e^T$ and
$\overline{a^{\dagger}_{\beta}}=e^{-T}a^{\dagger}_{\beta}e^T$ are respectively,  the
similarity-transformed annihilation and creation operators. \red{Here $| \Phi \rangle$ is the CC reference 
state as in Eq.~\ref{CC_ansatz}}, 
while $\langle \Phi_{L}|$ is the left eigenvector associated with the ground state of $\overline{H}$ ($\overline{H}$ being not Hermitian  has left- and right-eigenvectors \cite{bartlett2007,coupled_cluster}).
By definition, the parameter $\eta$ is such that $\eta \rightarrow 0$ in the
physical limit.
In practice,  the Green's function (\ref{gfcc})  is calculated by working in the complex
Berggren basis \cite{berggren1968,3} which (i) enables the description of  bound, resonant and scattering states
of the $A$ and $A\pm1$ nuclei 
and (ii) removes numerical instability associated with the poles of the Green's function.

\begin{figure}[t]
\begin{center}
 \begin{minipage}[*]{0.5\textwidth}
\includegraphics[width=0.85\textwidth]{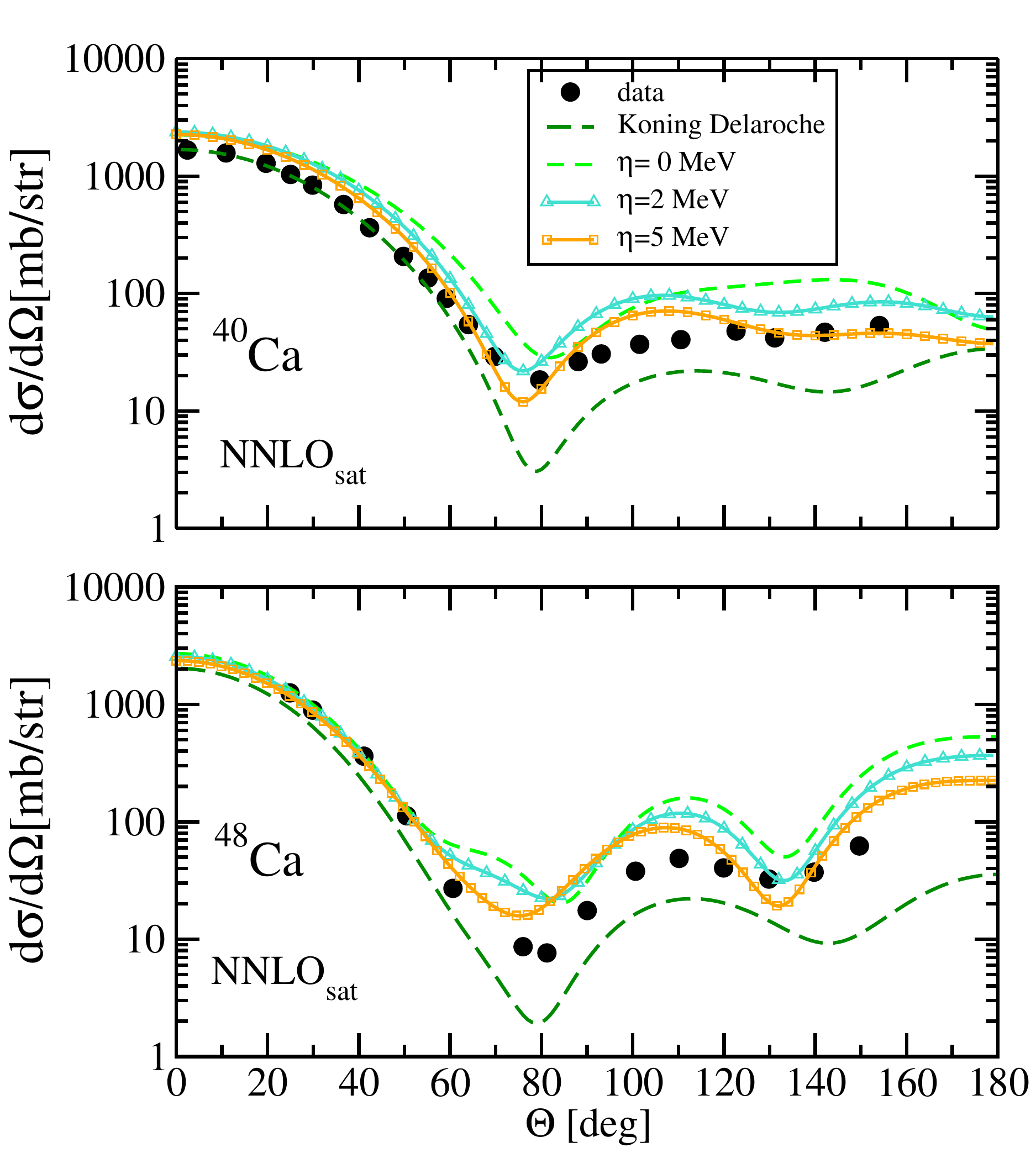}
\end{minipage}
 \begin{minipage}[*]{0.49\textwidth}
\caption{\label{fig4048}
CCSD differential elastic cross section for  $^{40}$Ca(n,n)$^{40}$Ca at 5.17 MeV (top)  and
  $^{48}$Ca(n,n)$^{48}$Ca at 7.81 MeV (bottom) calculated with the $\rm{NNLO_{sat}}$  interaction \cite{gfccpap2}.
  Calculations are shown for  $\eta=0,2,5$ MeV.  Results obtained
  using the phenomenological Koning-Delaroche  potential are shown for comparison.
Data points are taken from \cite{koning:2003zz} (errors on the data
are smaller than the symbols).
\newline
\red{
{\it Source:} Figure adapted from J. Rotureau et al. (2018) Phys. Rev. C 98, 044625 $\copyright$~APS. Reproduced with permission. \url{http://dx.doi.org/10.1103/PhysRevC.98.044625}.
}
}
\end{minipage}
\end{center}
\end{figure}

Figure \ref{fig4048} shows calculations for the neutron elastic cross sections \red{on} $^{40}$Ca and $^{48}$Ca calculated with
the optical potential  obtained by inverting  the Dyson equation fulfilled by the Green's function calculated at the  CCSD level \cite{gfccpap2}. 
Calculations were performed using the $\rm {NNLO_{sat}}$ chiral interaction \cite{n2losat} (which contains both NN+\red{3N}
terms) which reproduces the binding energy and charge radius for both systems~\cite{hagen2015,garciaruiz2016}.
Results are shown for  respectively 5.17~MeV and 7.81~MeV in order to compare  with available experimental data.
In both cases, one expects the calculated optical potentials
to have a finite imaginary part which reflects the loss of flux in the elastic channel.
More precisely, in the case of neutron scattering on $\rm{^{40}}$Ca at $E=5.17$~MeV, there is a potential
absorption due to excitation of $\rm{^{40}Ca}$
to  its first excited state \red{at} $E(0^+)= 3.35$ MeV  or second excited state \red{at} $E(3^-)= 3.74$ MeV.
However, the calculated  optical potentials yield  a negligible value for the absorption in all partial waves \cite{gfccpap2},
%
\red{which} indicates that correlations beyond the singles and doubles truncation
level in the CC method are needed to account for the absorption due to the target excitation.
A similar situation occurs for  the neutron scattering off  $\rm{^{48}Ca}$ at 7.81~MeV.

As shown in  Fig.~\ref{fig4048}, one could artificially increase absorption by
considering finite values of $\eta$ instead of taking the limit $\eta \rightarrow 0^+$. When $\eta$ increases, the
elastic scattering cross section decreases with a more pronounced
relative reduction at larger angles and the agreement with data improves.
Angular distributions calculated with the phenomenological
Koning-Delaroche  potential \cite{koning:2003zz} are also shown for comparison in  Fig.~\ref{fig4048}.
%
%
%



\subsubsection{Intermediate energies:  multiple scattering
method plus NCSM}

\label{sec:MultiScatt}

\red{As in the previous section, here we focus on nucleon-nucleus scattering}.
The spectator expansion is constructed within a multiple scattering theory predicated
upon the idea that two-body interactions between the projectile and the target nucleons
inside the nucleons play a dominant role. 
Thus, the leading-order term involves two-body interactions between the projectile and one
of the target nucleons (represented by one-body nuclear density), the second order term
involves the projectile interacting with two target nucleons and so forth. 
While the original spectator expansion~\cite{Siciliano:1977zz} referred to expanding the
many-body transition amplitude,  in pursuit of deriving an effective
interaction between the projectile and the target nucleus it is more natural to expand the effective
potential operator~\cite{Chinn:1993zza} in terms of active particles. 
\begin{figure}[h]
\begin{center}
\includegraphics[scale=0.35]{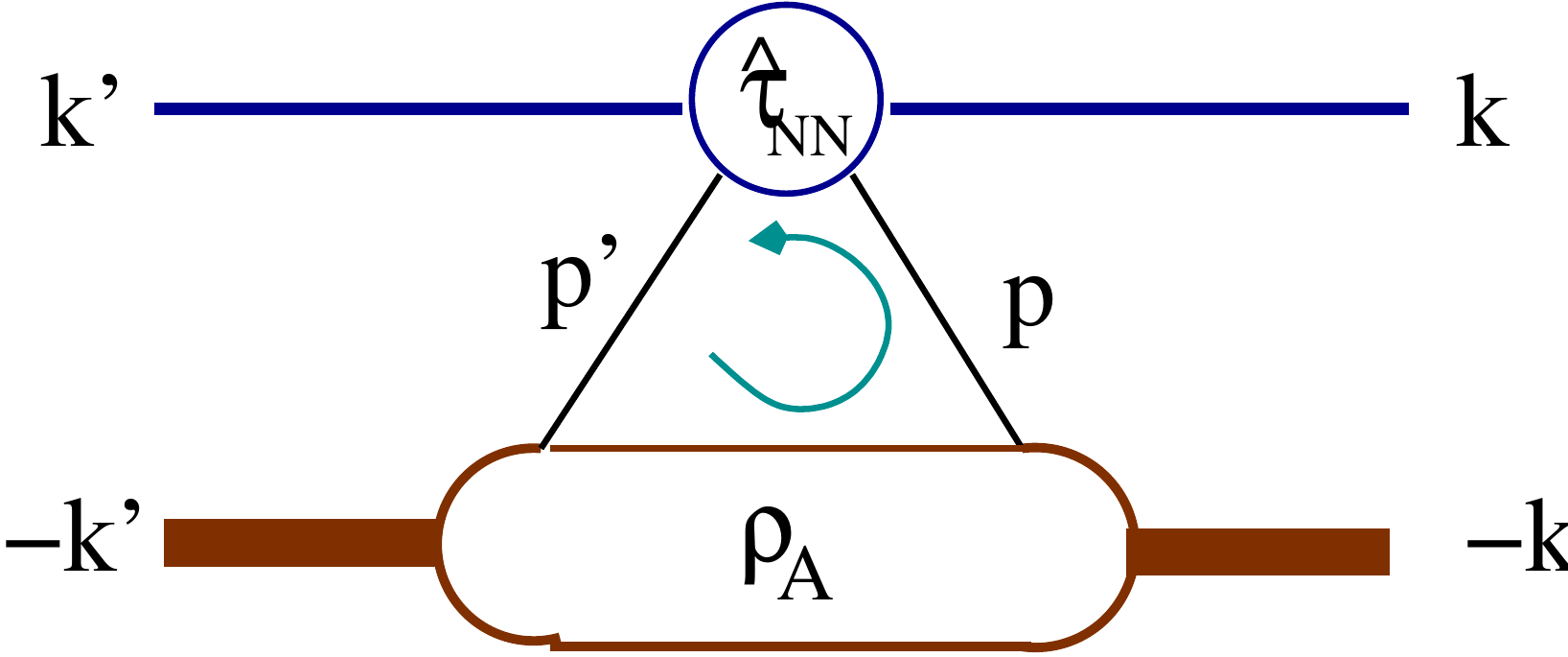}
\caption{Schematic illustration of the effective (optical) potential $\hat{U}\left({\bf
k^{\prime}},{\bf k}\right)$ for the single scattering term in the the multiple scattering
approach, where the momenta ${\bf k}$ and ${\bf k'}$ are the initial  and final momenta
of the projectile in the frame of zero total nucleon-nucleus momentum, $\hat{\tau}$ is
the NN t-matrix. The same nucleon-nucleon NN interaction is used to calculate the nuclear
density $\rho_A$ of the target.
\newline
\red{
{\it Source:} Figure from M. Burrows et al. (2019) Phys. Rev. C 99, 044603 $\copyright$~APS. Reproduced with permission. \url{http://dx.doi.org/10.1103/PhysRevC.99.044603}.
}
}
\label{Tmatrix}
\end{center}
\end{figure}

\begin{figure}[h]
\begin{center}
\begin{minipage}[]{0.57\textwidth}
\includegraphics[width=1.1\textwidth]{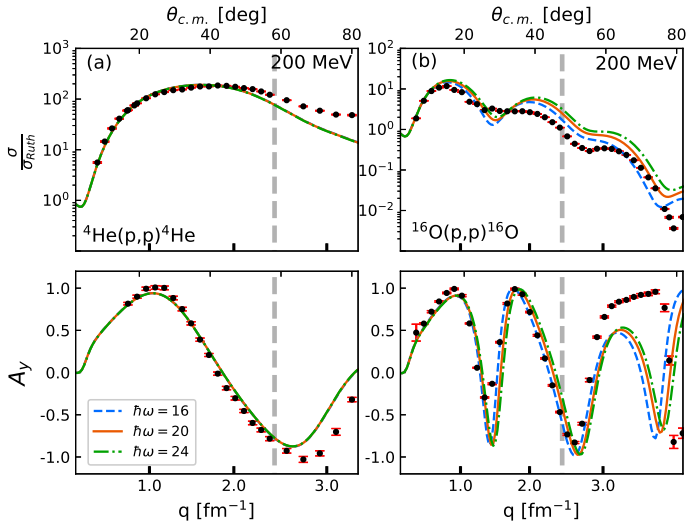}
\end{minipage}
\begin{minipage}[]{0.42\textwidth}
\caption{The angular distribution of the differential cross section divided by the
Rutherford cross section and 
of the analyzing power for elastic
proton scattering from $^{4}$He (left panels) and $^{16}$O (right panels) 
at 200~MeV laboratory kinetic energy
as function of the momentum transfer $q$ or the c.m. angle, calculated with the
NNLO${_{\rm{opt}}}$ chiral interaction~\protect\cite{Ekstrom13} and $N_{\rm{max}}$~=~18(10) for
$^4$He ($^{16}$O).
The data are from~Refs.~\cite{Moss:1979aw,Glover:1985xd}.
The dashed vertical line indicates 
$q=2.45$~fm$^{-1}$ corresponding to the energy 
of the $np$ system up to which the NNLO${_{\rm{opt}}}$ was fitted.
}
\label{pA-Observables}
\end{minipage}
\end{center}
\end{figure}

Current implementations of this expansion
are based  on two active particles leading to an effective potential
schematically given in Fig. \ref{Tmatrix} and meaning that one-body nuclear density and a
two-nucleon transition amplitude determine the leading-order {\it ab initio} effective nucleon-nucleus 
interaction, which is nonlocal as well as energy-dependent.
 We want to emphasize that now
the one-body nuclear density and the two-nucleon transition amplitudes are not only derived from the same
underlying nucleon-nucleon interaction~\cite{Burrows:2018ggt} but also enter on the same footing in the structure and reaction part of the calculation~\cite{Burrows-new}. Figure~\ref{pA-Observables} shows the angular
distribution of the differential cross section and the analyzing power for \red{protons on } $^4$He and
$^{16}$O \red{targets} at 200~MeV laboratory projectile kinetic energy with the effective interaction
calculated as described in~\cite{Burrows:2018ggt} and consistently based in the
NNLO$_{\rm{opt}}$ chiral interaction from~\cite{Ekstrom13}. 

It should also be pointed out that the many-body
character of the free Green's function treated in \cite{Burrows:2018ggt} in the extreme
closure approximation in principle connects even the first order of the effective
potential to two-body nuclear density,  if one wants to include the effect of other
nucleons in the nucleus on the struck target nucleon. In a mean-field picture this effect
was estimated in~\cite{Chinn:1995qn} and found to be important only at energies below
100~MeV projectile kinetic energy. Similarly, in the lowest order two-body antisymmetry
is achieved through the use of
two-body t-matrices which are themselves antisymmetric in the two
``active'' variables (corresponding to the weak binding limit in~\cite{GoldbergerWatson}).
For the next
order, requiring two-body densities, three ``active'' variables need to be
antisymmetrized, an effect which has been estimated in~\cite{Crespo:1992zz} and found
small in the regime of 200~MeV projectile energy. Genuine three-nucleon force effects
will only enter in the next order of the spectator expansion and also require two-body
densities as well as solving a three-body problem for the three active nuclei. However,
useful insights into the size and energy dependence 
 of those contributions may already be obtained by an approximate solution.

\subsection{Response functions and sum rules }

\begin{figure}[ht]
 \begin{minipage}[]{0.5\textwidth}
 \includegraphics[width=1\textwidth]{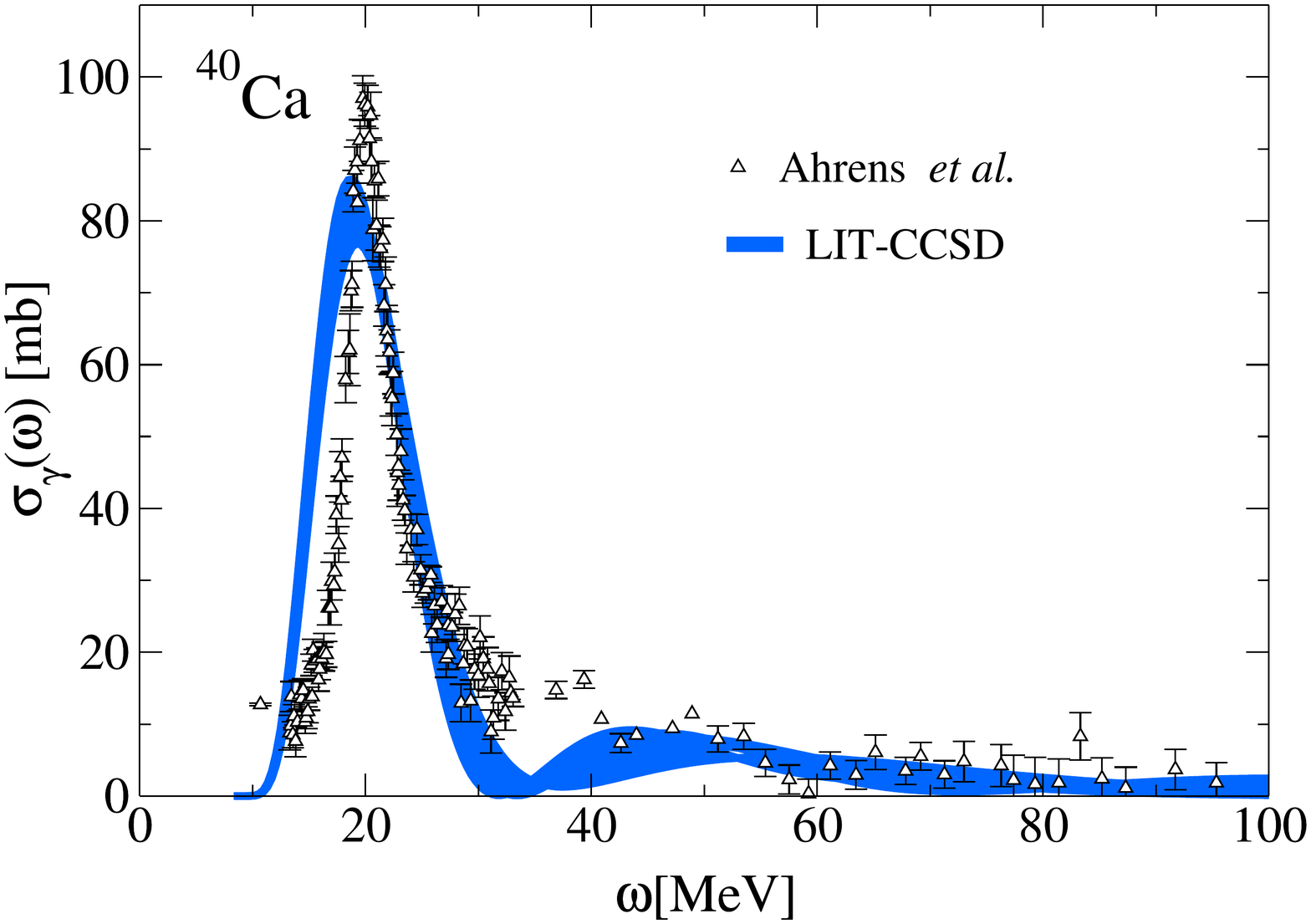}
 \end{minipage}
  \begin{minipage}[]{0.49\textwidth}
\caption{\label{fig_Ca40}$^{40}$Ca photonuclear cross section from the LIT method used in conjunction with coupled cluster theory \cite{Bacca2014} (in the coupled cluster singles and doubles approximation, CCSD) compared to experimental data~\cite{ahrens1975}. Result obtained with a two-body Hamiltonian from chiral effective field theory.
\newline
\red{
{\it Source:} Figure adapted from S. Bacca et al. (2014) Phys. Rev. C 90,  064619 \cite{PhysRevC.90.064619} $\copyright$~APS. Reproduced with permission. \url{http://dx.doi.org/10.1103/PhysRevC.90.064619}.
}
}
\end{minipage}
\end{figure}

\red{Unlike reactions mediated by the strong force, electroweak reactions are perturbative and 
well-described by Fermi's Golden Rule. This means for electroweak reactions, one only needs 
the response function,}
\begin{equation}\label{eq:S_omega}
R(\omega)=\int \sum_f |\bra{\psi_f} \hat{O}  \ket{\psi_0}|^2 \delta\left(E_f-E_0-\omega\right),
\end{equation}
where  $\ket{\psi_0}$ and $ \ket{\psi_f }$ are the initial and final state of the nucleus, with energy $E_0$ \red{and} $E_f$, respectively,   $\hat{O}$ is the excitation operator inducing the transition and $\omega$ is the excitation energy.
While the final states $\ket{\psi_f}$ typically include several break-up channels, the
response functions, as well as the corresponding moments, also called sum rules, can be obtained without explicitly solving \red{the many-body Schr\"odinger equation (or equivalent)} for the complicated final states in the continuum, by utilizing an integral transform approach.  A prominent example is the Lorentz integral transform (LIT) method~\cite{Efros:1994_PLB,Efros:2007_JPG}, where one applies an integral transformation on the response function using a Lorentzian kernel as
\begin{eqnarray}
L (\sigma, \Gamma) =  \frac{\Gamma}{\pi} \int d\omega \frac{R(\omega)}{(\omega - \sigma)^2 + \Gamma^2} =\langle \tilde{\psi} | \tilde{\psi} \rangle.
\end{eqnarray}
Here,  $\sigma$ and $\Gamma$ are the centroid energy and the width of the Lorentzian kernel, respectively. The Lorentz integral transform $L (\sigma, \Gamma)$ can be calculated as  the squared norm of the state $\ket{\tilde{\psi}}$, which is found as a unique solution to the bound-state-like equation
$\left( \hat{H} - z \right)  \ket {\tilde{\psi}}=\hat{O}\ket{\psi_0}$,
where  $z=E_0 + \sigma + i\Gamma$. The response function $R(\omega)$ is then recovered by numerically inverting the integral transform.
This method essentially reduces the problem to one that can be tackled by any suitable many-body bound-state technique.
Few- and many-body approaches previously used include hyperspherical harmonics expansions~\cite{Bacca2002,Gazit2006}, the no-core shell model~\cite{Stetcu:2007_NPA,Quaglioni2007}, coupled cluster theory~\cite{Bacca:2013_PRL,Bacca2014} (Fig. \ref{fig_Ca40}), and the symmetry-adapted no-core shell model~\cite{BakerSOTANCP42018,Baker2019}. 

The LIT method allows to obtain response functions by inverting transforms calculated with a finite width $\Gamma$. In  many-body calculations the convergence of such transforms in terms of the model space size is much faster than the convergence of a discretized response  computation, which corresponds to a calculation in the limit of $\Gamma \rightarrow 0$. Moreover, after the inversion of a finite-width LIT one takes the final state interaction in the continuum properly into account, as shown in benchmark calculations with few-body methods that explicitly calculate the final states~\cite{Golak:2002mw}. 
The LIT method has been applied to study electroweak response functions of various kinds, see e.g.~\cite{Efros:2007_JPG} and references therein, and more recently also to the study of  nuclear polarization effects for muonic atoms~\cite{Ji_PRL_2013}.
However, if one is interested only in integral properties \red{(moments)} of the response functions, namely in sum rules, the Lanczos sum rule (LSR) method (see, e.g., \cite{Dagotto:1994_RMP,NND_PRC_2014} and references therein) is especially suitable.
It has been shown that the convergence pattern in terms of the model space size is similar to that of the LIT transform~\cite{NND_PRC_2014}. This method has been used 
 for computing sum rules of low-energy electromagnetic transitions,  leading  for example to  interesting finding regarding the nuclear electric dipole polarizability~\cite{hagen2015,Miorelli2016}. More recently  it has  also been used to compute  monopole, dipole and quadrupole sum rules with various weights (e.g., see Ref.~\cite{BakerSOTANCP42018,Baker2019}), some of which are particularly relevant to muonic atoms (see e.g.~\cite{NND_PRC_2014,NND:2016_PhysLettB,Ji2018}). 

\subsection{Time-dependent basis function method}



The time-dependent Basis Function (tBF)
method  solves non-perturbative and time-dependent scattering problems in quantum
mechanics~\cite{ Du:2018tce}. The tBF is designed to utilize NCSM solutions, 
or solutions from other 
many-body methods,  directly for a limited set of reactions (such as Coulomb excitation, or Coulex) 
in its initial formulation.  For the system being probed, one first solves 
for the bound states and states above breakup threshold 
that form a discretized representation of the system’s continuum.
Then, for Coulex, one evaluates all possible electromagnetic transition matrix 
elements among these states.  In the initial applications, the E1 transition is 
expected to dominate and is the only transition retained.  The impinging
nucleus is treated in the interaction representation and is time-evolved through
the strong external field of the target at the amplitude level
so that the final state is a coherent superposition of all available bound and 
breakup states.

For an initial demonstration problem, this method was applied to the
Coulomb excitation of the deuteron in a trap by an impinging heavy ion \cite{ Du:2018tce}.
Subsequently, the more realistic Coulomb excitation of the deuteron scattering on $^{208}$Pb
at energies below the Coulomb barrier was solved and shown to be consistent with
experimental data \cite{Yin:2019kqv}.  Highly non-linear effects are found leading to
significant amplitudes for final states that are not directly populated from the ground
state by the electric dipole operator.

The role of an electromagnetic polarization potential is found to grow with increasing
bombarding energy \cite{Yin:2019kqv}.  The strength of the polarization potential
is governed by the system’s electric dipole polarizability leading to the suggestion
that Coulex, coupled with the analysis provided by tBF,  could \red{yield} precision access
to this fundamental observable property of the incident system.

Efforts are underway to include a microscopic nuclear effective  potential between
the incident system and the target in order to extend tBF to higher bombarding
energies. Applications to other light nuclear beams, such as Coulex of $^6$He,
is also underway.  

Since the tBF solves for the total amplitude of the system in the final state,
it provides a fully entangled quantum description of all possible final states.
An experiment will then correspond to a projection of that coherent superposition
defined by the specifics of the experimental apparatus.  The theoretical results
offer the opportunity to evaluate the entanglement entropy associated with that
experiment which can be analyzed as an additional metric for the validity
of the underlying Hamiltonian dynamics.


\subsection{The continuum from discrete spectra}

 \label{finvol}
In principle asymptotic normalization coefficients are observables, but difficult to measure directly.
One often relies on their inherent connection to low-energy scattering
properties encoded, for example, in the parameters of the effective range
expansion, as found in halo EFT and other methods,
but without a link to microscopic interactions.

Instead, one can obtain direct theoretical predictions of ANCs from finite-volume calculations 
generated 
in nuclear lattice EFT, similarly to lattice QCD.
In this
approach, which now can include a sizable number of
nucleons as well as the Coulomb 
interaction~\cite{Elhatisari:2015iga,Elhatisari:2016hby,Elhatisari:2017eno}, it 
could be possible in the near future to extract ANCs  for reactions such as $\alpha+^3$He or 
$^7$Be$+$p. 

The key idea behind extracting ANCs from the volume dependence of bound states 
goes back to L\"uscher, who derived~\cite{Luscher:1985dn}
that an $s$-wave bound state of two interacting particles with reduced mass $\mu$,  generated by an interaction with finite range $R$, is shifted in 
energy when it is enclosed in a cubic box of length $L$ with periodic boundary conditions.  
The volume dependent binding energy 
shift of the state is
\begin{equation}
 \Delta B(L) = -3|C|^2\frac{\mathrm{e}^{-{\kappa L}}}{\mu L}
 + \mathcal{O}\big(\mathrm{e}^{-{\sqrt2\kappa L}}\big) \,,
\label{eq:DeltaB-Luscher}
\end{equation}
where $\displaystyle\kappa=\sqrt{2\mu B}$ is the binding momentum of the state 
and $C$ denotes the ANC.  Equation~(\ref{eq:DeltaB-Luscher}) is valid for 
$L \gg R$, and the exponential form as well as the ANC occurring in the 
expression for the energy shift directly reflect that it is the 
\emph{asymptotic} (long-range) properties of the state that govern the volume 
dependence, and not the short-range details of the interaction.
While we focus here on the cubic periodic case relevant to lattice
calculations,  the same basic principle applies  generally, e.g., to infrared extrapolations of harmonic oscillator basis, where the
model-space truncation has been shown to impose an effective hard-wall boundary 
condition~\cite{More:2013rma,Furnstahl:2013vda,Furnstahl:2014hca,Wendt:2015nba}.
 Eq.~(\ref{eq:DeltaB-Luscher}) is useful as it directly relates 
the \emph{infinite-volume} quantities $\kappa$ and $C$ to the volume 
dependent energies, allowing a combined extraction of 
both from a calculation at different volumes.

While Eq.~(\ref{eq:DeltaB-Luscher}) is limited to two-body $s$-wave states, it 
can be generalized to higher angular momentum ~\cite{Konig:2011nz,Konig:2011ti}. 
%
%
Further generalization to more than two particles~\cite{Konig:2017krd}
 shows that the exponential
scale governing the asymptotic volume dependence of a given state is associated
with the nearest breakup threshold, with the shift still overall proportional to
the associated ANC if the breakup is into two particles, regardless of whether or not the state is 
described in an effective two-body halo picture.



\begin{figure}[h]
 \begin{minipage}[]{0.5\textwidth}
 \includegraphics[width=0.95 \textwidth]{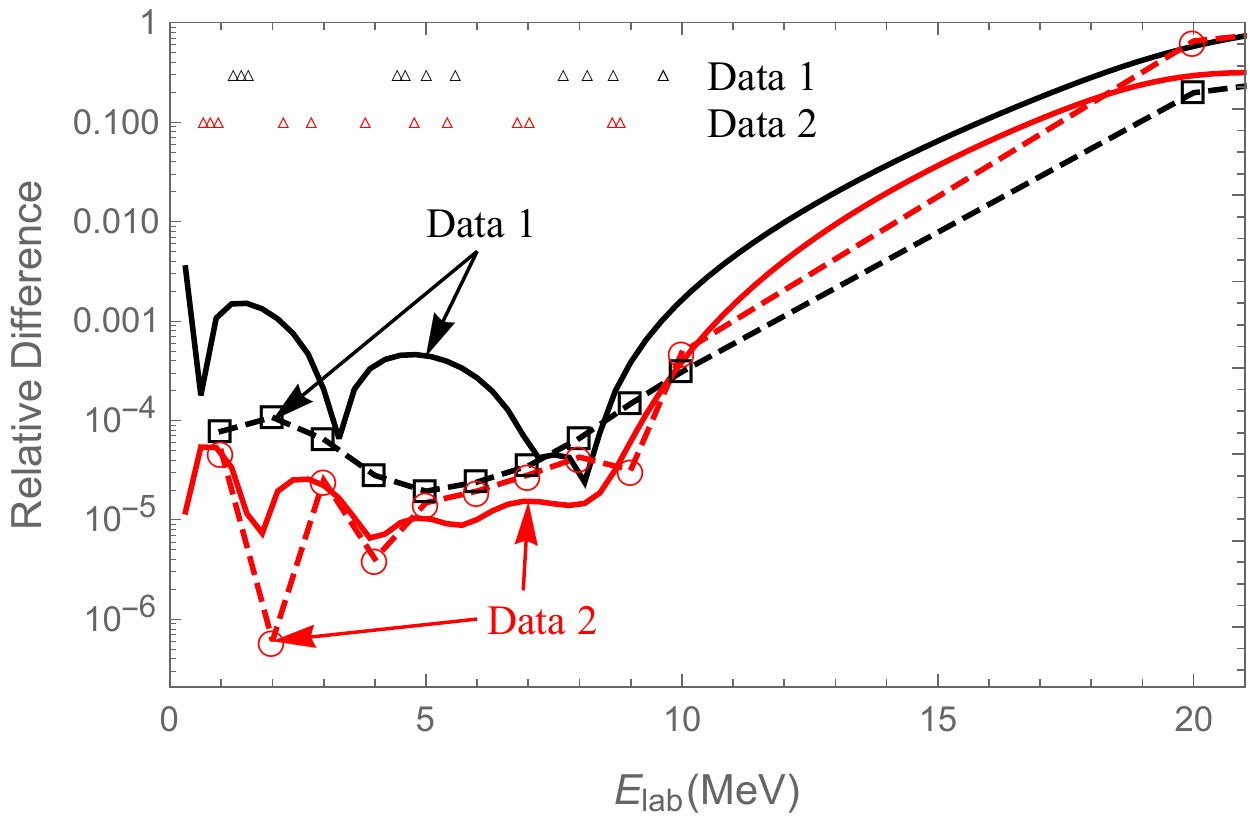}
 \end{minipage}
  \begin{minipage}[]{0.45\textwidth}
\caption{Relative error of the extracted NN $^1S_0$ scattering phase shifts (as compared to the exact ones) vs Lab energy. Two eigen-energy ``data'' sets are used: Data-1 with \red{trap frequency} $\omega=6,3, 1, 0.9, 0.8$ MeV and Data-2 with $\omega=1, 0.9, 0.8, 0.6$ MeV~\cite{Luu:2010hw,zhangvary}; the energies are marked by the triangles' x-axis values. The open symbols are the results (dashed \red{lines} to guide the eye), while the solid curves are the estimated 1-$\sigma$ error bar. }\label{XZfig:NN} 
\end{minipage}
\end{figure}

More broadly, the energy spectra of scattering/reacting hadrons, as discretized by periodic spatial boxes, have been used successfully to extract their scattering/reaction observables in lattice QCD~\cite{Luscher:1990ux,Briceno:2017max}. Other studies~\cite{Busch1998,Luu:2010hw,Rotureau:2011vf} show that this ``discrete-spectra to continuum'' strategy also works for trapping interacting particles or nuclear clusters in harmonic potential wells. Recent work~\cite{Zhang:2019cai} has improved the latter method to allow systematic control of theory errors. Based on this improved method, Fig.~\ref{XZfig:NN} demonstrates that the low-energy NN $^1S_0$ phase-shifts can be reliably extracted from the in-trap energy-levels~\cite{Luu:2010hw,zhangvary} as computed by NCSM, by showing the relative errors of the extractions as compared to the exact phase shifts. The errors are computed at the energies where the exact phase shifts are available~\cite{zhangvary}.   

This generalized L{\"u}scher method provides a compelling strategy to expand {\it ab initio} scattering/reaction calculations to medium-mass or even heavier nuclear systems, \red{as long as} their in-trap spectra can be computed using structure methods. This is demonstrated in a recent calculation of neutron-$\alpha$ and neutron-$^{24}\mathrm{O}$ scattering phase shifts based on this method~\cite{Zhang20201}. 
The next step is to apply this strategy to more complicated problems, including inelastic scattering, systems of charged particles, and systems with more than two clusters. 

\red{Applications so far have been limited to simple processes. Therefore, the first challenge is to generalize 
L\"uscher-type methods to coupled-channels and three-cluster scattering and reactions. 

A second challenge is error bars, which includes (1) errors caused by many-body-Hilbert-space truncations and (2)  errors propagated from underlying nucleon interaction, both of which are sources for error in other methods 
discussed in this paper.  Recent work~\cite{Zhang20201} has tried to extract errors in the context of L\"uscher-type methods,
but also illustrates the challenges in understanding these errors, demonstrated by large error bars in the paper.}


\subsection{Statistical reactions and level densities }

Most of this paper has focused on direct reactions, which can be understood as reactions
where the relevant states, whether bound or resonances, are well separated. But 
many relevant reactions, for astrophysics and in technological applications, involve 
the highly excited compound nucleus, where the bound states are closely spaced or resonances 
are overlapping. These are statistical reactions \cite{hofmann1975direct,PhysRevC.11.426}, and they 
can be successfully
understood only if the level density of the final states in the continuum 
is reliably evaluated \cite{voinov2007test}. 

The importance 
of the level density for understanding nuclear structure and reactions was realized in the first years of nuclear 
physics. Following Bethe, Landau and Frenkel, 
most approaches are based on the Fermi-gas models for the nucleus, where nuclear states are treated
as combinations of the particle-hole excitations \cite{gilbert1965composite}. 
Single-particle states were determined by 
semi-empirical  or density functional approaches, while collective states were added
by the random phase approximation and the construction of rotational bands,
leading 
to the so-called collective enhancement of the level density at relatively low 
energy. In most experimental analysis,  the level density of the ``back-shifted'' Fermi-gas  formula is  used,
a function of neutron $N$ and proton $Z$ numbers, excitation energy $E$, and total spin projection $M$,
\begin{equation}
\rho(E,N,Z,M)=\,\frac{1}{12\sqrt{2}a^{1/4}(E-\Delta)^{5/4}\sigma}\,e^{2\sqrt{a(E-\Delta)}-M^{2}/2\sigma^{2}}. \label{1}
\end{equation}
The level density for spin $J$ can be restored as the difference of $\rho$ for $M=J$ and
$M=J+1$.  The main parameter $a$, the back-shift $\Delta$ (supposedly due to pairing), and the
spin cut-off parameter $\sigma$ usually require empirical adjustments. 

Microscopic approaches are often based on the configuration-interaction shell model. 
These approaches either use exact diagonalization of very large Hamiltonian matrices, currently
with dimensions up to $10^{11}$, or statistical Monte Carlo methods 
(see, e.g., \cite{KooninDL97,AlhassidDKLO94,Alhassid16}). An inherent weak point 
here is the necessary truncation of orbital space. 
However,  experience shows that a shell-model exact diagonalization can predict well
the observed level density up to excitation energy about 15 MeV, or even further on, 
sufficient for  many applications. 
The total
level density in the shell-model space is the Gaussian function, as predicted by the random matrix theory, albeit for a finite space. 
One can extract thermodynamic
properties of the system (entropy, temperature, single-particle occupation numbers, etc.) and justify the ideas
of quantum chaos in real systems with no random elements (see, e.g., \cite{ZelevinskyBFH96,HoroiBZ99,HoroiGZ04,HoroiZ07,SenkovHZ11,SenkovHZ13,SenkovZ16}). 

Based on the chaotic properties of many-body dynamics with sufficiently strong interactions, 
the statistical procedures can avoid the diagonalization of prohibitively large
matrices. The ``moments'' method uses only \red{the} first two Hamiltonian moments which can be read from the matrix
itself \cite{HoroiKZ03,KotaH10,SviratchevaDV08,JOHNSON201572}.  This was successfully used for nuclei in $sd$ and $pf$ shells; with special precautions, 
the results coincide with those of complete diagonalization when the latter is practically possible. 

Most current theoretical and experimental level densities are well described by the 
``constant temperature model'' (CTM) with an exponential form rather than the Fermi-gas 
formula (\ref{1}),
\begin{equation}
\rho(E) = {\rm const}\,e^{E/T}.                                                                          \label{2}
\end{equation}
The parameter $T$ in Eq. (\ref{2}) is not an actual temperature 
kept constant while the system undergoes a phase transition. Instead, it is similar to the limiting temperature
at high energy with the exponentially rising number of resonances when the system has a crossover
to quark-gluon or string state. Such a behavior is known for the bag model of particle physics. In nuclei 
it marks the gradual transition to quantum chaos where the thermodynamic temperature of the initial stage,
\begin{equation}
T_{{\rm t-d}}=T[1-e^{-E/T}],                                                                        \label{3}
\end{equation}
reaches the value predicted by the bulk Gaussian curve. 

In practice, the parameter $1/T$ characterizes 
the rate of increase of the level density at low energy. 
As an example,  $^{28}$Si with $N=Z$ has the lowest temperature $T$ due to the presence of isospin-0 states. 
The temperature $T$ decreases if the single-particle levels are degenerate (the largest rate of chaotization 
for degenerate orbitals). The model predicts collective enhancement in the deformed case and the specific
$J$-dependence. The results for the best parameters for all $sd$ nuclei and all $J$-classes are 
tabulated \cite{KarampagiaSZ18}, while the generalization to heavier nuclei is in preparation. The main theoretical 
problems are (i)  to understand better the nature of the CTM and to give predictions for its parameters and their 
$A$- and $J$-dependence and (ii) to find ways for overcoming computational difficulties in going to heavier nuclei (see, e.g.,  
\cite{KarampagiaZ16,KarampagiaRZ17,KarampagiaSZ18,ZelevinskyKB18}).



\section{Where we have been and where we are going}


The past quarter century has seen a resurgence in microscopic  nuclear many-body theory, including the development and application
of a host of $A$-body techniques.   Many challenges 
remain related to merging  detailed many-body calculations
to reactions and other physics in the continuum. Of particular importance are applications to medium and heavy nuclei, especially unstable species, that build upon realistic inter-nucleon interactions and do not rely 
on data of stable nuclei, 
which  are and will be of interest to current and upcoming radioactive 
ion beam facilities. 


There remain many open questions \cite{Michel_2010}, and we cannot attempt to address 
them all. Two topics that threaded through many discussions were the need for 
more reliable effective inter-cluster interactions, more commonly called optical potentials, 
used for interpretation of experiments; and the importance of $A$-body calculations providing reasonable asymptotics and threshold energies.  Above all, however, we hope this work inspires and enables 
more collaboration across methods. Towards this end, in the Appendices we give more details about how to construct a Berggren basis and optical potentials.

\section*{Acknowledgements}

This work covers material discussed at a workshop sponsored by the Facility for Rare Isotope Beams Theory Alliance. 
%
%

This material is based upon work supported by the U.S.~Department of Energy, Office
of Science, Office of Nuclear Physics,  under the FRIB Theory Alliance
award DE-SC0013617, under Award Numbers DE-FG02-95ER-40934, DE-FG02-03ER41272, DE-FG02-93ER40756,  DE-FG02-97ER-41014, DE-FG02-87ER40371, 
DE-FG02-88ER40387, DE-SC0013365, DE-SC0009883, DE-SC0017887, DE-SC0018638,  DE-SC00018223, the 
NUCLEI SciDAC Collaboration under Award DE-SC0018083,  
by the U.S.~Department of Energy, National Nuclear Security Administration, under Award Numbers
 DE-NA0003883 and
DE-NA0003343,
and the National Institute for Nuclear Theory; 
by the National Science Foundation, Award Numbers PHY-1613362, PHY-1912643,  PHY-1811815, OIA-1738287, PHY-1913728, PHY-1614460,  PHY-1615092, PHY-1913620, PHY--1614460, PHY--1913069; 
by the Deutsche Forschungsgemeinschaft (DFG, German Research
Foundation) -- Projektnummer 279384907 -- SFB 1245;  by the 
International Scientific Cooperation Conicet-NSF 1225-17;  by the Russian Foundation of Basic Research grant No 20-02-00357;  by the Natural Science Foundation of China under Grants No. 11921006 and No.~11835001; and by the Royal Society and Newton Fund through the Newton International Fellowship No.~NF150402 and Crafoord foundation N.~20190607 and computing at DiRAC Data Intensive service at Leicester.
This research used resources of the National Energy Research Scientific Computing Center (NERSC), a U.S. Department of Energy Office of Science User Facility operated under Contract No. DE-AC02-05CH11231.
%





\appendix






\section{How to build a single-particle Berggren basis}\label{sec.berggren}


The general idea of the quasi-stationary formalism can be summarized as follows for the one-body problem: One starts with the time-independent Schr\"odinger equation for a given partial wave (\ref{eq_Schro}) and one looks for the solutions that are regular at the origin as defined in Eq.~(\ref{eq_reg_sol}) and with outgoing boundary conditions as defined in Eq.~(\ref{eq_out_bd_cond}).
\begin{equation}
	\frac{ {\partial}^{2} {u}_{\ell} (k,r) }{ \partial {r}^{2} } = \left( \frac{ \ell (\ell+1) }{ {r}^{2} } + \frac{2m}{ {\hbar}^{2} } V (r) - {k}^{2} \right) {u}_{\ell} (k,r).
    \label{eq_Schro}
\end{equation}

\begin{equation}
	{u}_{\ell} (k,r) \stackrel{\sim}{ r \sim 0 } {C}_{0} (k) {r}^{\ell+1}.
    \label{eq_reg_sol}
\end{equation}

\begin{equation}
	{u}_{\ell} (k,r) \stackrel{\sim}{ r \to \infty }{C}_{+} (k) {H}_{\ell,\eta}^{+} (kr) + {C}_{-} (k) {H}_{\ell,\eta}^{-} (kr)
    \label{eq_out_bd_cond}
\end{equation}
Two kinds of solutions come out of the quasi-stationary problem and both can have real or complex eigenenergies. The resonant solutions, also called Gamow or Siegert states, are associated with discrete energies and are poles of the $S$-matrix. They are often called just ``poles''. The scattering solutions correspond to the non-resonant (continuum) states and have continuous energies.

\begin{figure}[htb]
    \centering
	\includegraphics[width=0.7\textwidth]{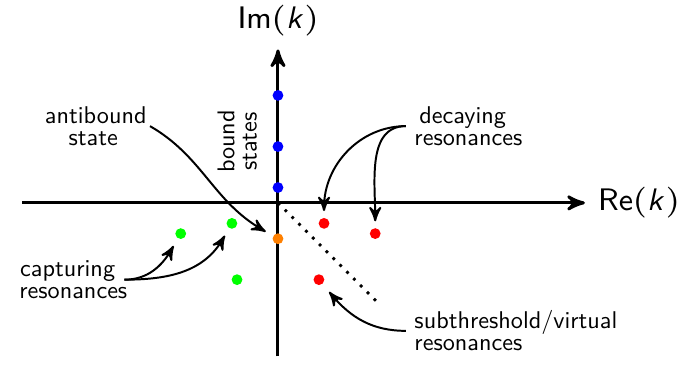}
    \caption{Zoology of $S$-matrix poles in the complex momentum plane. Bound states are on the positive part of the imaginary axis, while antibound or virtual states are on the negative part. Decaying resonances lie in the fouth quadrant and their time-reversal symmetric, the capturing resonances, lie in the \red{third} quadrant. We note that the decaying resonances below the -45 degree line (dotted line) cannot be interpreted as physical states like antibound states (negative energy, positive width), and are usually called subthreshold or virtual resonances.}
    \label{fig_poles}
\end{figure}

A convenient way to look at the resonant solutions of the quasi-stationary problem is to look at the position of their momenta in the complex momentum plane as shown in Fig.~\ref{fig_poles}.

At that point, a natural question is: Is it possible to use those states to expand any physical state similarly to the Mittag-Lefller expansion? This answer was found by T. Berggren in a groundbreaking work published in 1968 \cite{berggren1968} where he demonstrated that a single-particle completeness relation can be built using resonant states and scattering states. He later demonstrated the equivalence with the Mittag-Lefller approach \cite{berggren93_481}.

\begin{figure}[htb]
    \centering
	\includegraphics[width=0.7\textwidth]{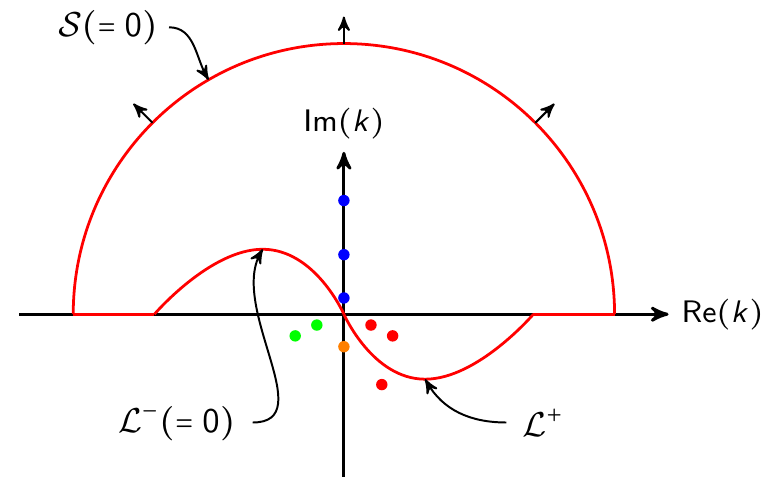}
    \caption{Construction of a typical Berggren basis using Cauchy's residue theorem. The shape of the contour could be changed to go around antibound states, providing that they are accouned for in the completeness relation.}
    \label{fig_BB_semi}
\end{figure}

The proof of the so-called Berggren basis is based on the Cauchy's residue theorem, where the Newton basis \cite{newton82_b6} that is only made of bound states and positive energy scattering states is deformed, as shown in Fig.~\ref{fig_BB_semi}, to form a complex contour that surrounds selected poles of the $S$-matrix (resonant states).

\begin{figure}[htb]
    \centering
	\includegraphics[width=0.6\textwidth]{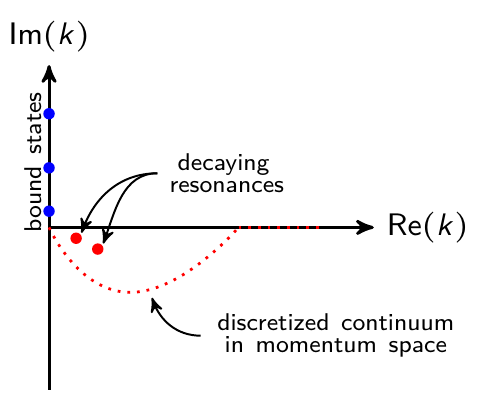}
    \caption{The non-resonant continuum states along the complex contour are discretized in momentum space, which allows to cover relatively high energies in the completeness relation.}
    \label{fig_BB}
\end{figure}

In the particular case shown in Fig.~\ref{fig_BB_semi}, only bound states, narrow decaying resonances and scattering states along the contour are considered, but the specific shape of the contour is unimportant providing that the poles inside are properly accounted for in the completeness relation. It is possible to show that if the contour is divided in three parts denoted $\mathcal{S}$, $\mathcal{L}^{-}$ and $\mathcal{L}^{+}$ in the Fig.~\ref{fig_BB_semi}, in the limit of an infinite radius only the $\mathcal{L}^{+}$ contour has a nonzero contribution in the completeness relation as defined in Eq.~(\ref{eq_BB}).

\begin{equation}
	\sum_{ n \in (b,d) } \ket{ {u}_{\ell} ( {k}_{n} ) } \bra{ \tilde{u}_{\ell} ( {k}_{n} ) } + \int_{\mathcal{L}^{+}} dk \, \ket{ {u}_{\ell} (k) } \bra{ \tilde{u}_{\ell} (k) } = {\hat{1}}_{\ell,j}.
    \label{eq_BB}
\end{equation}
The sum in Eq.~(\ref{eq_BB}) runs over the bound states \red{$b$} and decaying resonances \red{$d$} (radial part), while the integral along the contour $\mathcal{L}^{+}$ defines the non-resonant continuum. The tilde over the bras denotes the time reversal operation which is equivalent to the complex conjugate operation. As a consequence, the norm for decaying resonances is the rigged Hilbert space norm (square) as shown in Eq.~(\ref{eq_RHS_norm}), that reduces to the usual norm for bound states (square modulus).

\begin{equation}
	\mathcal{N}^{2} = \braket{ \tilde{u}_{\ell,\eta} | {u}_{\ell,\eta} } = \int_{0}^{\infty} dr \, \tilde{u}_{\ell,\eta}^{*} (r) {u}_{\ell,\eta}  (r) = \int_{0}^{\infty} dr \, {u}_{\ell,\eta}^{2} (r).
    \label{eq_RHS_norm}
\end{equation}
In practice, the Berggren basis is represented as in Fig.~\ref{fig_BB} and the contour representing the non-resonant continuum is discretized in momentum space using a quadrature technique such as the Gauss-Legendre quadrature with about 30-45 scattering states. The truncation of the contour along the real axis usually requires a maximal momentum of about ${ {k}_{\mathrm{max}} = 6 \, \mathrm{fm}^{-1} }$ to ensure the completeness.

%
The simplest method to generate a Berggren basis is based on the analytical continuation of spherical Bessel function in the complex plane. Spherical Bessel functions are solutions of the stationary Schr\"odinger equation without any potential as shown in Eq.~\ref{eq_Schro_no_pot}.

\begin{equation}
	\frac{ {\partial}^{2} {\psi}_{\ell} (k,r) }{ \partial {r}^{2} } = \left( \frac{ \ell (\ell+1) }{ {r}^{2} } - {k}^{2} \right) {\psi}_{\ell} (k,r).
    \label{eq_Schro_no_pot}
\end{equation}
Only the solutions of the first kind ${ {j}_{\ell} (kr) }$ that are regular at the origin are of interest to build basis states as defined in Eq.~\ref{eq_sph_Bessel_state}.

\begin{equation}
	{\phi}_{\ell} (kr) = \sqrt{\frac{2}{\pi}} k r {j}_{\ell} (kr).
    \label{eq_sph_Bessel_state}
\end{equation}
These basis states satisfy a special case of the Berggren basis when only positive energy scattering states are involved as in Eq.~\ref{eq_sph_Bessel_basis}.

\begin{equation}
	\int_{0}^{\infty} dr \, {\phi}_{\ell} (kr) {\phi}_{\ell} (k'r) = {\delta}_{k,k'}.
    \label{eq_sph_Bessel_basis}
\end{equation}
The trick is to deform the real axis to form contour in the complex momentum plane and to compute the basis states in Eq.~\ref{eq_sph_Bessel_state} of complex arguments $kr$. The analytical continuation of the spherical Bessel functions in the complex plane can be achieved using the recurrence relation refined in Eq.~\ref{eq_sph_Bessel_recu} (NIST 10.51(i)) and which is accurate up to $\ell = 6$.

\begin{equation}
	{f}_{n+1} (z) + {f}_{n-1} (z) = \frac{2n+1}{z} {f}_{n} (z)
    \label{eq_sph_Bessel_recu}
\end{equation}
Once the basis states are computed, any one-body potential can be diagonalized, and the resulting eigenstates form a new basis that might include one or more resonant states depending on the potential.

The second method based on the Jost function \cite{newton82_b6} requires more work but provides smaller bases for the same level of completeness, which is unimportant at that point but can be crucial in many-body applications. There are in fact two equivalent Jost function methods, each one having pros and cons. The first Jost function method presented hereafter is the one used in the Gamow shell model \cite{2,3}, while the second one can be found in Ref.~\cite{masui13_1944}.

The first Jost function method is basically a search of the zeroes of the outgoing Jost function for resonant states. One starts with the one-body stationary Schr\"odinger equation including the Coulomb potential as shown in Eq.~\ref{eq_Schro_Coul}.

\begin{equation}
	\frac{ {\partial}^{2} {u}_{\ell,\eta} (k,r) }{ \partial {r}^{2} } = \left( \frac{ \ell (\ell+1) }{ {r}^{2} } + \frac{2m}{ {\hbar}^{2} } V (r) - \frac{ 2 \eta k }{r} + {k}^{2} \right) {u}_{\ell,\eta} (k,r).
    \label{eq_Schro_Coul}
\end{equation}
One wants solutions that are regular at the origin and with outgoing boundary conditions, so at large distances, the solutions are the incoming and outgoing Hankel functions:

\begin{equation}
	{H}_{\ell, \eta}^{\pm} (z) = 
	\begin{cases}
		{F}_{\ell, \eta} (z) \mp i {G}_{\ell, \eta} (z)  \, \mathrm{for } \eta \neq 0, \\
		z [ {j}_{\ell} (z) \mp {n}_{\ell} (z)] \, \mathrm{for } \eta = 0,
	\end{cases}
    \label{eq_Hankel}
\end{equation}
which are expressed using either the Coulomb wave functions if there is a Coulomb potential, or the spherical Bessel functions if there is none. Consequently, a general solution of Eq.~\ref{eq_Schro_Coul} at large distances can be written as a linear combination of Hankel functions:

\begin{equation}
	{u}_{\ell,\eta} (k,r) = {C}^{+} (k) {H}_{\ell,\eta}^{+} (kr) + {C}^{-} (k) {H}_{\ell,\eta}^{-} (kr).
    \label{eq_sol_Hankel}
\end{equation}
The solution at large distances must match the solution at intermediate distances, and so the general solution of Eq.~\ref{eq_Schro_Coul} can then be written:

\begin{equation}
	{u}_{\ell,\eta} (k,r) = {C}^{+} (k) {u}_{\ell,\eta}^{+} (k,r) + {C}^{-} (k) {u}_{\ell,\eta}^{-} (k,r).
    \label{eq_sol_interm}
\end{equation}
The goal is then to determine the incoming and outgoing functions ${ {u}_{\ell,\eta}^{\pm} (k,r) }$ and the associated coefficients ${ {C}^{\pm} (k) }$. The Schr\"odinger equation can be integrated from zero to $r=R$ with $R$ large enough to reach the asymptotic region, and at that point one can use the matching conditions defined in Eqs.~(\ref{eq_match0},\ref{eq_match1}).

\begin{eqnarray*}
    	&\frac{d}{dr} \left( {C}^{+} (k) {H}_{\ell,\eta}^{+} (kR) + {C}^{-} (k) {H}_{\ell,\eta}^{-} (kR) \right) = \frac{ d {u}_{\ell} (k,R) }{dr}. \label{eq_match0} \\
		&{C}^{+} (k) {H}_{\ell,\eta}^{+} (kR) + {C}^{-} (k) {H}_{\ell,\eta}^{-} (kR) = {u}_{\ell} (k,R). \label{eq_match1}
\end{eqnarray*}
While these conditions are sufficient for scattering states, outgoing resonant states (bound states, decaying resonances) satisfy ${ {C}^{-} (k) = 0 }$ by definition, which means that the differentiability of the wave function is not ensured by Eqs.~(\ref{eq_match0},\ref{eq_match1}). In fact, the additional constrain for outgoing states comes from the outgoing Jost function. The incoming and outgoing Jost functions are defined as \cite{newton82_b6}:

\begin{equation}
	\mathcal{J}_{\ell}^{ \pm } (k) = W ( {u}_{\ell}^{\pm} (k,r) , {u}_{\ell} (k,r) ) = {u}_{\ell}^{\pm} (k,r) \frac{ d {u}_{\ell} (k,r) }{dr} - {u}_{\ell} (k,r) \frac{ d {u}_{\ell}^{\pm} (k,r) }{dr}.
\end{equation}
The presence of the Wronskian makes the Jost functions independent of $r$ by construction, and so numerically the differentiability of the wave function is enforced by varying $k$ until ${ \mathcal{J}_{\ell}^{+} (k) = 0 }$, which is just a search of zeroes. The advantage of this method is that the momenta of the resonant states is determined while solving the Schr\"odinger equation.

The second Jost function method \cite{masui13_1944} is relatively simpler to implement but requires to know the momenta of the resonant states to include in the Berggren basis beforehand. In practice, this information is obtained by first solving the problem using the method based on spherical Bessel functions, and then just keeping the poles of interest.

The idea of this method is to write the wave function at intermediate distances as a linear combination of Hankel functions as in Eq.~\ref{eq_sol_Hankel}, except that before the asymptotic region the coefficients ${ {C}^{\pm} (k) }$ are replaced by the "$r$-dependent" Jost functions ${ \mathcal{F}_{\ell, \eta}^{+} (r, k) }$. The trick is to have an equation for the derivative of the $r$-dependent Jost functions with initial conditions, which allows to integrate the Schr\"odinger equation from zero to any desired value.

Using the initial conditions for the $r$-dependent Jost functions and the wave function, the derivatives of the $r$-dependent Jost functions can be computed, and hence the values of the $r$-dependent Jost functions at the next point, which can be used to compute the wave function etc. At large distances, the $r$-dependent Jost functions must become constant and correspond to the Jost functions ${ \displaystyle \lim_{r \to \infty} \mathcal{F}_{\ell, \eta}^{+} (r, k) = \mathcal{J}_{\ell, \eta}^{+} (k) }$, which implies that the incoming and outgoing coefficients are given by: ${ \displaystyle {C}^{\pm} (k) = \mathcal{J}_{\ell, \eta}^{\pm} (k) / 2 }$. These coefficients are the asymptotic normalization coefficients  and can then be used to compute phase-shifts for instance.

Once all necessary resonant and scattering states have been obtained, the normalization can be achieved for scattering states \cite{michel08_146} by satisfying Eq.~\ref{eq_norm_scat} while for resonant states a regularization method must be used to compute the norm as defined in Eq.~(\ref{eq_RHS_norm}).

\begin{equation}
	{C}^{+} (k) {C}^{-} (k) = \frac{1}{2\pi}
    \label{eq_norm_scat}
\end{equation}
Several regularization methods can be used in principle to normalized a decaying resonance state, but the exterior complex-scaling method \cite{dykhne61_1041,gyarmati71_38,simon79_436} appears as a relatively convenient approach as it can be applied on any wave function. The norm is then computed as follows:

\begin{eqnarray*}
	\mathcal{N}^{2} &= \int_{0}^{R} dr \, {u}_{\ell,\eta}^{2} (r) + {({C}^{+} (k))}^{2} \int_{R}^{\infty} dr \, {( {H}_{\ell,\eta}^{+} (kr) )}^{2}, \nonumber \\
    &= \mathcal{I}_{R} + {({C}^{+} (k))}^{2} \int_{0}^{\infty} dx \, {( {H}_{\ell,\eta}^{+} (k[R + x{e}^{i\theta}]) )}^{2} {e}^{i\theta},
    \label{eq_norm_pole}
\end{eqnarray*}
where the tail of the wave function is rotated in the asymptotic region by a given angle until the integral is regularized. Once this is achieved, the result does not depend on the angle $\theta$ and the position $R$ around which the rotation is done.

As a final remark, one of the biggest difficulty with the Jost function methods is to compute the Hankel functions in the complex plane. So far, it seems that such functions are only available in two publications \cite{thompson86_616,michel07_145}.

\section{How to build an effective nucleon-nucleus (`optical') potential via Green's functions}
\label{App:Green}

In principle, Green's functions can be generated post-hoc using the many-body method of choice (cf. also Sec. \ref{Sect:Green}). In this appendix we outline some notable points and the necessary formalism to generate the self-energy from a many-body calculation.

The Green's function,
\begin{equation}
G(\alpha, \beta; t-t_0 ) = - \frac{i}{\hbar} \int \sum u_n (\alpha) u^*_n (\beta) 
e^{-\frac{i}{\hbar}\varepsilon_n (t-t_0)},
\end{equation}
represents a non-local, one-body, time-dependent Green's function relevant for a single-particle problem. The integral sum runs over eigenstates \red{indexed by $n$} of the spectrum with eigenvalues $\varepsilon_n$ and eigenfunctions $u_n$, and $\alpha, \beta$ are relevant complete sets of quantum numbers (\textit{e.g.} $\bm{r}, \bm{r}'$, $\bm{k}, \bm{k}'$ or harmonic oscillator quantum numbers $N,N', \ell,j$). 
Imposing $t > t_0$, that is causality in time-forward propagation, the time-dependent Green's function can be rewritten through the Fourier transform of the Heaviside $\Theta (t-t_0)$. This generates the spectral representation of the forward Green's function,
\begin{equation}
G(\alpha, \beta; E) = \int \sum \frac{u_n (\alpha) u^*_n (\beta)}{E-\varepsilon_n + i\eta},
\end{equation}
with $\eta \rightarrow 0$.
	
In the nuclear many-body problem, the reference state is usually the ground state of a doubly closed shell nucleus. Therefore, a particle can be both added to and removed from the ground state. It is important to consider both forward propagating and backward propagating Green's functions, reducing to the familiar K\"all\'en-Lehmann spectral representation \cite{kallen52,Lehmann54}, that is, for a nucleus of $A$ nucleons,
\begin{equation}
G(\alpha, \beta; E) = \sum \frac{u_p (\alpha) u^*_p (\beta)}{E - (E^{A+1}_p - E^A_0) + i\eta} + \frac{v_h (\alpha) v^*_h (\beta)}{E - (E^A_0 - E^{A-1}_h) - i\eta} .
\label{eq:prop}
\end{equation}
The eigenvalues $E^{A+1}_p$ and $E^{A-1}_h$ correspond to the eigenstates of the particle ($A+1$) and hole ($A-1)$ configurations, respectively.
The particle and hole amplitudes $u_p, v_h$ are given by 
\begin{equation}
u_p (\alpha) = \langle \psi^A_0 | a_\alpha | \psi^{A+1}_p \rangle, \qquad v_n (\alpha) = \langle \psi^A_0 | a^\dagger_\alpha | \psi^{A-1}_h \rangle,
\label{eq:dens}
\end{equation}
with $a_\alpha$ and $a^\dagger_\alpha$ representing the addition and removal operator for the chosen basis $\{\alpha\}$ and $| \psi^{A}_h \rangle$ are the relevant many-body $A$ particle wavefunction of the state $h$, making $u_p (\alpha)$ the overlap function of basis state $\alpha$.

Consequently, the effective nucleon-nucleus interaction can be calculated through the inversion of the Dyson equation (\ref{eq:Dyson}),
\begin{equation}
\Sigma^\star = (G^0)^{-1} - G^{-1},
\label{eq:DysonInv}
\end{equation}
where $G^0$ and $G$ are the bare and dressed propagators respectively.

An example of procedure to construct the self energy from a general many-body method can be:
\begin{enumerate}
\item {\it Construct the bare single-particle Green's function in the given many-body context}. That is, consider Eq. (\ref{eq:prop}) and $E^{A}_0$ as the unperturbed ground state of the system with $A$ particles (e.g. harmonic oscillator filled to some $sd$ shell closure) and $E^{A+1}_n$ as the possible states of the system with $A+1$ particles (e.g. harmonic oscillator filled to $sd$ shell closure, $+1$ particle in each of the possible states in the space under consideration, $1p_{3/2}$ state, $1p_{1/2}$ state, etc...). Densities are given from the chosen basis states. In the unperturbed case this will give the fully occupied or unoccupied states of the naive shell model in harmonic oscillator basis.

\item {\it Construct, in the same basis, the dressed single-particle Green's function from the given many-body calculation}. That is, considering Eq. (\ref{eq:prop}), using the densities (\ref{eq:dens}) and ground $A$-particles and excited $A+1$-particles states coming from the many-body wavefunction. In this example one can consider the wavefunction of the ground and excited states in the $sd$-shell. Calculating all excited states to build $E^{A+1}_n$, and the corresponding overlap function.

\item {\it Invert the Dyson equation} (\ref{eq:DysonInv}) to obtain the self-energy, which is the non-local, generalized, nucleon-nucleus optical potential in the chosen basis.

\item Eventually, {\it solve the Dyson equation} (\ref{eq:Dyson}) for the obtained self-energy, at the same many-body expansion to verify convergence of the propagators.
\end{enumerate}

Once the self-energy is obtained, it is possible to exploit its correlation content to explore the single-particle spectrum and spectral density, nonlocality, volume integrals of the imaginary part, and to provide a check for dispersion relations.

\bibliographystyle{h-physrev5}
\bibliography{mainRefs}


\end{document}